\documentclass[preprint2]{aastex}

\usepackage{graphics}
\usepackage{mathptmx}
\usepackage{natbib}
\usepackage{amssymb}
\usepackage[usenames,dvips]{color}

\newcommand{\feh}{[\rm{Fe}/\rm{H}]}

\newcommand{\msun}{$M_\odot$}

\newcommand{\hi}{H\,{\textsc i}\rm}
\newcommand{\mgi}{Mg\,{\textsc i}}
\newcommand{\mgii}{Mg\,{\textsc {ii}}}
\newcommand{\mgiii}{Mg\,{\textsc {iii}}}
\newcommand{\fei}{Fe\,{\textsc i}}

\newcommand{\Mg}[5]{\mbox{$#1\,^#2{\rm #3}^{{\rm #4}}_{\rm #5}$}}
\newcommand{\Teff}{T_{\rm eff}}
\newcommand{\EW}{W_{\lambda}}
\newcommand{\mA}{{\rm m\AA}}
\newcommand{\Elow}{E_{\rm low}}
\newcommand{\Eup}{E_{\rm up}}
\newcommand{\Vmic}{\xi_{\rm t}}

\newcommand{\kms}{km s$^{-1}$}

\slugcomment{To appear in Astrophysical Journal}

\shorttitle{Mg NLTE effects}
\shortauthors{Bergemann et al.}

\begin{document}


\title{Red Supergiant Stars as Cosmic Abundance Probes. III. NLTE effects in 
J-band Magnesium lines}


\author{Maria Bergemann}
\affil{Max-Planck Institute for Astronomy, 69117, Heidelberg, Germany}
\email{bergemann@mpia-hd.mpg.de}
\author{Rolf-Peter Kudritzki}
\affil{Institute for Astronomy, University of Hawaii, 2680 Woodlawn Drive,
Honolulu, HI 96822}
\email{kud@ifa.hawaii.edu}
\author{Zach Gazak}
\affil{Institute for Astronomy, University of Hawaii, 2680 Woodlawn Drive,
Honolulu, HI 96822}
\email{zgazak@ifa.hawaii.edu}
\author{Ben Davies}
\affil{University of Liverpool, UK}
\email{bdavies@ast.cam.ac.uk}
\and
\author{Bertrand Plez}
\affil{Laboratoire Univers et Particules de Montpellier, Universit\'e
Montpellier 2, CNRS, F-34095 Montpellier, France}
\email{bertrand.plez@univ-montp2.fr}



\begin{abstract}
Non-LTE calculations for \mgi\ in red supergiant stellar atmospheres are 
presented to investigate the importance of non-LTE for the formation of \mgi\ 
lines in the NIR J-band. Recent work using medium resolution spectroscopy of 
atomic lines in the J-band of individual red supergiant stars has demonstrated 
that technique is a very promising tool to investigate the chemical composition 
of the young stellar population in star forming galaxies. As in previous work, 
where non-LTE effects were studied for iron, titanium and silicon, substantial 
effects are found resulting in significantly stronger \mgi\ absorption lines. 
For the quantitative spectral analysis the non-LTE effects lead to magnesium 
abundances significantly smaller than in LTE with the non-LTE abundance 
corrections varying smoothly between $-$0.4 dex and $-$0.1 dex for 
effective temperatures between 3400 K and 4400 K. We discuss the 
physical reasons of the non-LTE effects and the consequences for extragalactic 
J-band abundance studies using individual red supergiants in the young massive 
galactic double cluster h and $\chi$ Persei.
\end{abstract}





\section{Introduction}

Over the last years the quantitative spectroscopic analysis of medium 
resolution (R $\sim$ 3200) J-band spectra of red supergiant stars (RSGs) has 
been established as a very promising tool to investigate the chemical evolution 
of star forming galaxies. RSGs emit most of their enormous luminosities of 
$10^{5}$ to $\sim 10^{6}$ L$_{\odot}$ at infrared wavelengths and can be 
easily identified as individual sources through their brightness and colors 
(Humphreys and Davidson, 1979, Patrick et al., submitted). Their J-band spectra 
are characterized by strong and isolated atomic lines of iron, titanium, 
silicon and magnesium, ideal for medium resolution spectroscopy, in particular 
because the  molecular lines of OH, 
H$_{2}$O, CN, and CO which otherwise dominate the H- and K-band are weak. 
Detailed recent studies of RSGs in the Milky Way and the Magellanic Clouds 
(Davies et al., 2013, 2014, Gazak et al., 2014b), in the Local Group dwarf 
galaxy NGC 6822 (Patrick et al., 2014) and in the Sculptor group galaxy NGC 300 
at 1.9 Mpc (Gazak 2014c) demonstrate that this new medium resolution 
J-band technique has an enormous potential and yields stellar metallicities 
with an accuracy of $\sim$ 0.10 dex per individual star. With present-day NIR 
multi-object spectrographs attached to large telescopes such as MOSFIRE/KECK 
and KMOS/VLT, galaxies up to 10 Mpc can be studied in this 
way to determine metallicities and metallicity gradients providing an important 
alternative to the use of blue supergiant stars (see for instance Kudritzki et 
al., 2012, 2013, 2014) or HII regions (see as examples Bresolin et al., 2011, 
2012). In addition, Gazak et al. (2013, 2014a) have shown that the integrated 
J-band light of young massive super star clusters (SSCs) is dominated by their 
population of RSGs as soon as they are older than 7 Myr and that the same 
analysis technique can be applied increasing the potential volume for 
metallicity determinations in the local universe by a factor of thousand. 
With the use of future adaptive optics (AO) MOS IR spectrographs at the next 
generation of extremely large telescopes the J-band method will become even 
more powerful and will render the possibility to measure stellar metallicities 
of individual RSGs out to the enormous distance of 70 Mpc (Evans et al., 2011).

A crucial aspect of the spectroscopic J-band analysis technique is to account 
for the effects of departures from local thermodynamic equilibrium (LTE) which 
if neglected at the low densities of RSG atmospheres could introduce 
significant 
systematic errors. In two previous papers (Bergemann et al., 2012 and 2013 - 
hereafter Paper I and II) we have carried out non-LTE (NLTE) line formation 
calculations for iron, titanium and silicon and investigated the 
strengths of NLTE effects which were found to be moderate for iron, but 
substantial for titanium and silicon. In this third paper we extend this work 
to 
magnesium which shows two strong absorption line features in the J-band arising 
from highly excited levels which can provide important information 
on stellar metallicity and the ratio of $\alpha$ to iron elements. We 
describe the atomic model and details of the magnesium line formation 
calculations in Section 2 and present a discussion of the basic NLTE-effects in 
Section 3. In Section 4 we calculate NLTE abundance corrections. In   
Section 5 we compare with observations for a few selected RSG objects in Per 
OB1 and discuss the consequences of including \mgi\ lines for the J-band 
technique.

\section{Model atmospheres, line formation calculations, model atom and 
spectrum synthesis}

\subsection{Model atmospheres and line formation}

The NLTE line formation calculations require an underlying model atmosphere 
which provides the temperature and density stratification together with the 
number densities of the most important atomic and molecular species 
contributing 
to the continuous and line background opacities which affect the radiation 
field 
in the magnesium radiative transitions. As in Paper I and II we use MARCS model 
atmospheres (Gustafsson et al., 2008) for this purpose and calculate a small 
grid of models assuming a stellar mass of 15 \msun with five effective 
temperatures (T$_{\rm eff}$ = 3400, 3800, 4000, 4200, 4400~K), three gravities 
($\log g = 1.0$, $0.0$, $-0.5$ (cgs)), three metallicities 
([Z]\footnote{Hereafter we adopt the notation of [Z] to represent 
stellar metallicity following our series of papers (Bergemann 2012, 2013); this 
notation is identical to [Fe/H], the relative abundance of iron.}$\,\equiv\,$ 
log Z/Z$_{\odot}$ $= -0.5$, $0.0$, $+0.5$). Two values are adopted for the 
microturbulence, $\xi_{t} = 2$ and $5$ km/s, respectively. As discussed in 
Papers I and II, this grid covers the range of atmospheric parameters expected 
for RSG's and allows us to assess the importance of NLTE effects over this 
range. In addition, we also compute model atmospheres for the Sun and Arcturus 
as an additional test of our magnesium model atom.

The NLTE occupation numbers for magnesium are then calculated using the NLTE 
code DETAIL \citep{butler85}. The final J-band spectrum synthesis is carried 
out with the separate code SIU (Reetz 1999) in a slightly modified version as 
described in Paper I. For all further details we refer the reader to 
Papers I and II. It is important to use the most up-to-date 
linelists in our spectroscopic diagnostics methods. In this work, we have thus 
also implemented the new $^{12}$C$^{14}$N linelist \citep{brooke}. However we 
note that molecular contamination in the J-band is minimal (e.g. Davies et al. 
2009) and this improvement will not impact our previous results 
presented in Papers I and II.
\subsection{Mg model atom and statistical equilibrium 
calculations}{\label{sec:atom}} 

\begin{figure*}
\includegraphics[width=0.6\textwidth,angle=90]{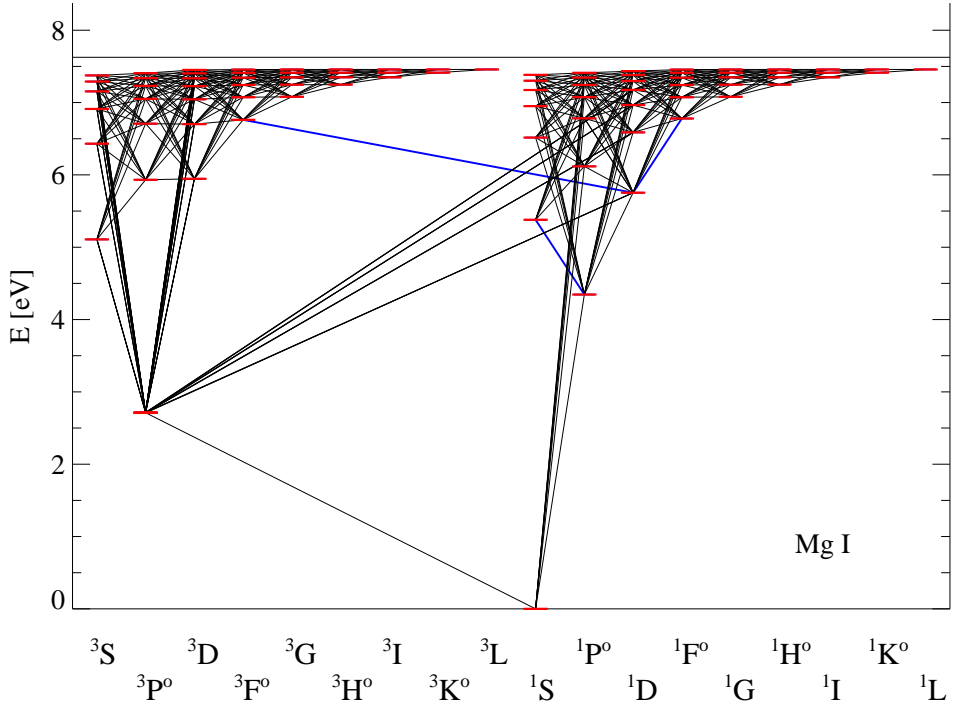}
\caption{The atomic model for the first ionisation stage of magnesium. 
The observed J-band line transitions are highlighted in blue.}
\label{grotrian}
\end{figure*}

Our atomic model consists of three ionisation stages \mgi, \mgii, and \mgiii, 
represented by 85, 6, and 1 levels respectively. The number of radiative 
transitions in the 1st and 2nd stages is 422 and 8. This model was 
first described in \citet{zhao} and Zhao \& Gehren (2000), and updated by 
\citet{mash13}. Electron-impact excitation is computed from the rate 
coefficients by Mauas et al. (1988), where available, otherwise \citet{zhao} is 
used for the remaining transitions. Ionisation by electronic collisions was 
calculated from the Seaton (1962) formula with a mean Gaunt factor set equal to 
g $= 0.1$ for \mgi\ and to $0.2$ for \mgii. For 
\hi\ impact excitations and charge transfer processes, rate coefficients were 
taken from the detailed quantum mechanical calculations of \citet{barklem12}. 
The transition probabilities for radiative bound-bound transitions were taken 
from Opacity project (Butler et al. 1993). The same source provides 
photoionisation cross-sections from the lowest \mgi\ levels; for the remainder, 
we use the quantum defect formulae of Peach (1967). Fig. \ref{grotrian} shows 
the atomic model of neutral magnesium with the observed J-band line transitions 
highlighted 
in blue. 
\subsection{Atomic data and spectrum synthesis of J-band magnesium lines}
\begin{deluxetable}{cccccccc}
\tabletypesize{\scriptsize}
\tablecolumns{8}
\tablewidth{0pt}
\tablecaption{\mgi\ J-band lines}
\label{tab1}

\tablehead{
\colhead{Elem.} &
\colhead{$\lambda$} &
\colhead{$\Elow$} &
\colhead{lower} &
\colhead{$\Eup$} &
\colhead{upper} &
\colhead{$log~gf$} &
\colhead{$log~C_6$} \\
\colhead{} &
\colhead{\AA} &
\colhead{[eV]} &
\colhead{conf.} &
\colhead{[eV]} &
\colhead{conf.} &
\colhead{}      &
\colhead{} \\[1mm]
\colhead{(1)}	&
\colhead{(2)}	&
\colhead{(3)}	&
\colhead{(4)}	&
\colhead{(5)}	&
\colhead{(6)}	&
\colhead{(7)}  &
\colhead{(8)} }	
\startdata
\\[-1mm]
    & 11828.185 & 4.346096 & 3p $^1$P$_1^{\mathrm o}$  & 5.394090 & 4s $^1$S$_0$          & $-$0.305 & $-$29.7 \\
    & 12083.278 & 5.753635 & 3d $^1$D$_2$           & 6.779505 & 4f $^3$F$_3^{\mathrm o}$ & $-$1.500 & $-$29.3 \\
    & 12083.346 & 5.753635 & 3d $^1$D$_2$           & 6.779498 & 4f $^3$F$_2^{\mathrm o}$ & $-$1.500 & $-$29.3 \\
    & 12083.662 & 5.753635 & 3d $^1$D$_2$           & 6.779472 & 4f $^1$F$_3^{\mathrm o}$ & ~~0.050 & $-$29.3 \\
\enddata
\end{deluxetable}

The basic information about the observed magnesium lines in the J-band is
given in Table 1. The four lines belong to multiplets 175, 224, and 225 
(multiplet numbers taken from the NIST online database). The 
line at 11828.185 \AA\ forms in the transition between the  
\Mg{3p}{1}{P}{\circ}{} and \Mg{4s}{1}{S}{}{} levels. The three lines around 
12083  \AA\ originate in the transitions between \Mg{3d}{1}{D}{}{} - 
\Mg{4f}{1}{F}{\circ}{}, and \Mg{3d}{1}{D}{}{} - \Mg{4f}{3}{F}{\circ}{} levels. 
In a typical observed spectrum of a late-type star, the three lines from 
multiplets $224$ and $225$ merge and are thus unresolved. Hereafter, we refer 
to 
them as a single line at 12083 \AA. However, in the line formation and spectrum 
synthesis calculations they are treated correctly as three individual lines 
(see Fig. \ref{prof_rsg_12083}).
%
%
%
\begin{deluxetable}{cccccccc}
\tabletypesize{\scriptsize}
\tablecolumns{8}
\tablewidth{0pt}
\tablecaption{Stellar parameters of the reference and Per OB1 stars}
\tablehead{
\colhead{Star} &
\colhead{$\Teff$} &
\colhead{$\Delta \Teff$} &
\colhead{$\log g$} &
\colhead{$\Delta \log g$} &
\colhead{[Z]} &
\colhead{$\Delta$ [Z]} &
\colhead{$\xi_t$} 
\\[1mm]
\colhead{} &
\colhead{K} &
\colhead{} &
\colhead{dex} &
\colhead{} &
\colhead{dex} &
\colhead{} &
\colhead{km$/$s} 
\\[1mm]
\colhead{(1)} &
\colhead{(2)} &
\colhead{(3)} &
\colhead{(4)} &
\colhead{(5)} &
\colhead{(6)} &
\colhead{(7)} &
\colhead{(8)} 
}
\startdata
\\[-1mm]
Sun        & 5777  &  1   & $~$4.44  &  0.00  &  $~$0.00  &  0.05  &  1.00  \\
Arcturus   & 4286  &  35  & $~$1.64  &  0.06  &  $-$0.52  &  0.08  &  1.50  \\
BD +56 595 & 4060  &  25  & $~$0.20  &  0.70  &  $-$0.15  &  0.13  &  4.00  \\
BD +56 724 & 3840  &  25  & $-$0.40  &  0.50  &  $~$0.08  &  0.09  &  3.00  \\
BD +59 372 & 3920  &  25  & $~$0.50  &  0.30  &  $-$0.07  &  0.09  &  3.20  \\
HD 13136   & 4030  &  25  & $~$0.20  &  0.40  &  $-$0.10  &  0.08  &  4.10  \\
HD 14270   & 3900  &  25  & $~$0.30  &  0.30  &  $-$0.04  &  0.09  &  3.70  \\
HD 14404   & 4010  &  25  & $~$0.20  &  0.40  &  $-$0.07  &  0.10  &  3.90  \\
HD 14469   & 3820  &  25  & $-$0.10  &  0.40  &  $-$0.03  &  0.12  &  4.00  \\
HD 14488   & 3690  &  50  & $~$0.00  &  0.20  &  $~$0.12  &  0.10  &  2.90  \\
HD 14826   & 3930  &  26  & $~$0.10  &  0.20  &  $-$0.08  &  0.07  &  3.70  \\
HD 236979  & 4080  &  25  & $-$0.60  &  0.30  &  $-$0.09  &  0.09  &  3.10  \\   
\\
\enddata
\end{deluxetable}

Other than for the optical spectral range where we use oscillator strengths 
from Chang \& Tang (1990) and Aldenius et al. (2007), there are no good 
experimental $\log gf$ values for the IR Mg I lines. For the $11828.185$ \AA~ 
\mgi\ line,  Civis et al (2013) calculated $\log gf = -0.292$, while Tachiev \& 
Froese Fischer 
(2003)\footnote{\url{http://physics.nist.gov/cgi-bin/ASD/lines1.pl}} provide 
$\log gf = -0.333$. We adopt the mean of the two values from the two references, 
i.e $\log gf = -0.305$.  

\begin{figure*}
\hbox{
\includegraphics[width=0.7\columnwidth, angle=-90]{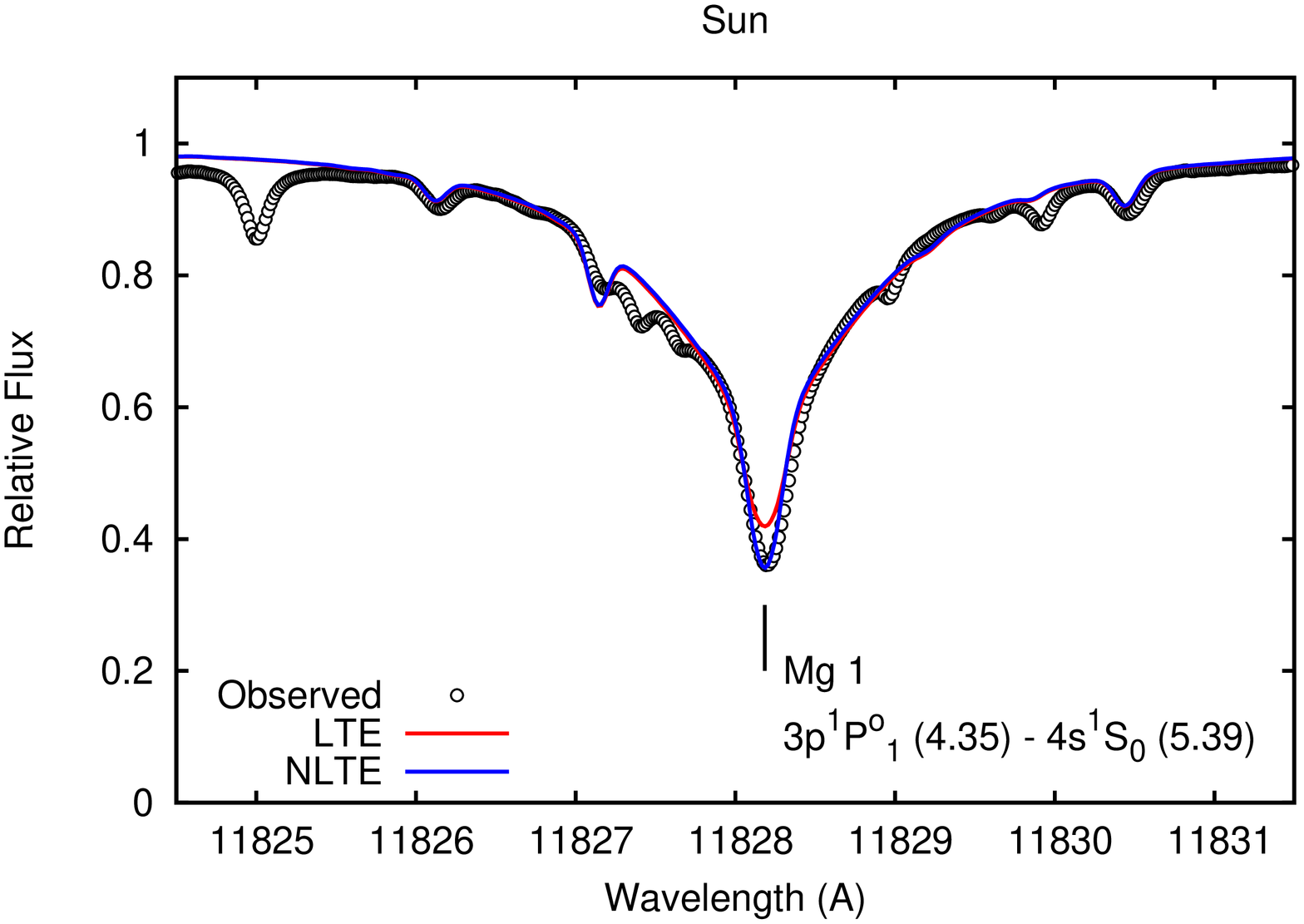}
\includegraphics[width=0.7\columnwidth, angle=-90]{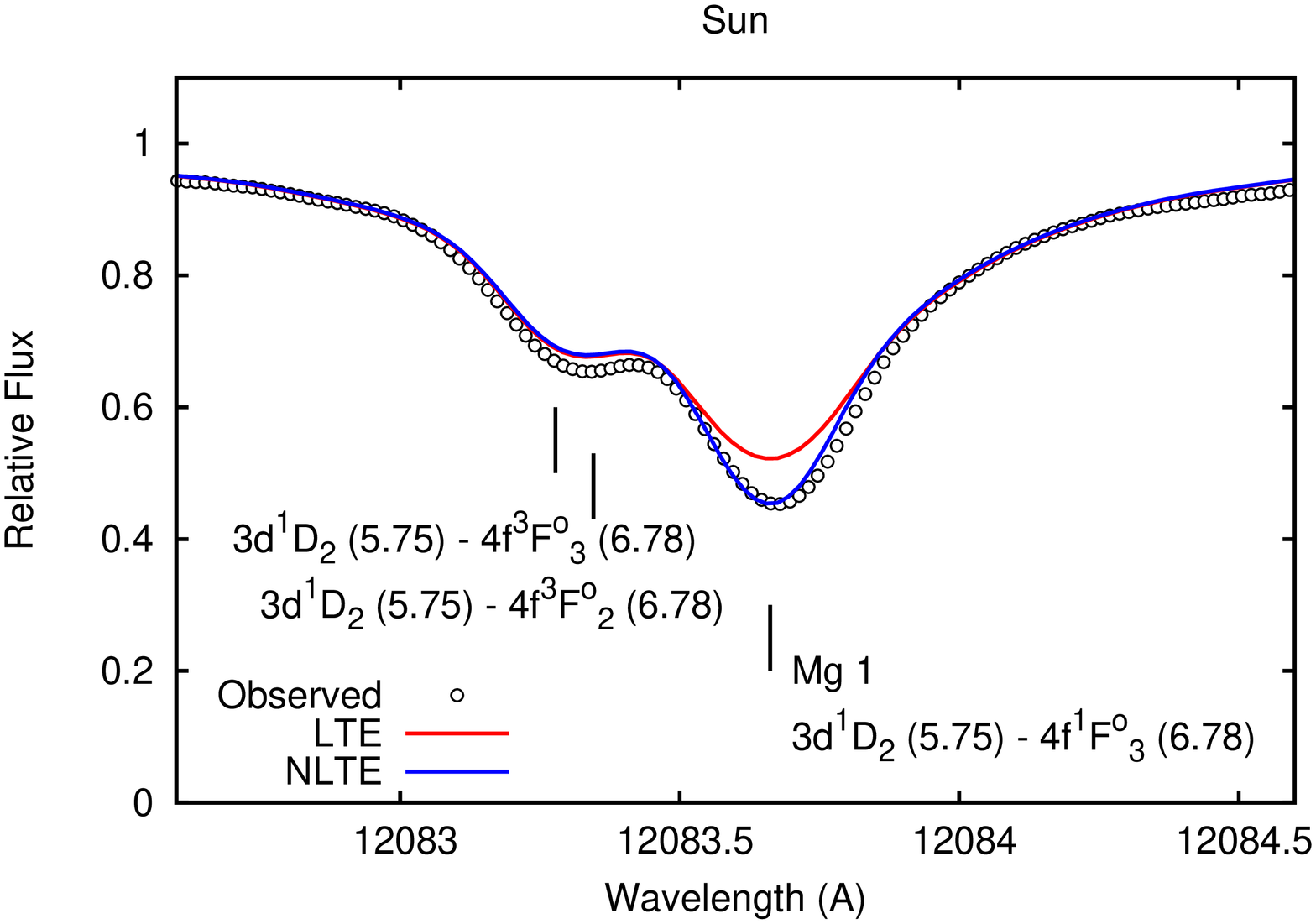}
}
\hbox{
\includegraphics[width=0.7\columnwidth, angle=-90]{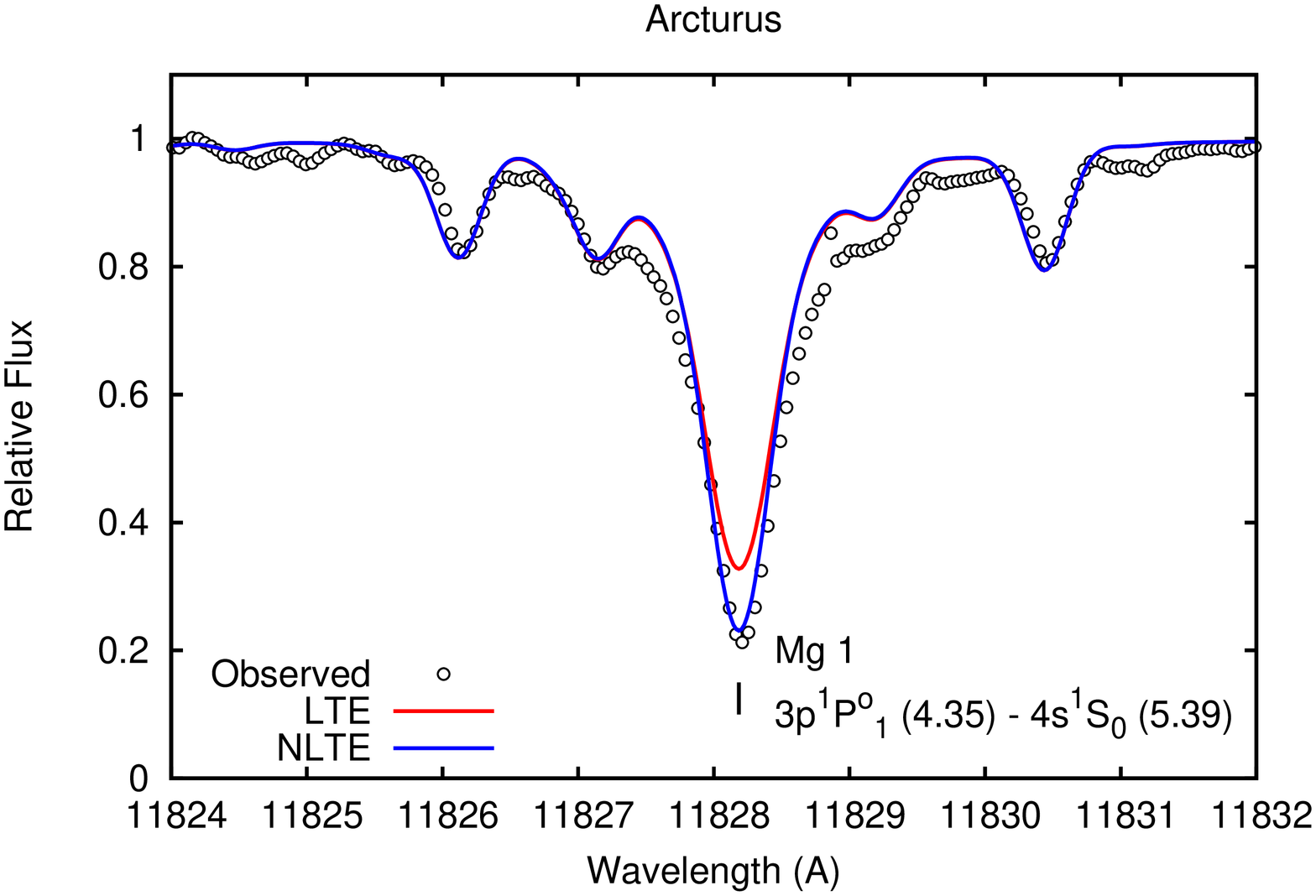}
\includegraphics[width=0.7\columnwidth, angle=-90]{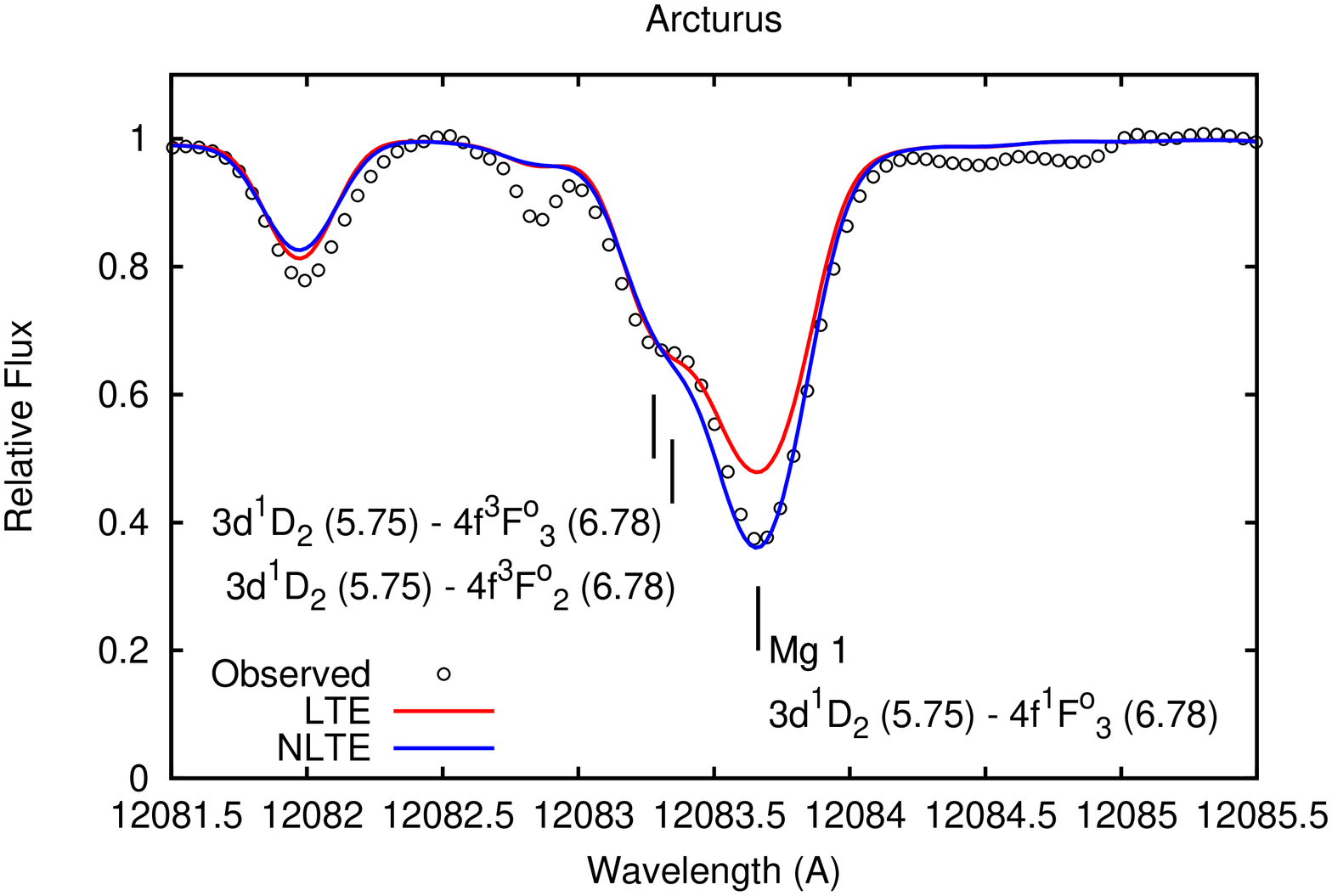}
}
\caption{NLTE and LTE line profiles of the \mgi\ 11828 \AA\ 
(left) and 12083 \AA\ (right) computed for the Sun (top panels) and Arcturus 
(bottom panels).}
\label{profiles}
\end{figure*}

For the triplet at $12083$ \AA, the available data are contradictory. In the 
Kurucz online 
database\footnote{\url{http://kurucz.harvard.edu/atoms/1200/gf1200.pos}}, 
we find  $-2.155$,  $-9.300$, $0.553$ for the three lines $12083.278$, 
$12083.346$, $12083.662$, respectively. For the transition at 12083.662 \AA,  
theoretical calculations by Chang \& Tang (1990) and Civis et al. (2013) 
provide $\log gf = 0.41$ and $\log gf = 0.44$, respectively. Our previous RSG 
synthetic grid (see Paper I) included the following values $0.45$,  $-0.79$, 
$0.415$ for the $12083.264$, $12083.346$, $12083.662$, respectively. These 
datasets come from the VALD database \citep{kupka00}\footnote{also to be found 
under 
\url{http://www.pmp.uni-hannover.de/cgi-bin/ssi/test/kurucz/sekur.html}},  
however, as our test calculations have shown, they hugely over-estimate the 
line depths in the spectrum of the Sun. The values appear 
to be wrong, e.g. the $\log gf$ value for the semi-forbidden line at 12083.278 
\AA\ is as large as the value of the allowed counterpart at 12083.662 \AA. In 
addition, there is a large differences between the Kurucz' quantum-mechanical 
values listed in different online tables for the 12083.278 and 12083.346 lines. 
Given these largely conflicting $gf$-values we decided to extend our NLTE 
calculations to the Sun with the well established magnesium abundance of $\log 
(Mg/H) + 12 = 7.53$ (Asplund et al. 2009). We then iterate the $\log gf$ values 
until we obtain a satisfactory fit of the 12083 \AA\ Mg triplet line 
complex. We obtain as best fitting values $-1.5$\footnote{The same $\log gf$ 
values are adopted for the two bluer components because they are 
indistinguishable in the observed spectra.} for the bluer weak components and 
$0.05$ for the strong red component, respectively. We then carry out Mg NLTE 
calculations for the red giant star Arcturus using the stellar parameters from 
\citet{bergemann12b} as given in Table 2. The observed spectra for the 
Sun and Arcturus were taken from \citet{kur84} and \citet{hinkle95}, 
respectively. For Arcturus, we adopt a slight $\alpha$-enhancement, 
[Mg/Fe]$=0.2$, which is also confirmed by the 
earlier work on abundances by Ramirez \& Allende Prieto (2011). The comparison 
with observed J-band Mg lines for the Sun and Arcturus is shown in Fig. 
\ref{profiles} and indicates reasonable agreement.

Calculating the relatively strong J-band magnesium lines as in Fig. 
\ref{profiles} also requires the inclusion of spectral line broadening. 
We account for broadening caused by various mechanisms: microturbulence 
(see Table 2), macro-turbulence, and broadening due to elastic collisions with 
\hi\ atoms. We tested the $\alpha$ and $\sigma$ coefficients from the 
\citet{barklem00} database of quantum-mechanically calculated values, however 
these data produce 
spectral lines which are too weak compared with the observed spectra of our 
reference stars (Sun, Arcturus). We thus scale these values by $+0.3$ dex. The 
adopted atomic data are given in Table 1. For macroturbulence
we assume the radial-tangential profile, and fit this parameter independently
of instrumental profile and rotation. For the red supergiants, we fit a total 
FWHM which consists of instrumental FWHM and macroturbulence, as the broadening 
effects cannot be separated in the spectra. The values of the FWHM were taken 
from \citet{gazak14b}.
\section{NLTE effects in J-band Mg lines}

In general, the NLTE effects which we encounter for our grid of RSG atmospheres 
are very similar to those found by previous studies of cool FGK stars (see 
\citealt{zhao}, \citealt{shim}, \citealt{merle} and references therein). The 
driving mechanism is normal 
photospheric photoionisation of neutral magnesium governed by the super-thermal 
radiation field which escapes the deep photospheres essentially unchanged once 
its optical depth has dropped below unity. This mechanism works independently 
of the model atom, of course as long as photoionisation cross-sections are not 
set to zero. We thus shall not repeat the complete analysis as done in Zhao et 
al (1998), but only summarise the most important aspects relevant for the 
atmospheres of RSGs. For the discussion of NLTE effects, it is convenient to 
employ the concept of energy level departure coefficients, which are defined as
\begin{equation}
b_i = n_i^{\rm NLTE}/n_i^{\rm LTE}
\end {equation}
where $n_i^{\rm NLTE}$ and $n_i^{\rm LTE}$ are NLTE and LTE atomic level 
populations [cm$^{-3}$], respectively.

\begin{figure*}
\hbox{
\includegraphics[width=0.3\textwidth, angle=-90]{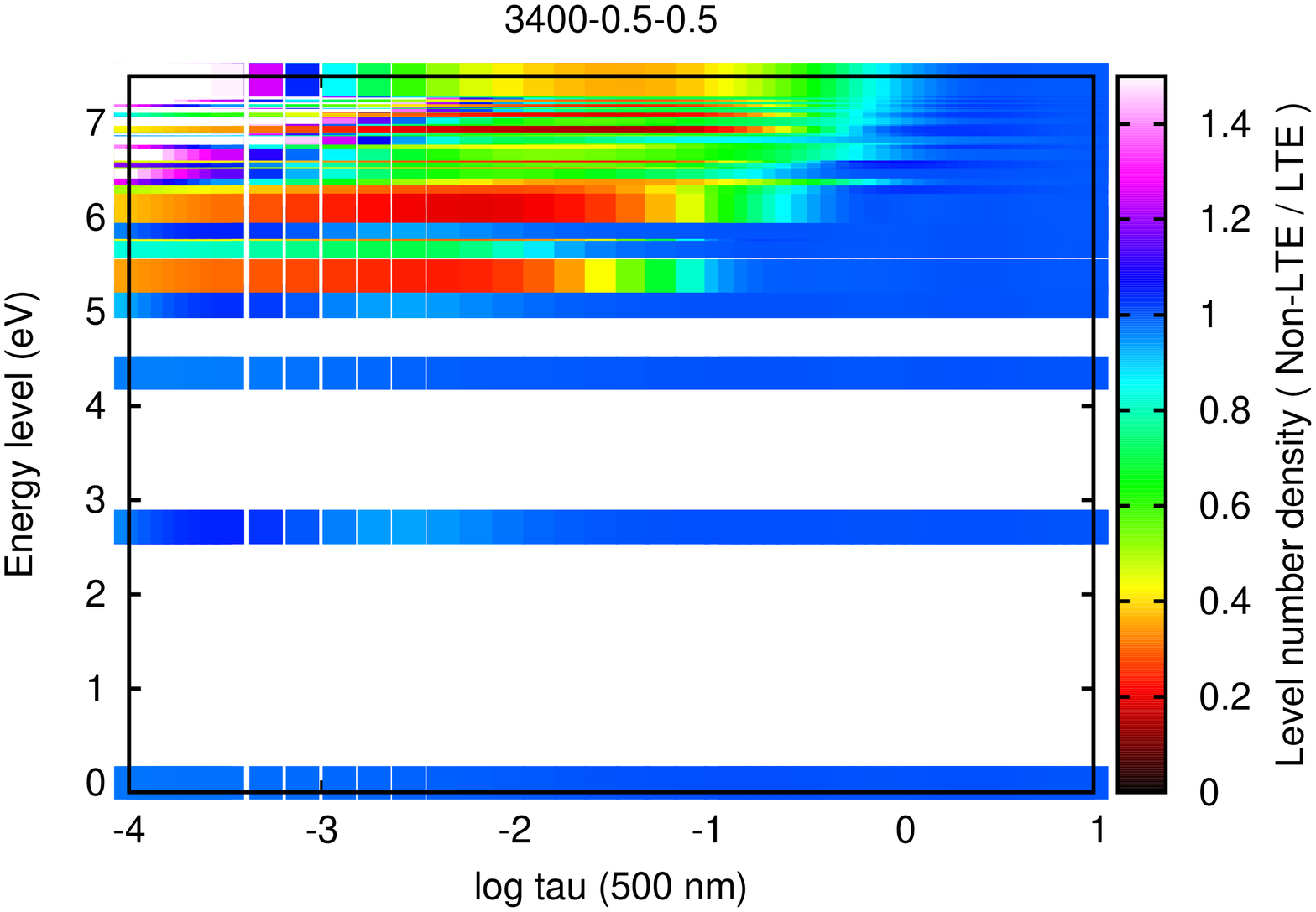}
\includegraphics[width=0.3\textwidth, angle=-90]{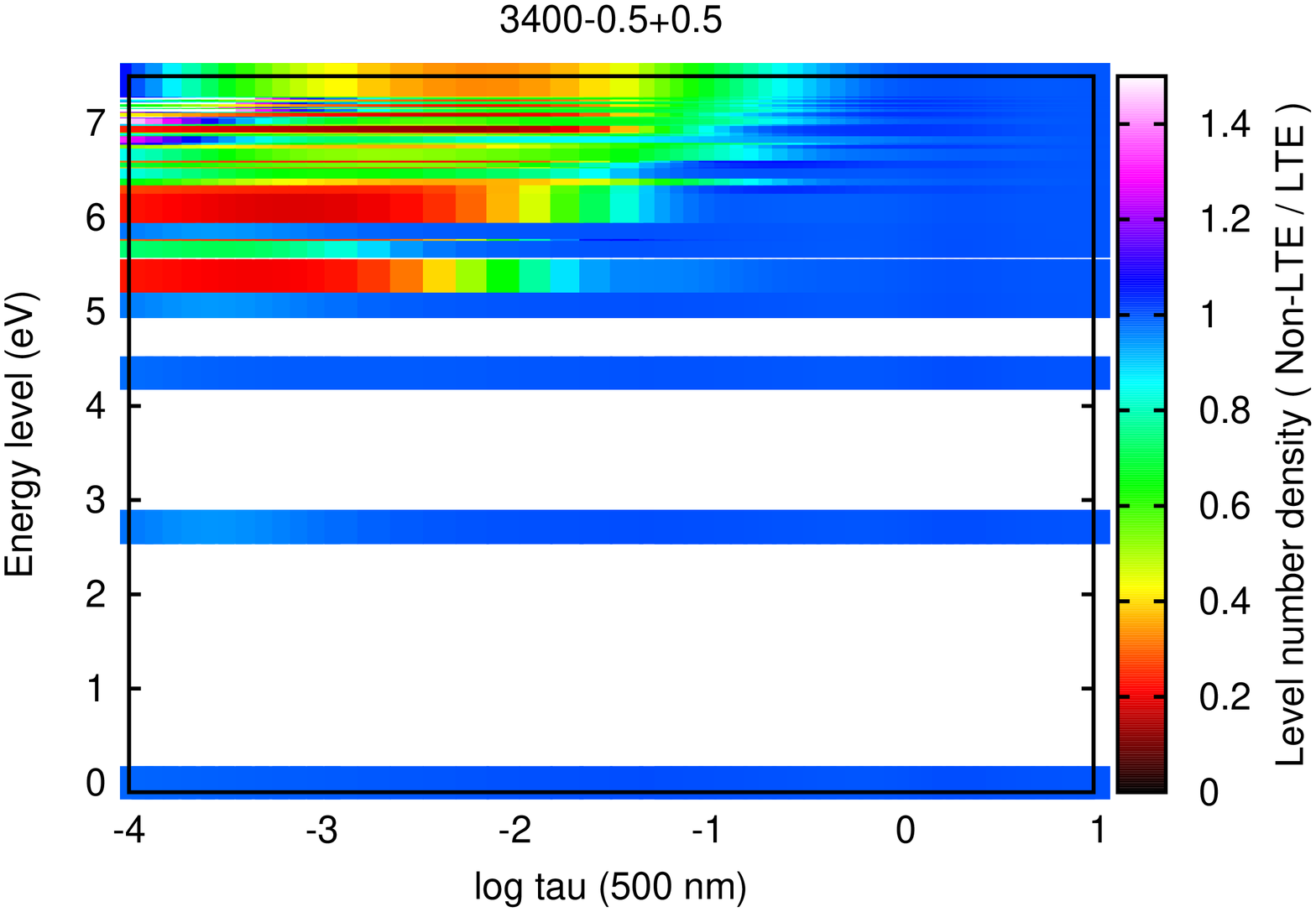}}
\hbox{
\includegraphics[width=0.3\textwidth, angle=-90]{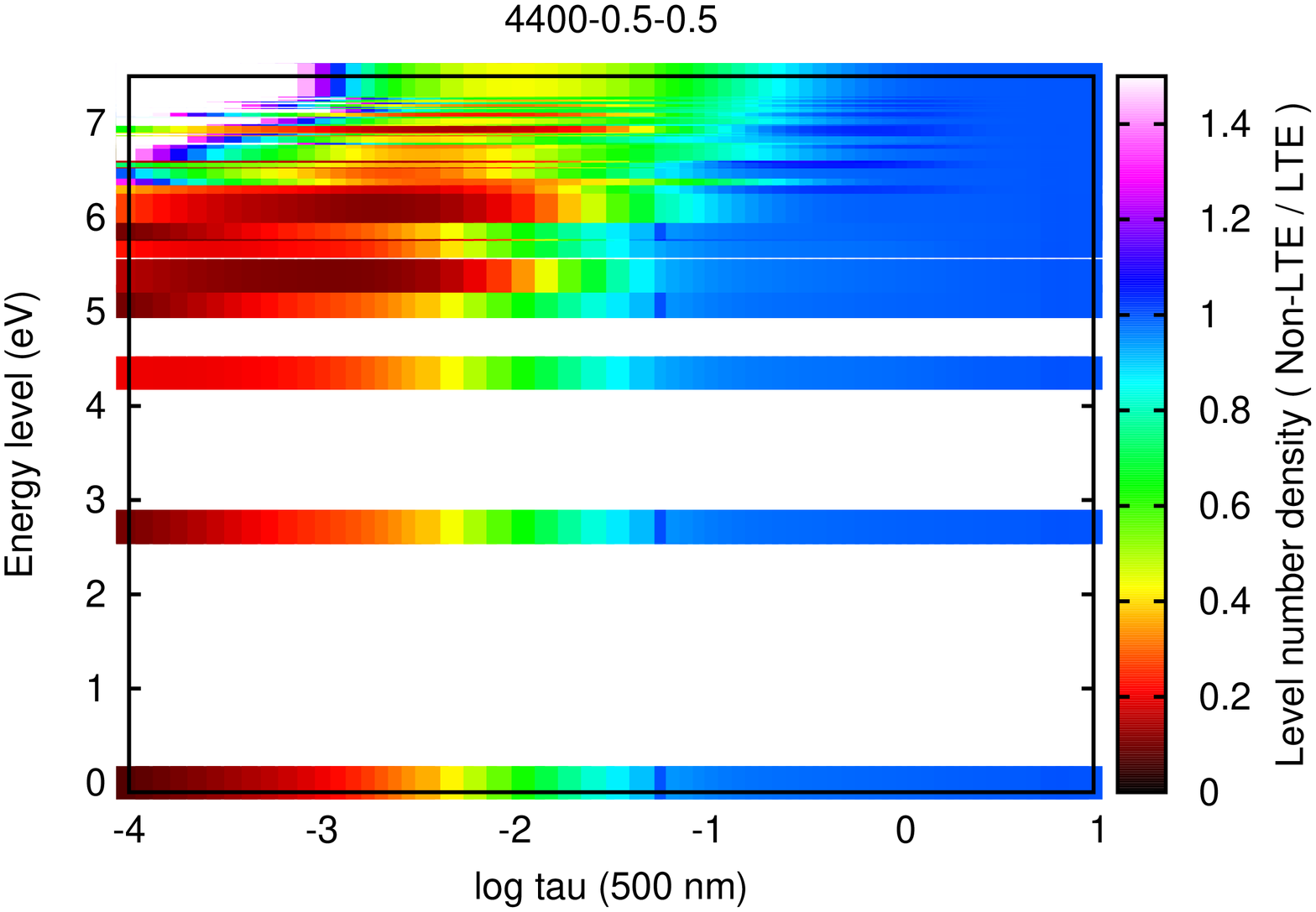}
\includegraphics[width=0.3\textwidth, angle=-90]{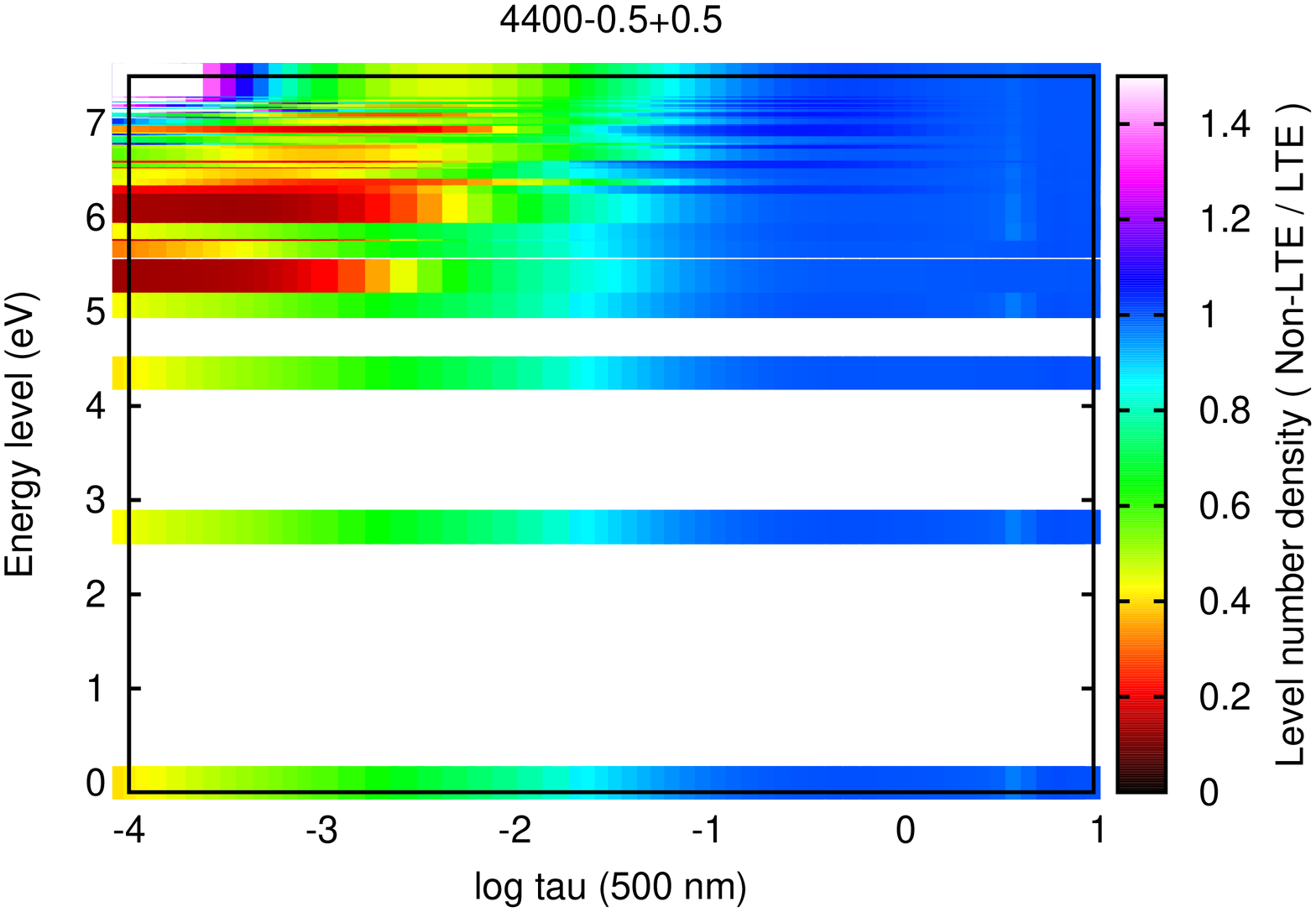}}
\caption{\mgi\ NLTE departure coefficients in RSG atmospheres shown for a few 
examples of the model grid. Top: T$_{\rm eff} = 3400$K, log g $= -0.5$ and [Z] 
$= -0.5$ (left) and  $0.5$ (right). Bottom: same as top but T$_{\rm eff} = 
4400$K. The value of the departure coefficients is color coded. The y-axis is 
the excitation energy of the levels and the x-axis is continuum optical depth 
at 500 nm.}
\label{departures1}
\end{figure*}

\begin{figure*}
\hbox{
\includegraphics[width=0.3\textwidth, angle=-90]{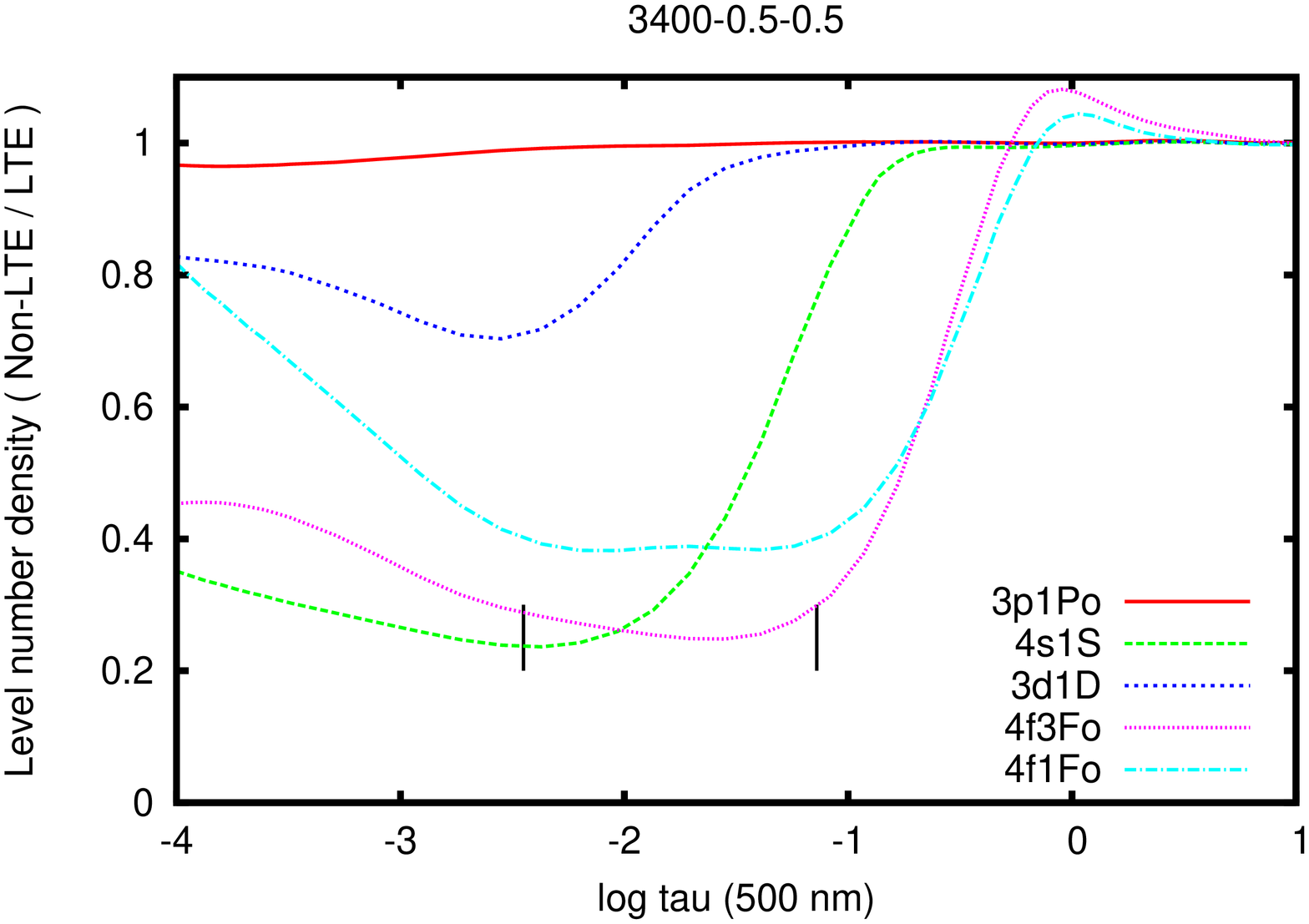}
\includegraphics[width=0.3\textwidth, angle=-90]{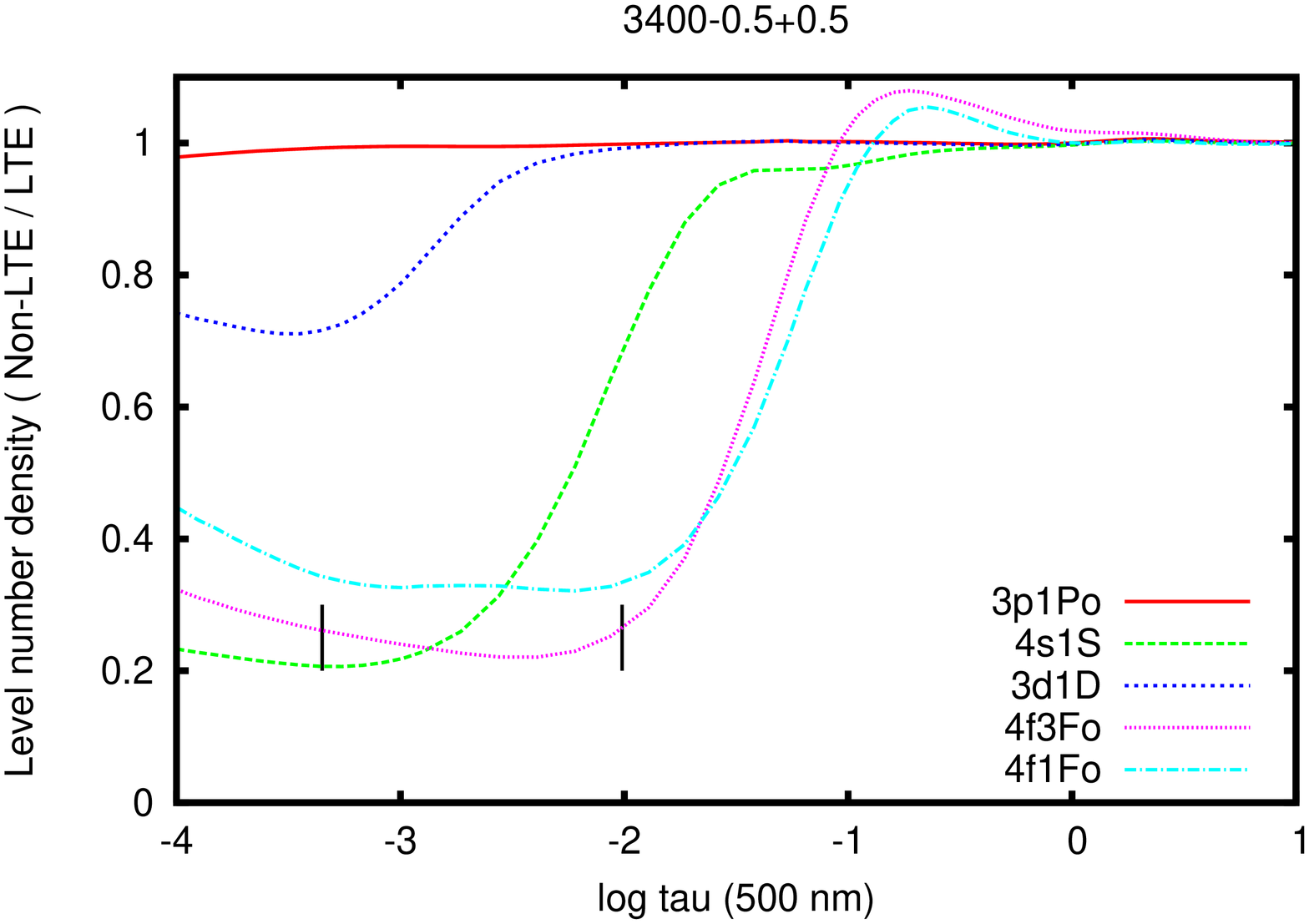}}
\hbox{
\includegraphics[width=0.3\textwidth, angle=-90]{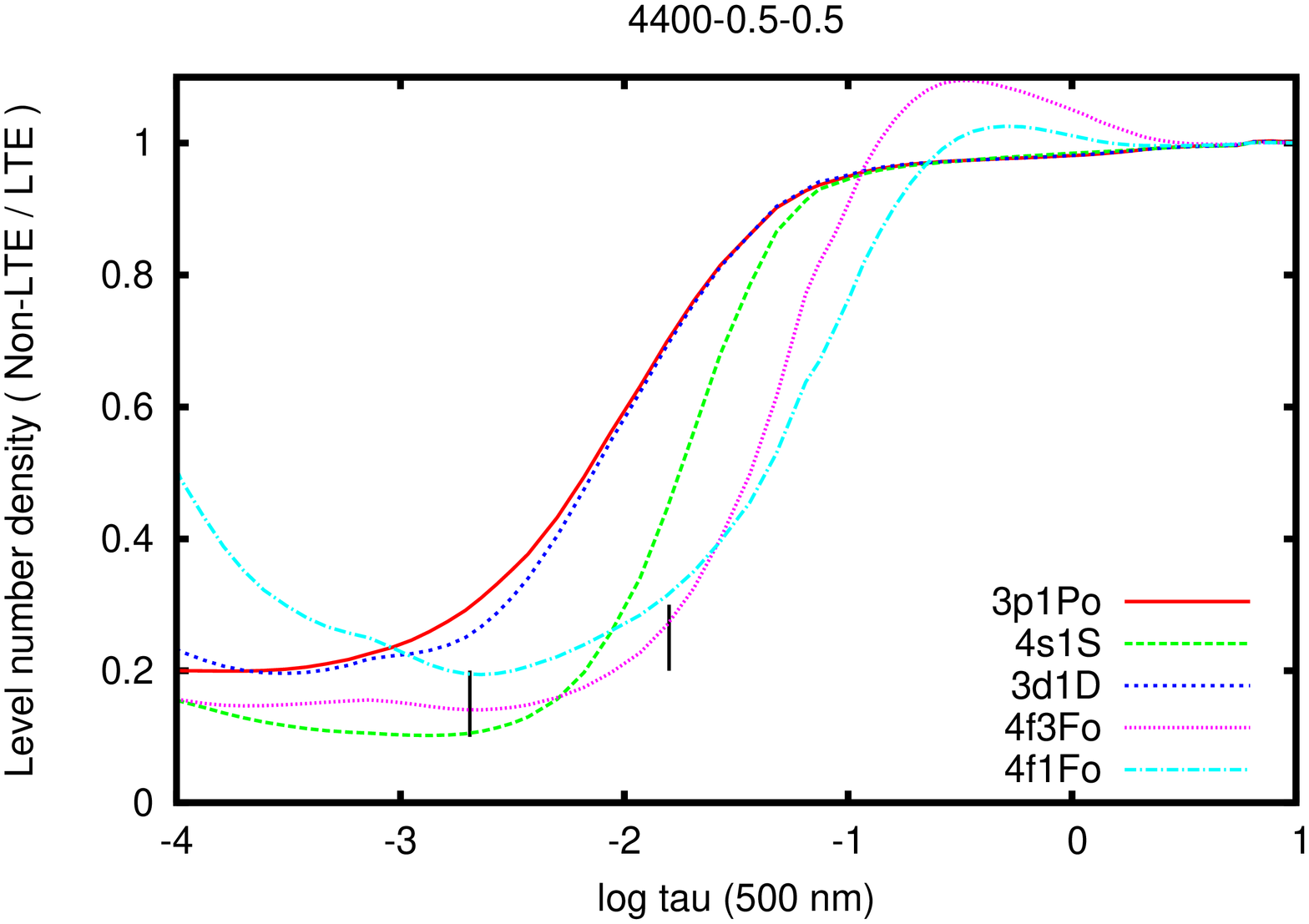}
\includegraphics[width=0.3\textwidth, angle=-90]{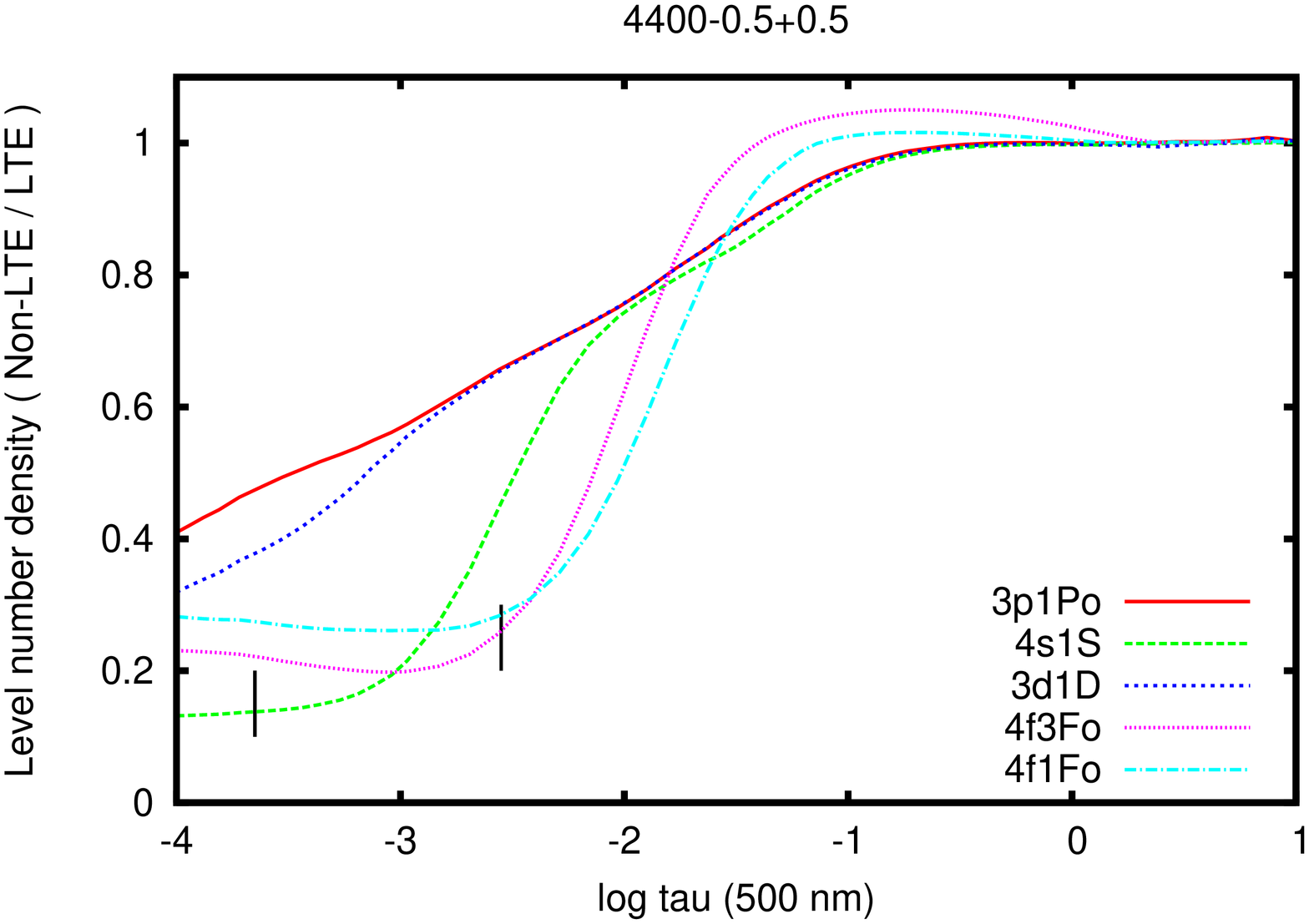}}
\caption{\mgi\ NLTE departure coefficients of the lower and upper levels of the 
J-band transitions shown for the same models as in Fig. \ref{departures1}. The 
NLTE line core
optical depths $\log \tau (11828.185$ \AA$) = 0$ and $\log \tau (12083.662$ 
\AA$) = 0$ are also indicated by black vertical marks. The 12083.662 line forms 
deeper in the atmosphere.}
\label{departures2}
\end{figure*}

Fig. \ref{departures1} shows the departure coefficients $b_i$ for a selection 
of models from our RSG model grid. The diagrams are somewhat different from the 
conventional 1D representation of $b_i$ as a function of optical depth 
$\tau$ in that on the y-axis we show the level energy, and the colour code 
indicates the departure coefficient. In this way we obtain a nice overview of 
the general trend of the NLTE effects as a function of excitation energy. Fig. 
\ref{departures2} also shows the conventional representation of the departure 
coefficients of the upper and lower levels of the J-band transition for the 
same models as Fig. \ref{departures1}.

The effects of NLTE are clearly seen in this diagram, in particular for energy 
levels at 4  - 5 eV (\Mg{3p}{1}{P}{\circ}{}, \Mg{4s}{3}{S}{}{}, 
\Mg{4s}{1}{S}{}{}) with large photo-ionisation cross-sections. Ionization by 
the super-thermal UV radiation field becomes important with proximity to the 
outer atmospheric boundary, and number densities of energy levels are much 
lower 
than the LTE predicts. Conceptually, the behaviour of \mgi\ atomic number 
densities bears a strong resemblance with that of \fei\ as discussed in detail 
in \citet{bergemann12c} and in Paper I.

\begin{figure*}
\hbox{
\includegraphics[width=0.15\textwidth, angle=90]{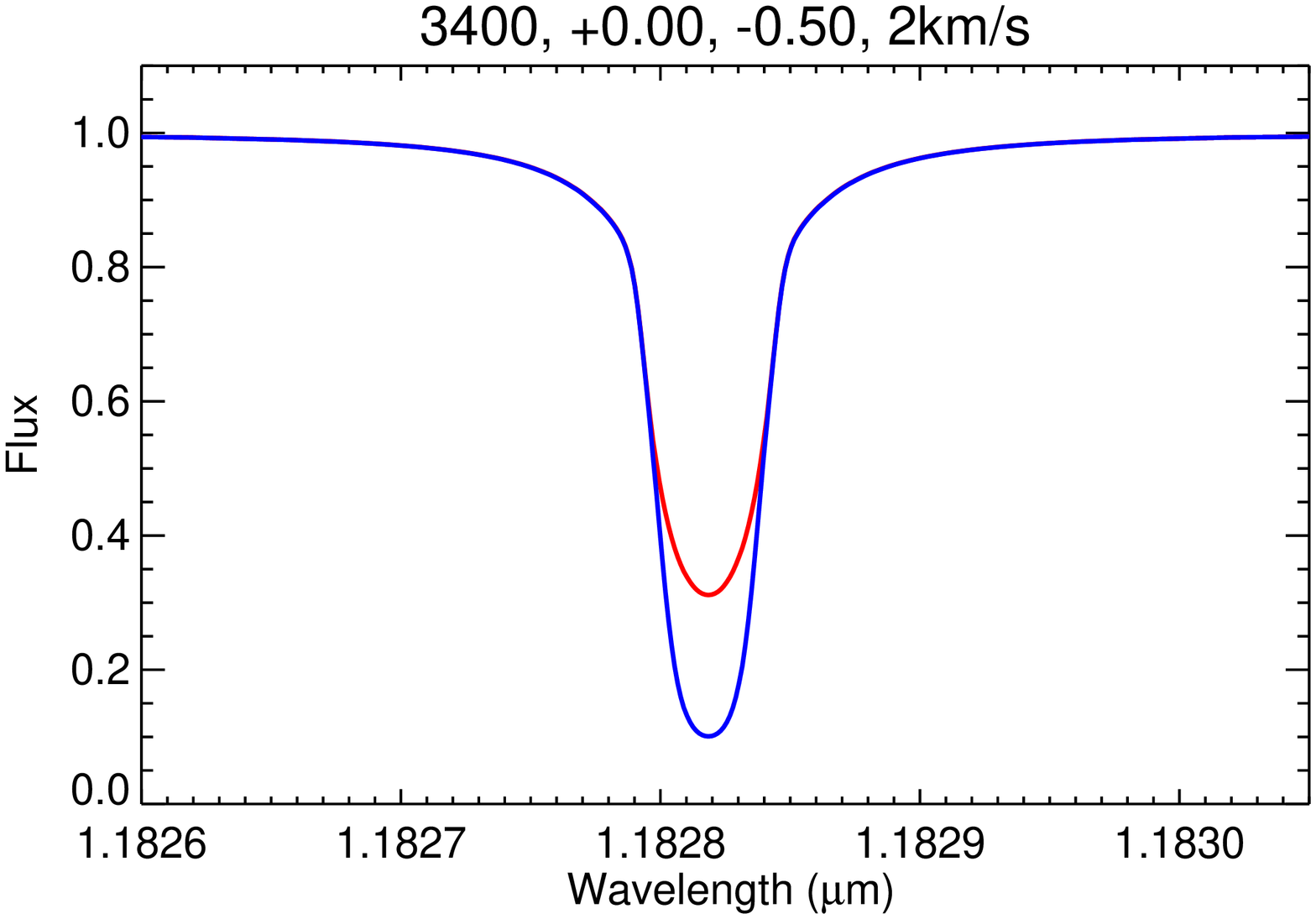}
\includegraphics[width=0.15\textwidth, angle=90]{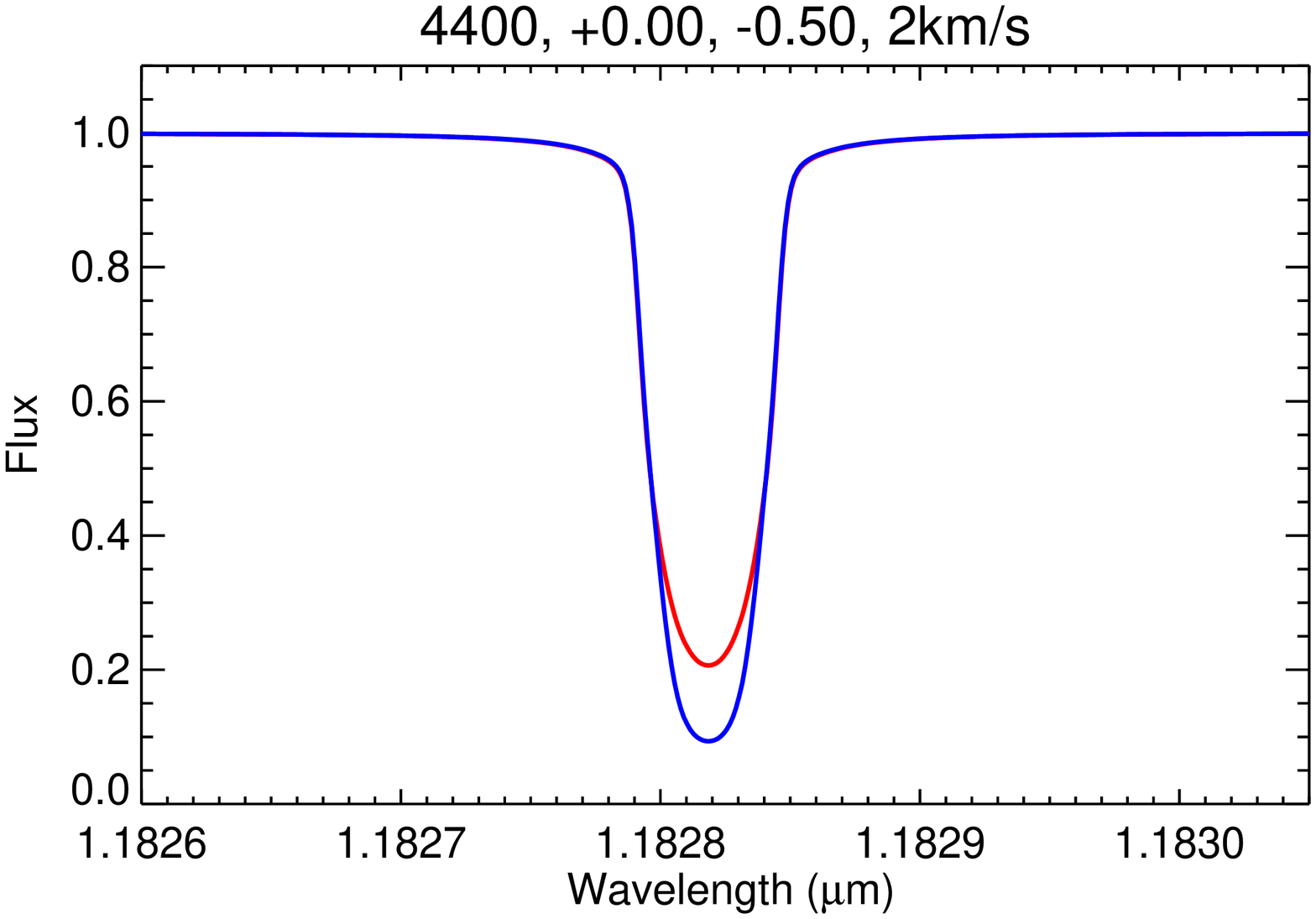}
\includegraphics[width=0.15\textwidth, angle=90]{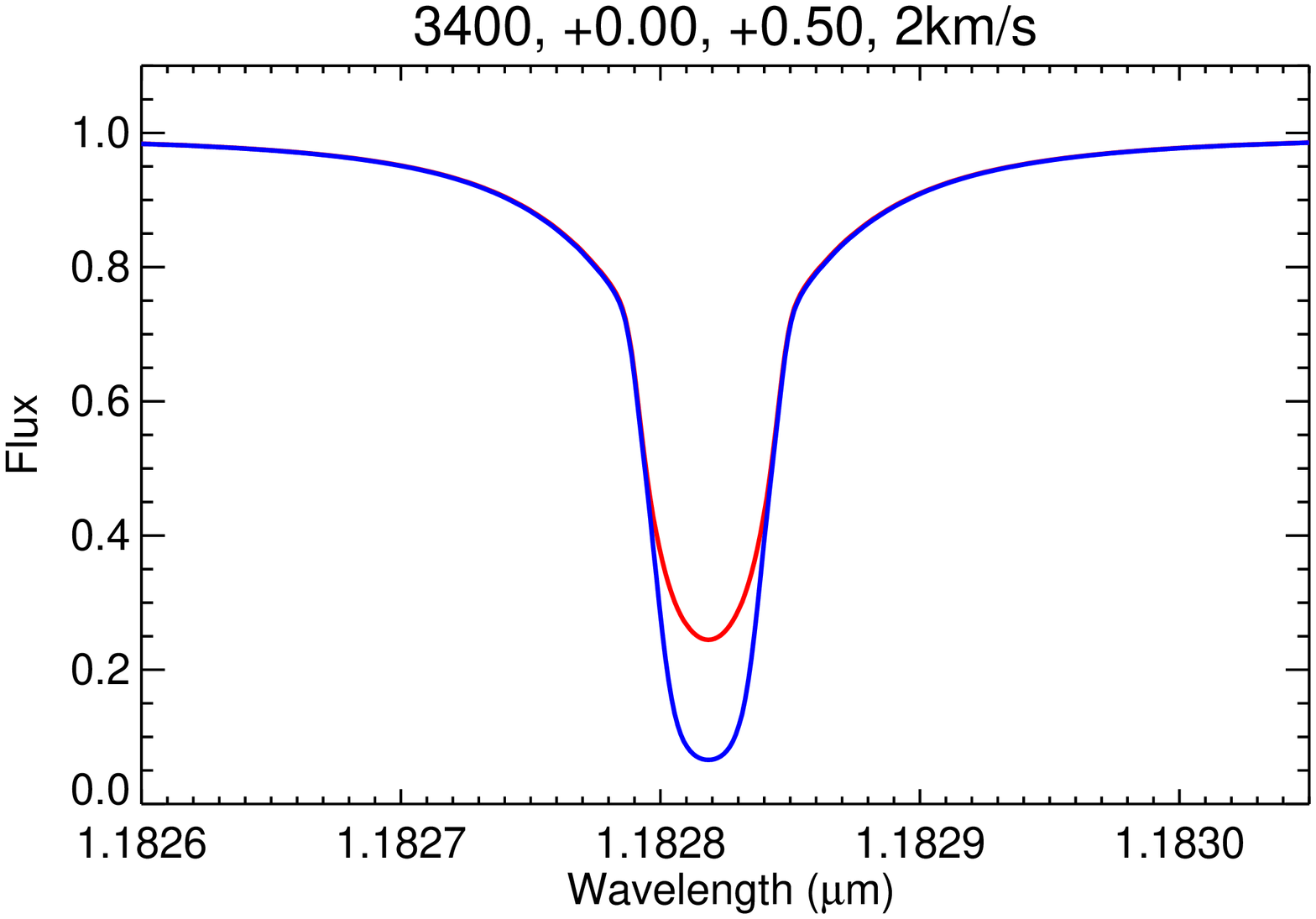}
\includegraphics[width=0.15\textwidth, angle=90]{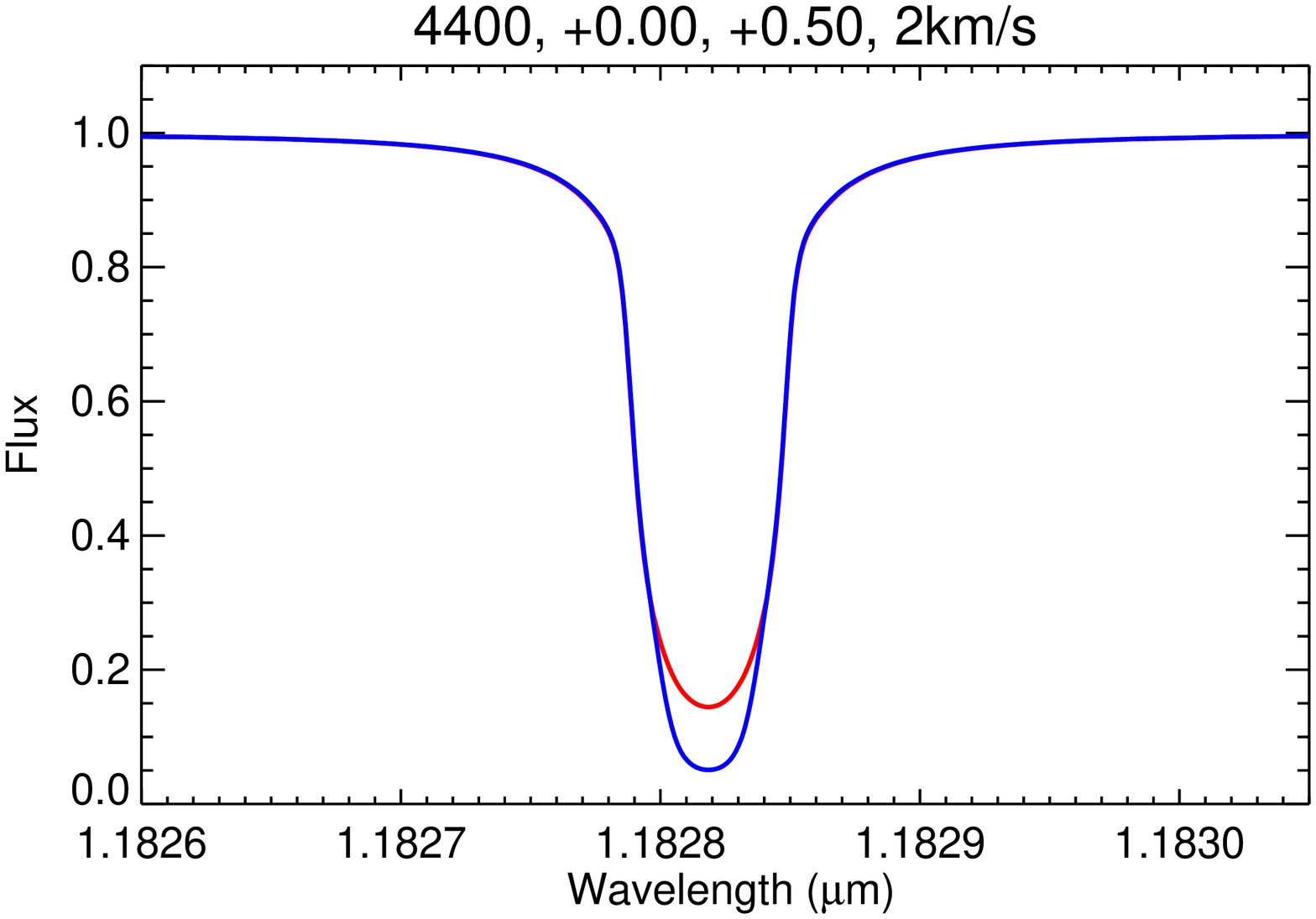}}
\hbox{
\includegraphics[width=0.15\textwidth, angle=90]{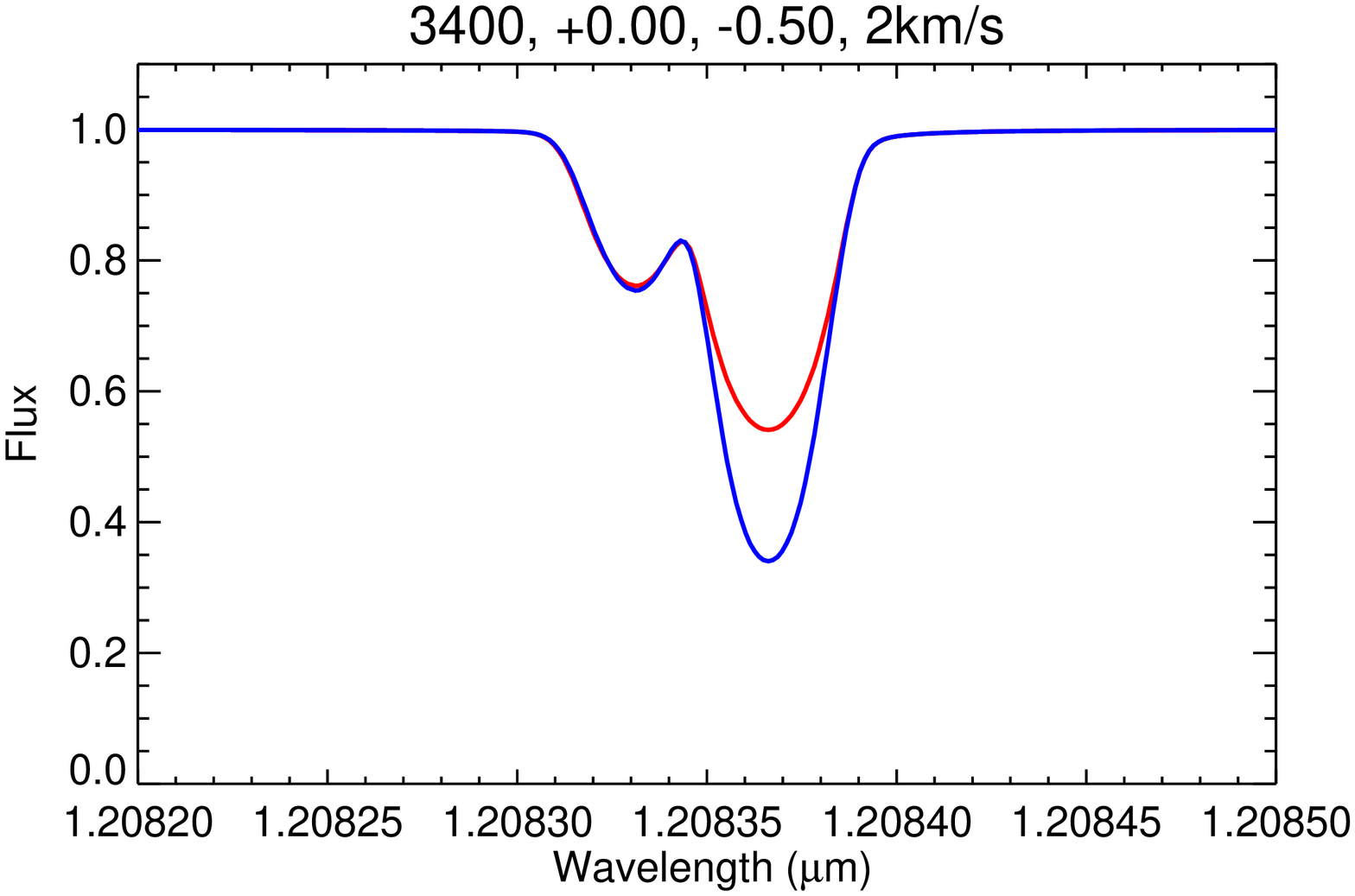}
\includegraphics[width=0.15\textwidth, angle=90]{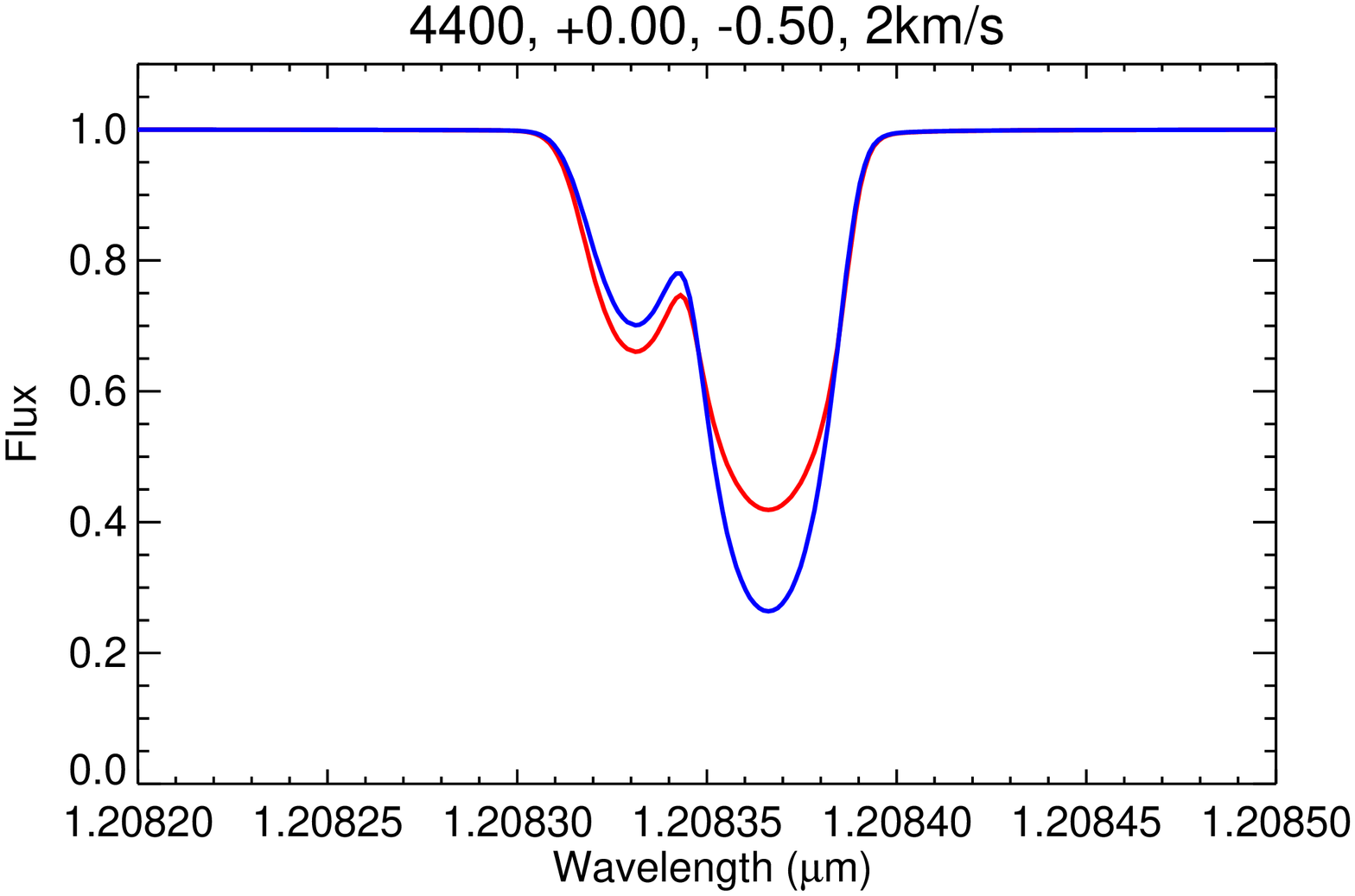}
\includegraphics[width=0.15\textwidth, angle=90]{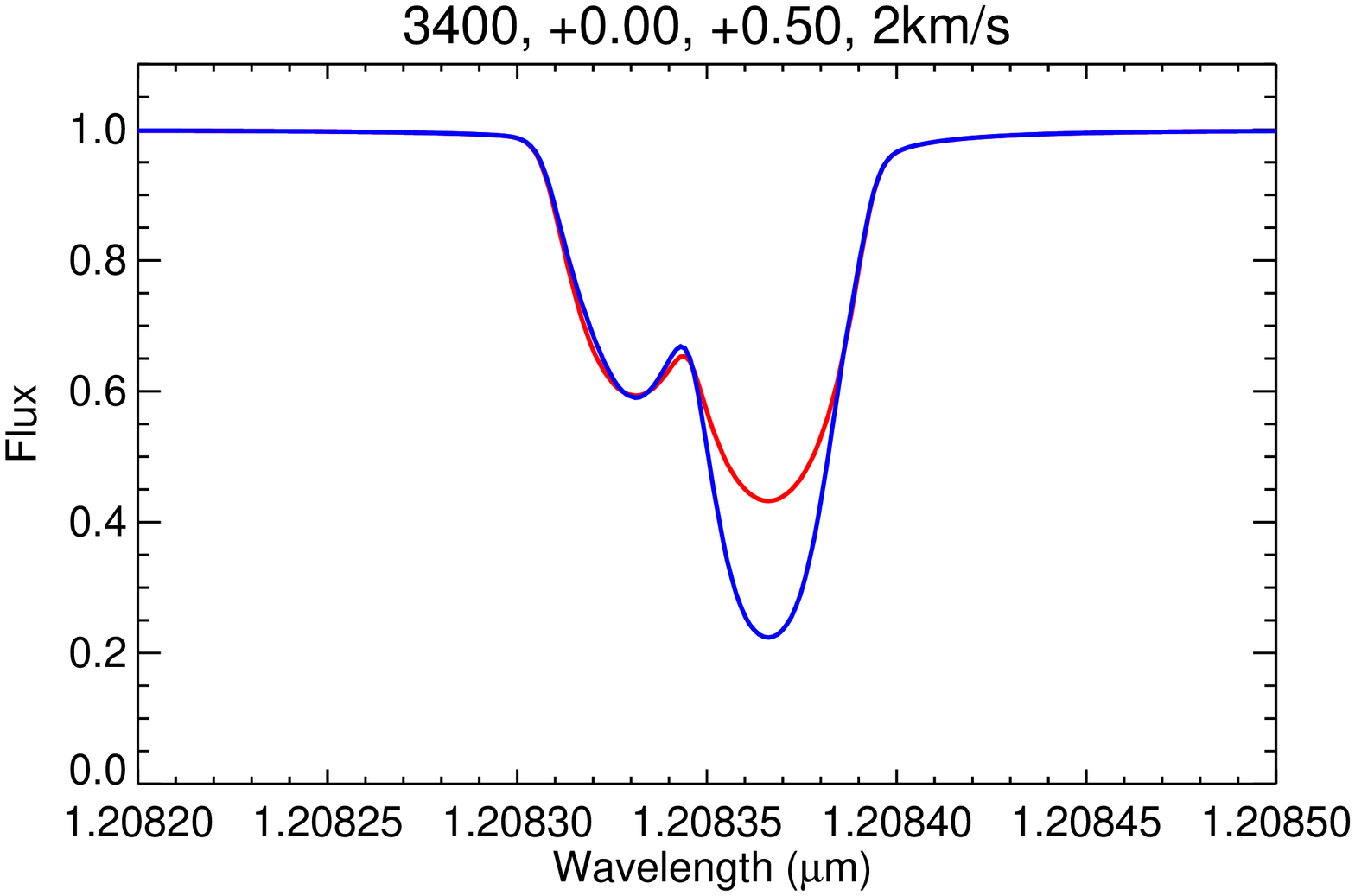}
\includegraphics[width=0.15\textwidth, angle=90]{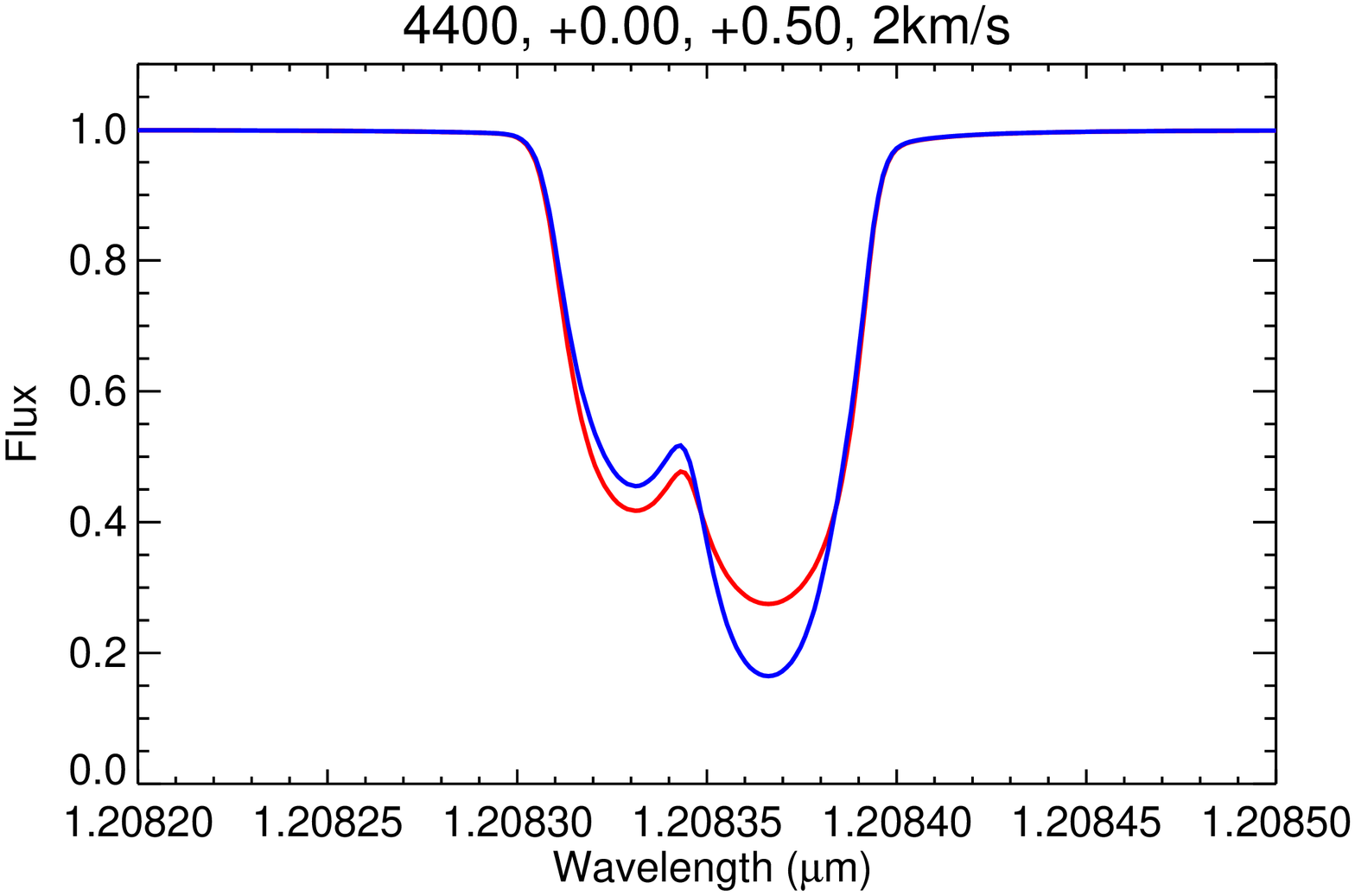}}
\caption{Profiles of the J-band \mgi\ lines computed in LTE (red) and NLTE 
(blue) for a few examples of the model atmosphere grid. All models are for 
$\log g  = 0.0$ and $\Vmic = 2$ km$/$s. The upper row is for the line at 
11828 \AA\ and the bottom row for 12083 \AA.}
\label{prof_rsg}
\end{figure*}

For the IR J-band transitions, the main additional NLTE effect is through 
radiation transport in the spectral lines themselves. The lower level of the 
12083.660 \AA~line is \Mg{3d}{1}{D}{}{} at 5.75 eV, and its only connection to 
the 
lower-lying levels, \Mg{}{1}{P}{\circ}{}  (4.36 eV) and \Mg{}{3}{P}{\circ}{} 
(2.71 eV), is via the forbidden magnetic-dipole transitions at $\sim 4070 - 
4080$ \AA~ 
(\Mg{}{3}{P}{\circ}{}  -  \Mg{}{1}{D}{}{}) and the allowed transition in the 
near-IR (at $8806$ \AA, \Mg{}{1}{P}{\circ}{1}  -  \Mg{}{1}{D}{}{2}) 
respectively 
(see the energy level diagram, Fig. 1).  The lower level of the line at 11823 
\AA\ is \Mg{3p}{1}{P}{\circ}{}. The level is connected with the ground state 
through a resonance line at 2852 \AA\ (\Mg{}{1}{S}{}{}  -  
\Mg{}{1}{P}{\circ}{}), which is optically thick throughout the formation depths 
of the J-band \mgi\ lines and does not effectively participate in the NLTE 
radiation transport. This configuration explains the fact that the both J-band 
lines have a nearly two-level-atom line source function S$_{ij}$. Once the 
lines have become optically thin, downward cascading electrons over-populate 
the lower levels of the transitions relative to the upper levels, and the 
source functions become sub-thermal, S$_{ij} <$ B$_{\nu}$ (see Fig. 
\ref{departures2}). Therefore the J-band \mgi\ lines are stronger in NLTE than 
in LTE, and the NLTE effect further increases with decreasing $\Teff$ and $\log 
g$. This is a very similar situation to the formation of the J-band 
silicon lines discussed in Paper II. The resulting differences between the 
J-band spectral lines profiles in the LTE and non-LTE case are illustrated in 
Fig. \ref{prof_rsg}.

\begin{figure}
\includegraphics[width=0.3\textwidth, angle=90]{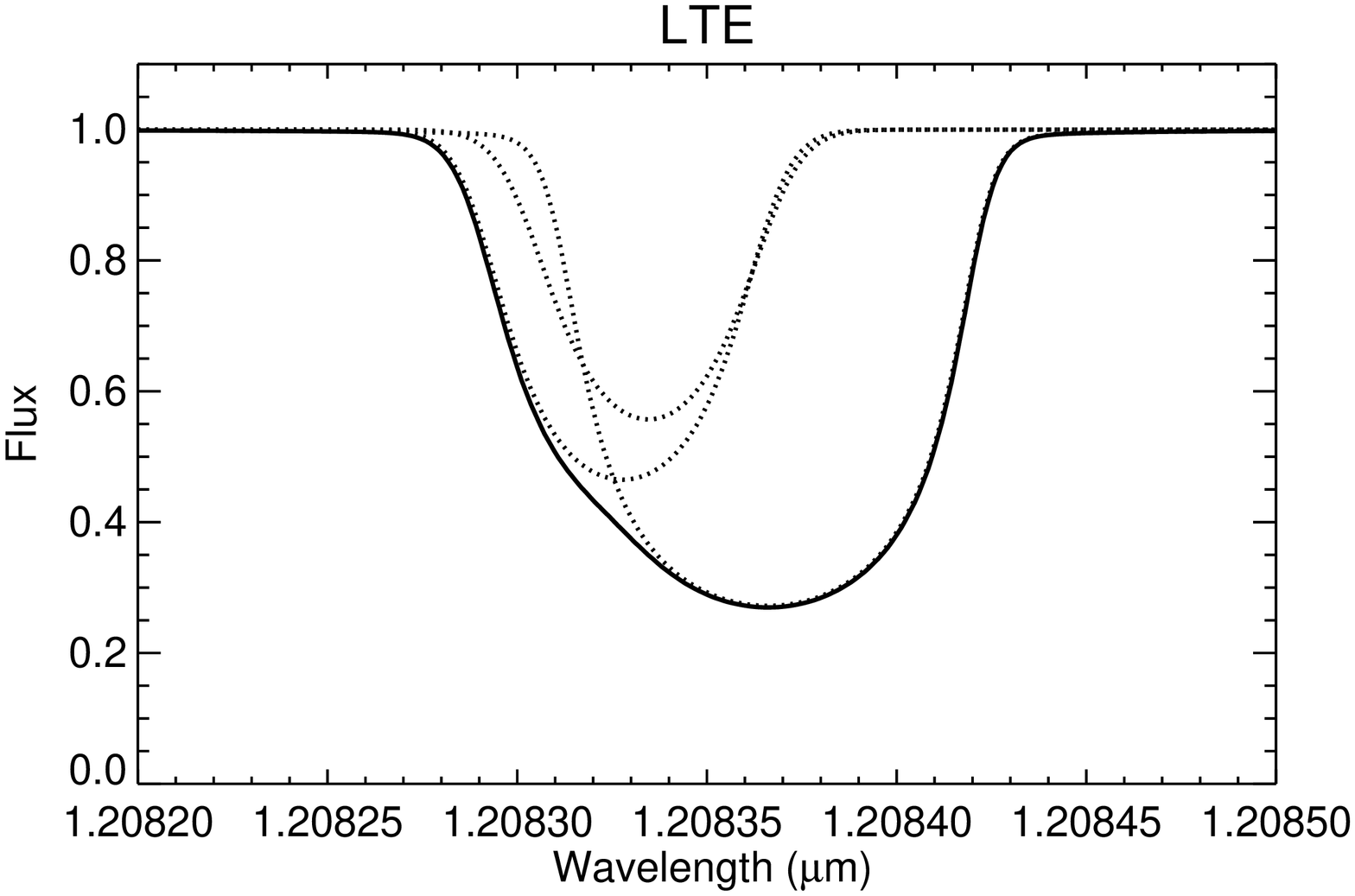}
\includegraphics[width=0.3\textwidth, angle=90]{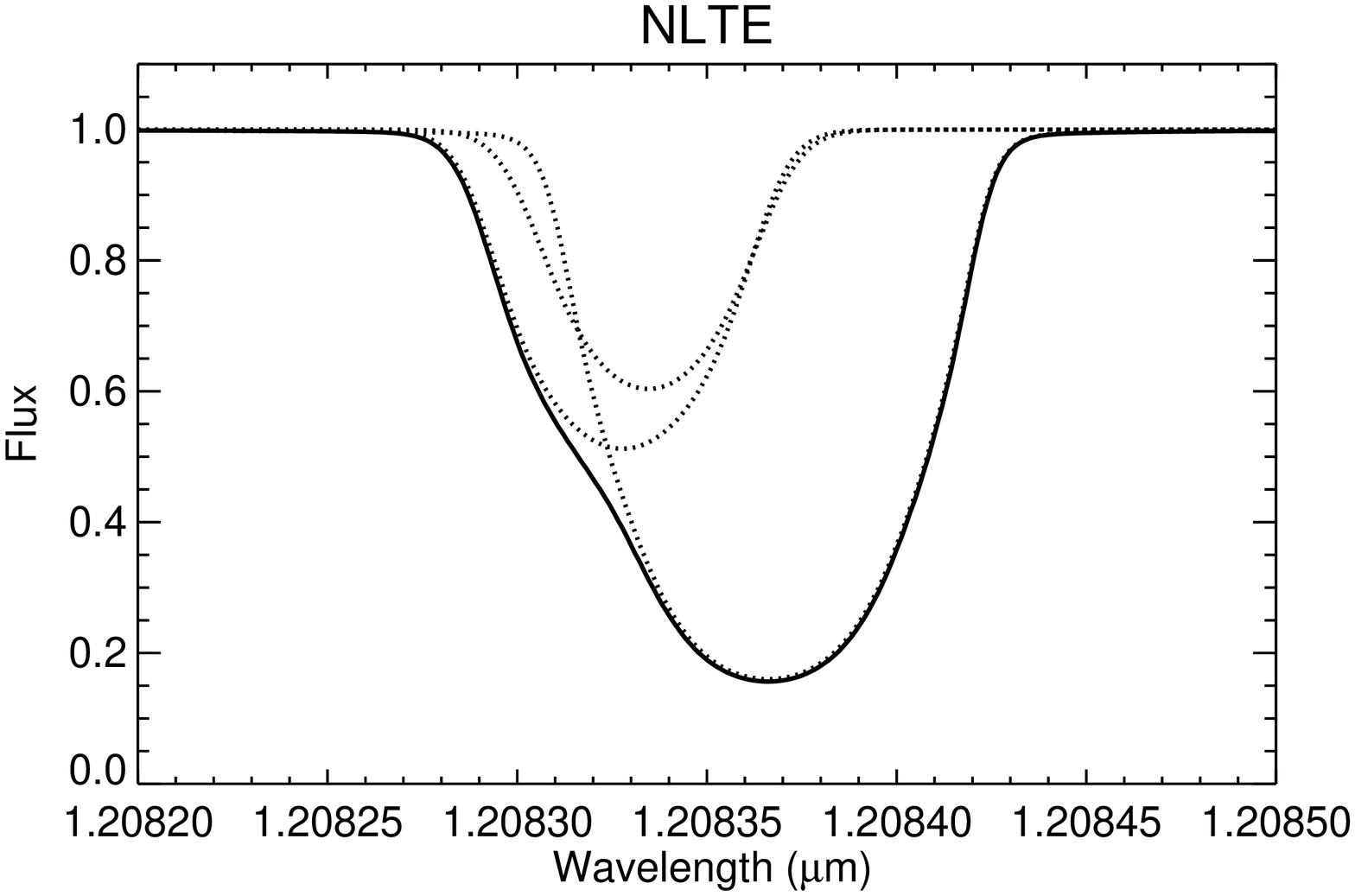}
\caption{Individual line components (dashed) contribution to the total profile 
(solid) of the J-band \mgi\ line at 12083 \AA\ computed with $\Teff = 4400$, 
$\log g  = 0.0$, and $\feh = 0.5$ in LTE (top) and NLTE 
(bottom).}
\label{prof_rsg_12083}
\end{figure}

Beyond this more general discussion of the J-band NLTE effects the formation of 
the line profiles of the 12083 \AA\ super-line are affected by an additional 
complication. Since the observed feature is a blend of three Mg I components 
from different multiplets, the result is a highly asymmetric shape, which 
varies with stellar parameters (see Fig. \ref{prof_rsg}, bottom four panels). 
Each of the components suffers from its own NLTE effect. To illustrate this 
peculiar effect, we provide the profiles of each component computed in LTE and 
NLTE for the model with $\Teff = 4400$, $\log g  = 0.0$, and $\feh = 0.5$ in 
Fig. \ref{prof_rsg_12083}. The bluest line at 12083.278 \AA\, which forms in 
the 
3d $^1$D$_2$ - $^3$F$_3^{\mathrm o}$ transition, is weaker in NLTE than in LTE. 
With its very small oscillator strength this line forms in the deeper 
atmospheric layers where the departure coefficient of the upper level is larger 
than the one of the lower level. The upper levels of the 
12083 transitions are very close to the \mgii\ continuum at 7.64 eV 
and are sensitive to recombination cascades from \mgii, which causes their 
overpopulation in deeper atmospheric layers.  As a consequence, the NLTE source 
function 
is super-thermal, S$_{ij} >$ B$_{\nu}$.
The reddest component at 12083.662 \AA\ has a much larger oscillator strength 
and forms much 
further out in the atmosphere, where the departure coefficient of the upper 
level is always smaller than the one of lower level as already described above. 

\section{NLTE \mgi\ abundance corrections}
Over the whole RSG grid the J-band \mgi\ absorption lines are stronger in 
NLTE than in LTE. Quantitatively, this information is summarised in Table 3 
which compiles the equivalent widths for two values of microturbulence, 2 
and 5 km$/$s. As a consequence, magnesium abundances obtained from a LTE fit of 
observed J-band lines are systematically too high. This can be quantitatively 
assessed by introducing NLTE abundance corrections $\Delta$.
\begin{equation}
\Delta_{\rm Mg~I} = \log \rm{A (Mg)}_{\rm NLTE} - \log \rm{A (Mg)}_{\rm LTE}
\end {equation}

\begin{figure*}
\includegraphics[width=0.8\textwidth, angle=0]{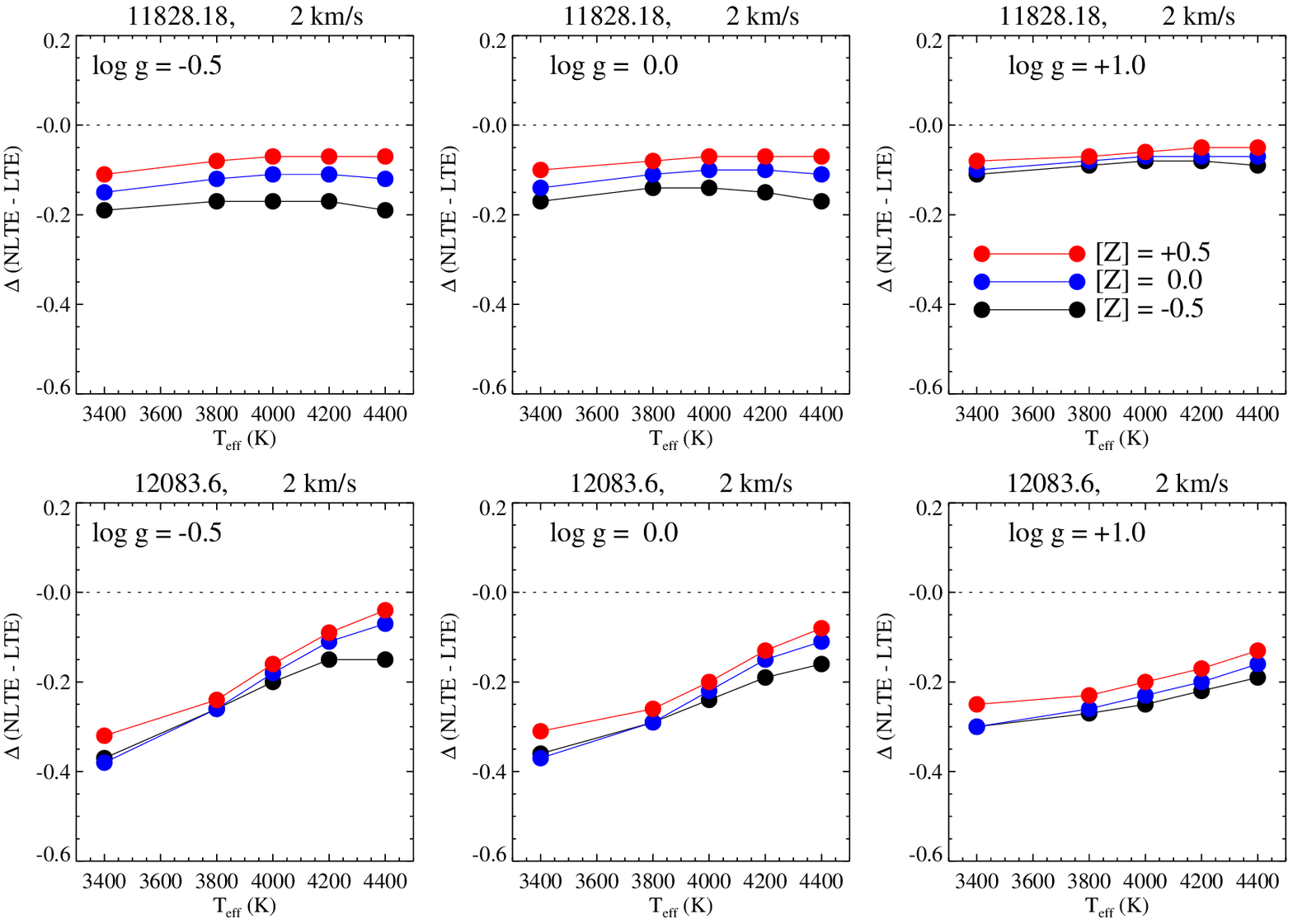}
\includegraphics[width=0.8\textwidth, angle=0]{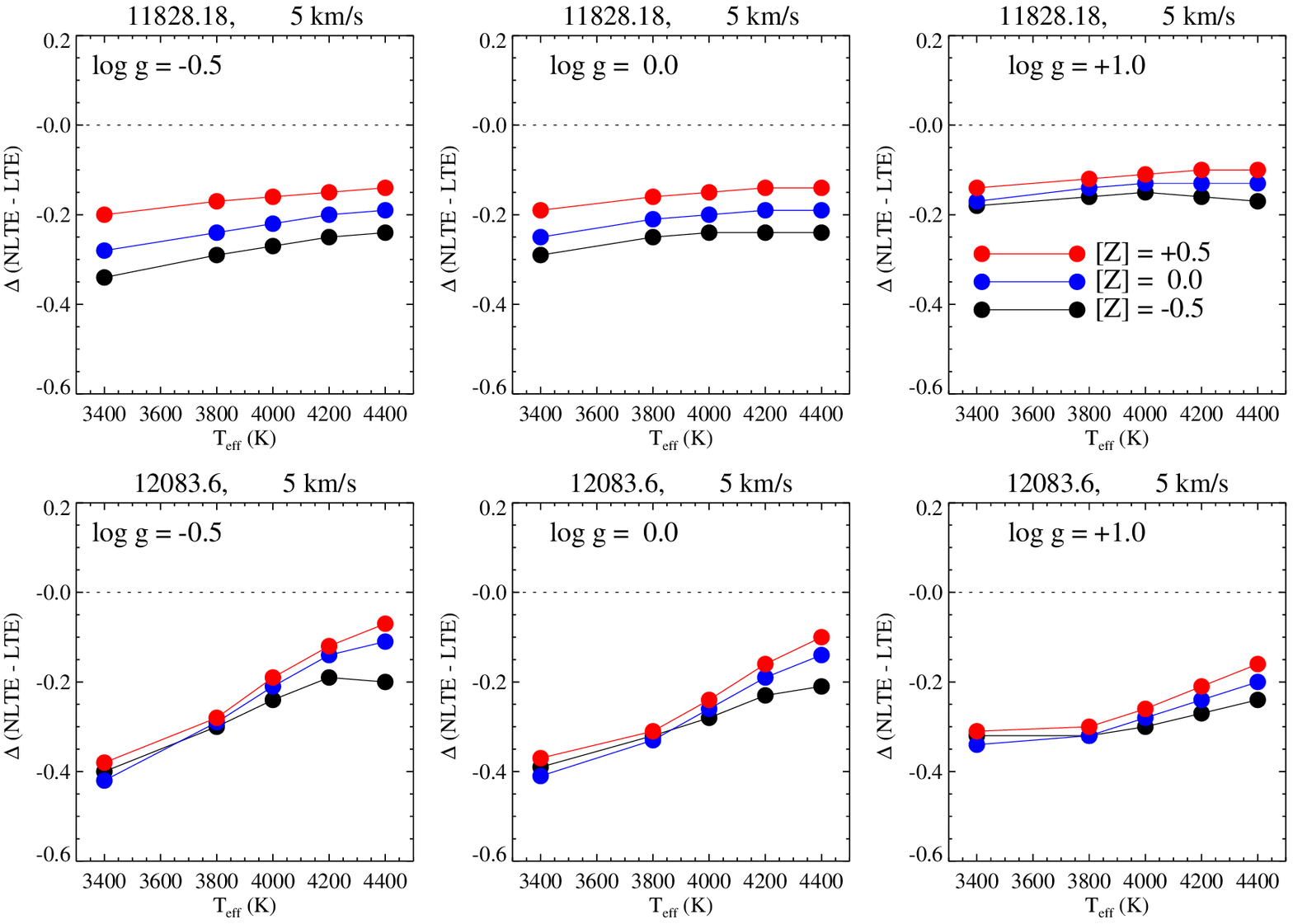}
\caption{NLTE abundance corrections for the 11828 (top) and 12083 (bottom) \AA\ 
\mgi\ lines as a function of effective temperature and metallicity for log g = 
$-$0.5 (left), 0.0 (middle), and 1.0 (right). Top panels: $\Vmic = 2$ \kms; 
bottom panels: $\Vmic = 5$ \kms.}
\label{nltegrid1}
\end{figure*}

\begin{figure*}
\hbox{
\includegraphics[width=0.3\textwidth, angle=90]{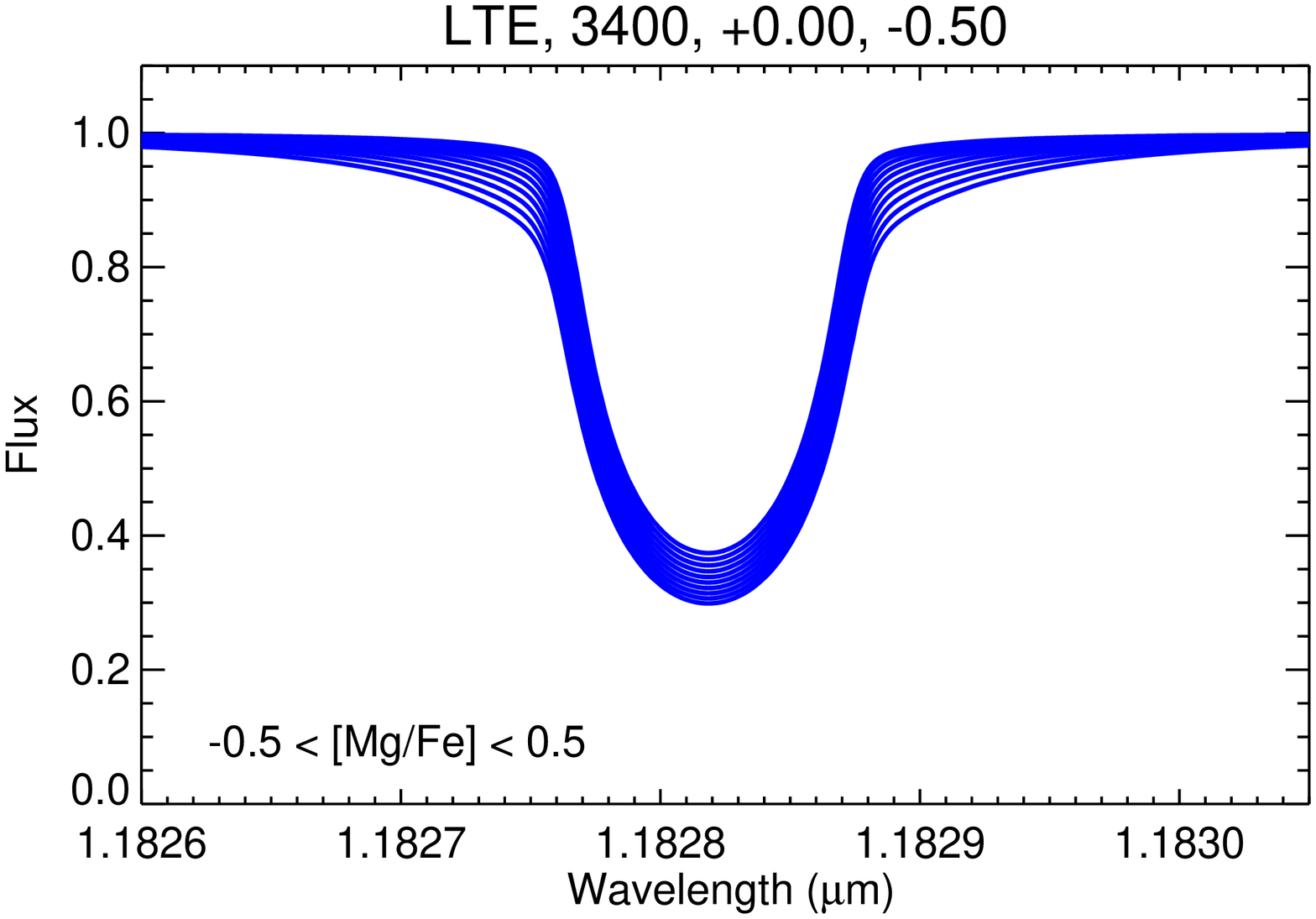}
\includegraphics[width=0.3\textwidth, angle=90]{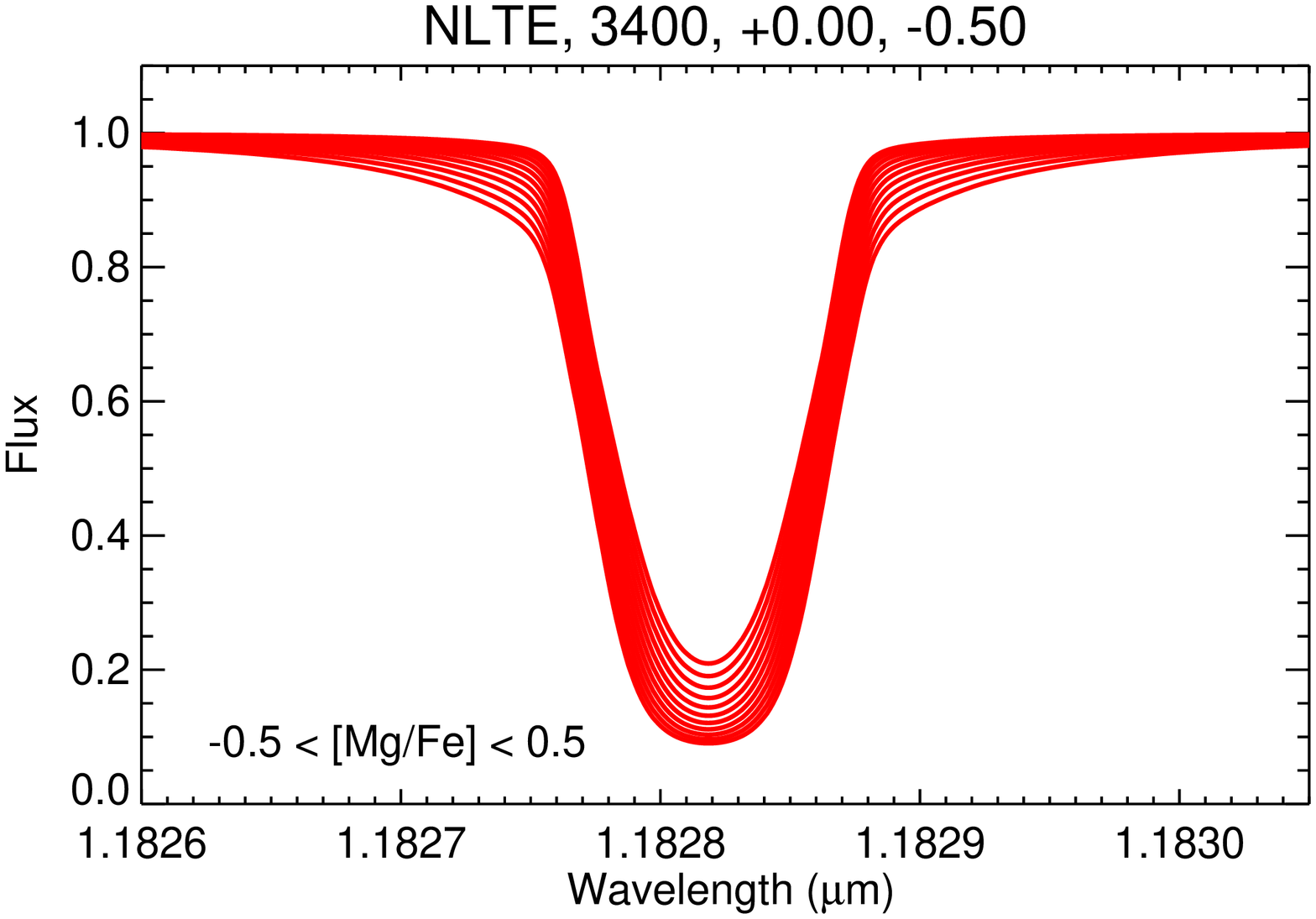}}
\hbox{
\includegraphics[width=0.3\textwidth, angle=90]{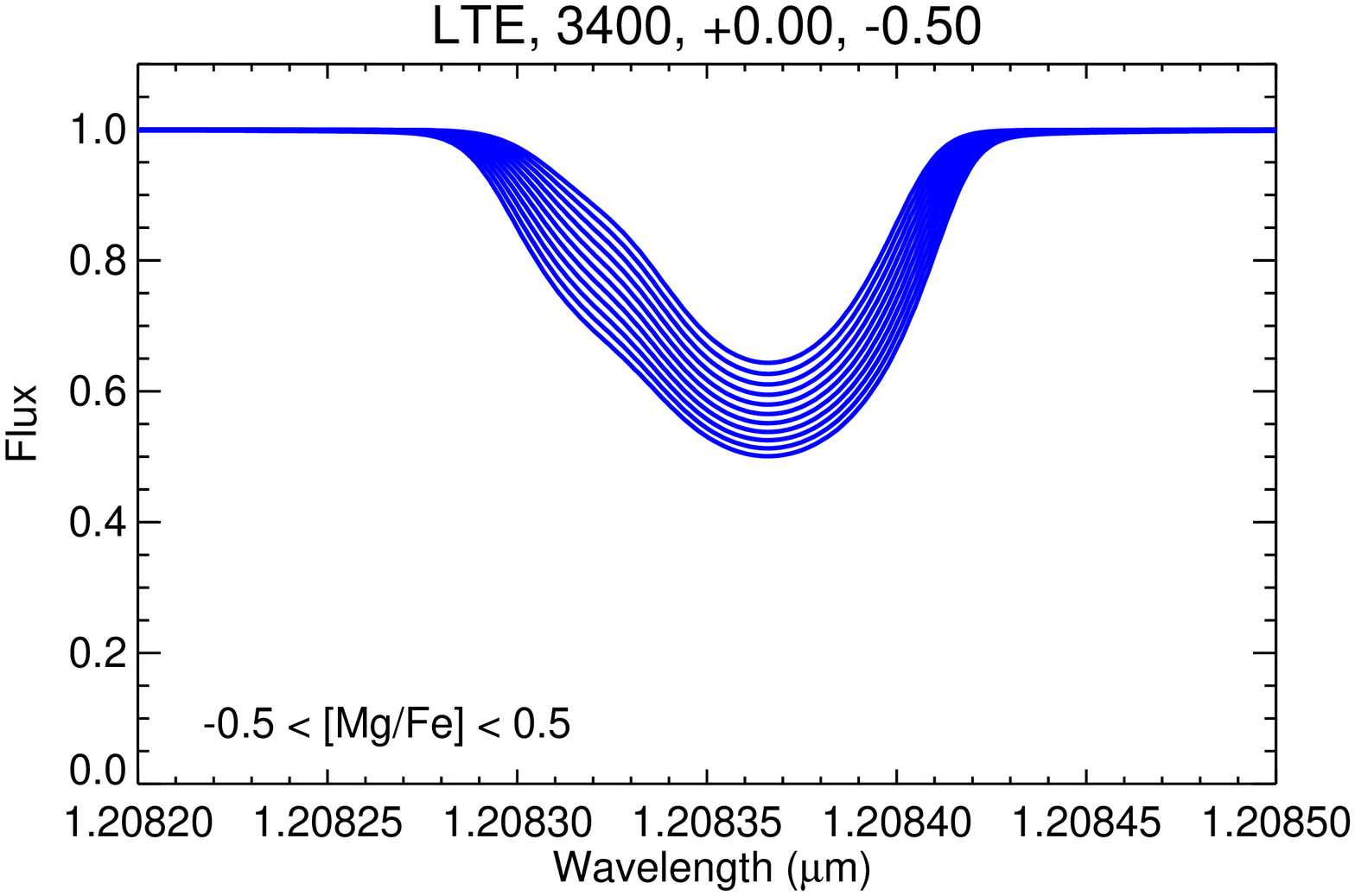}
\includegraphics[width=0.3\textwidth, angle=90]{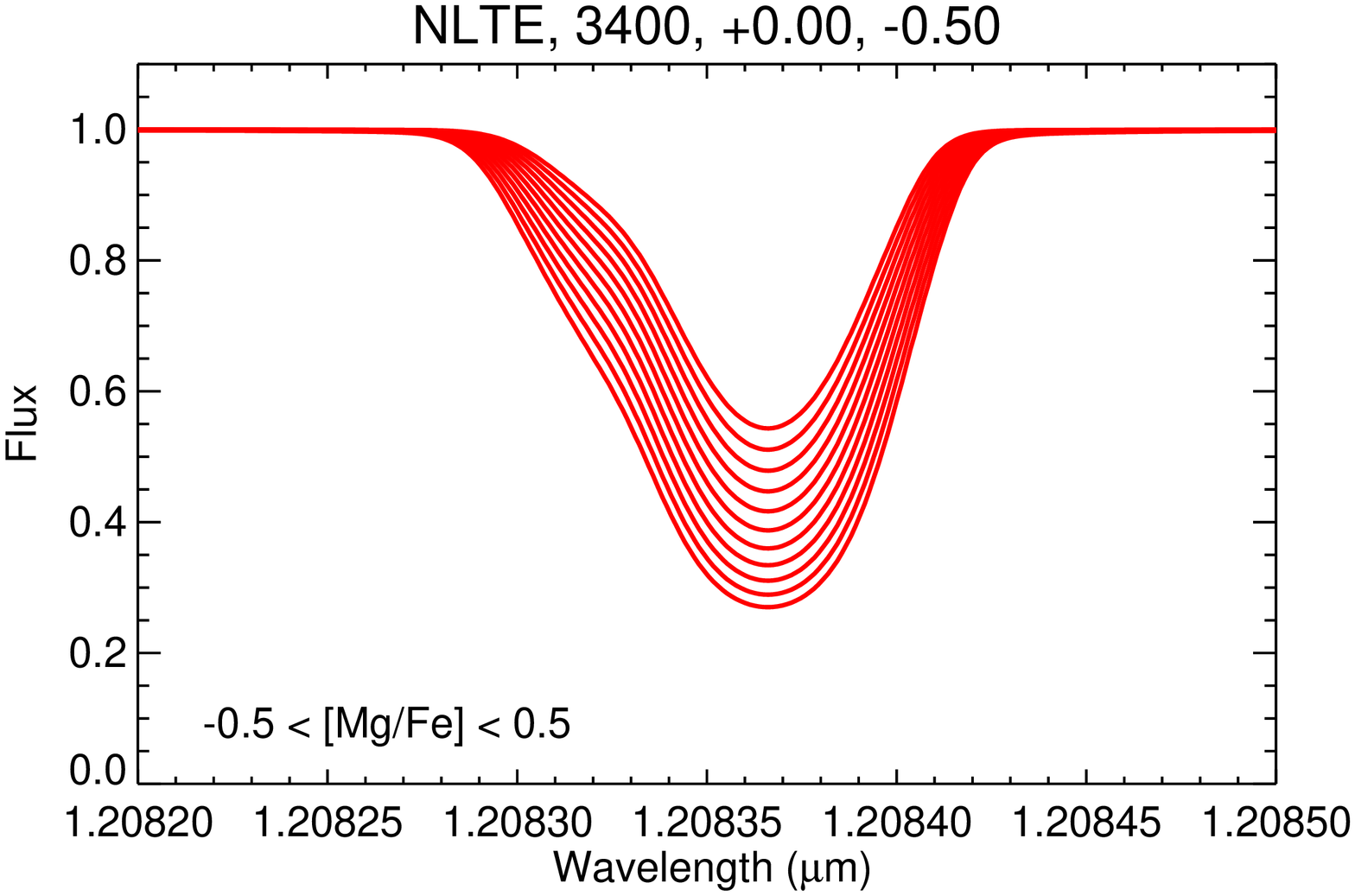}}
\caption{LTE (left) and NLTE (right) lines profiles for the 11828 (top) and 
12083 (bottom) \AA\ Mg I lines as a function of Mg abundance. [Mg$/$Fe] varies 
from $-$0.5 to +0.5 with a step of 0.1 dex.}
\label{abundance}
\end{figure*}

\begin{figure*}
\hbox{
\includegraphics[width=0.3\textwidth, angle=90]{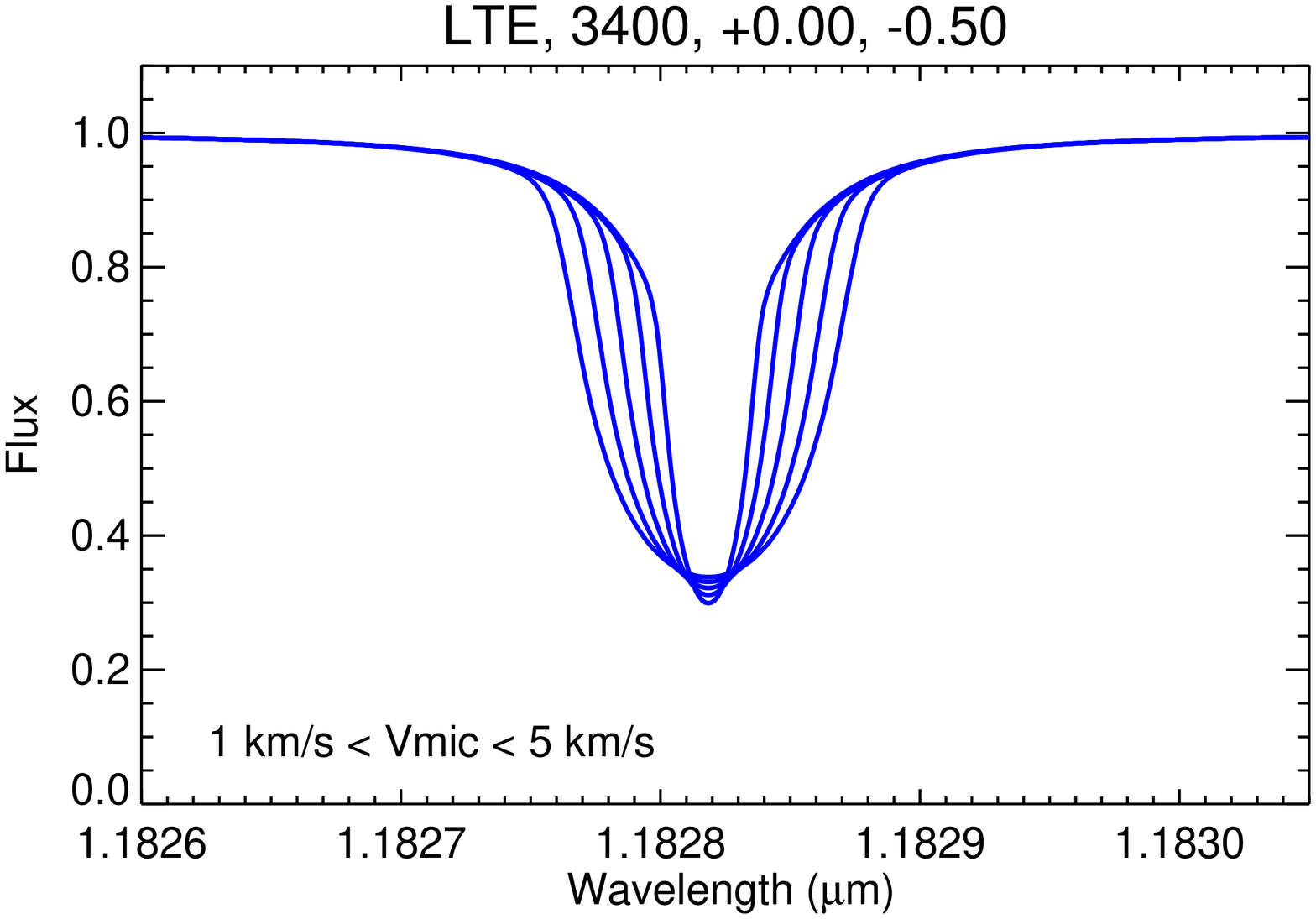}
\includegraphics[width=0.3\textwidth, angle=90]{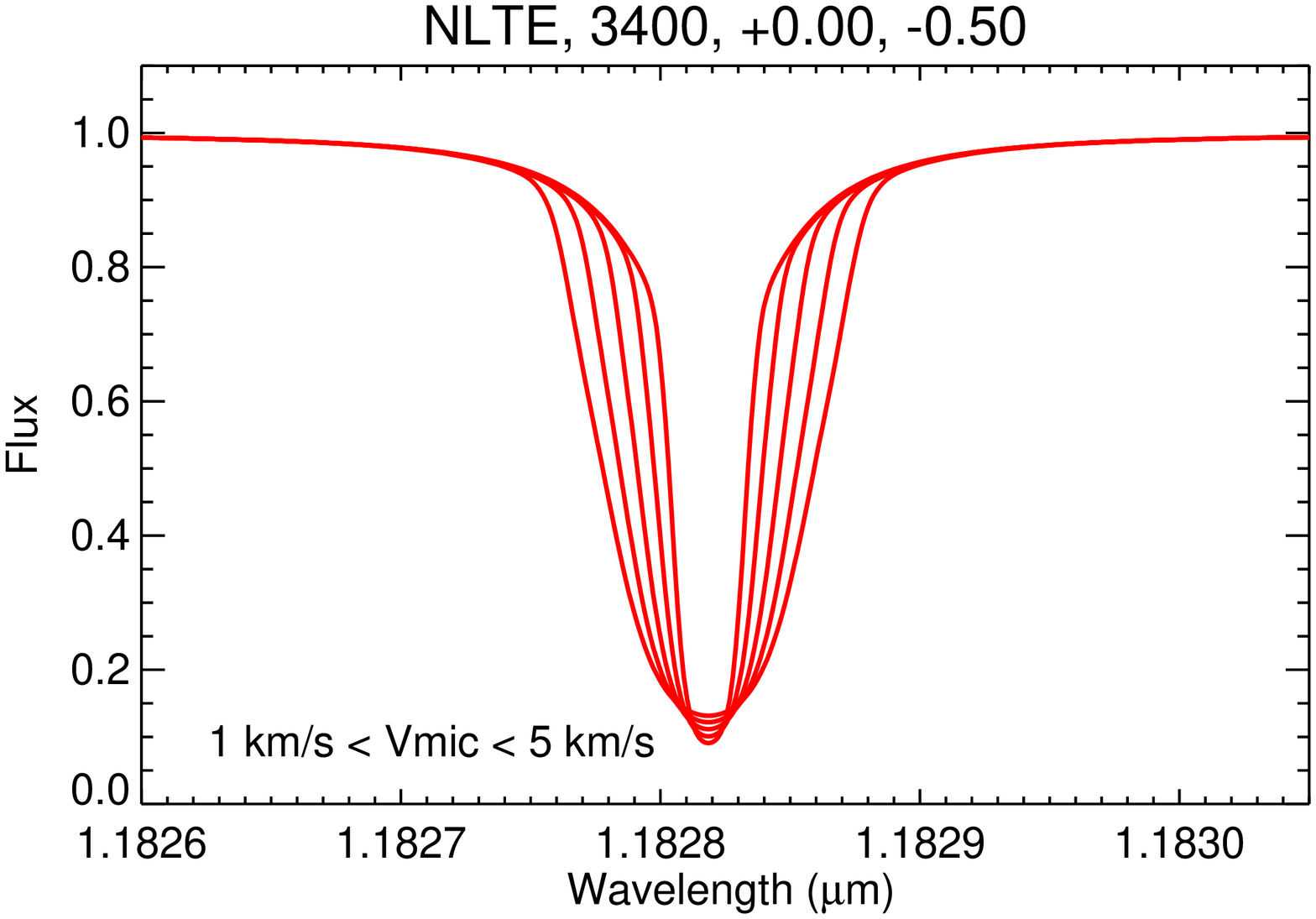}}
\hbox{
\includegraphics[width=0.3\textwidth, angle=90]{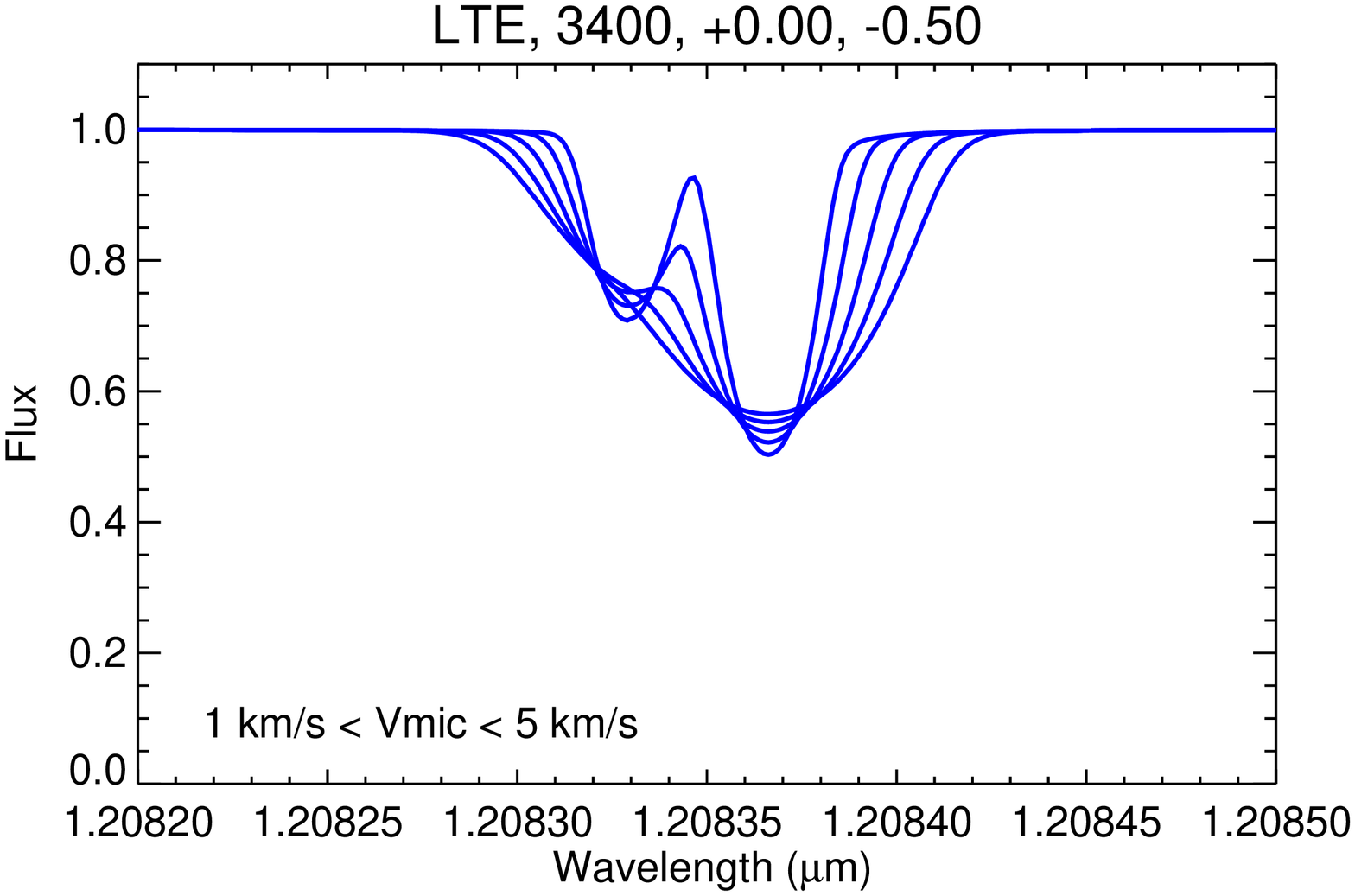}
\includegraphics[width=0.3\textwidth, angle=90]{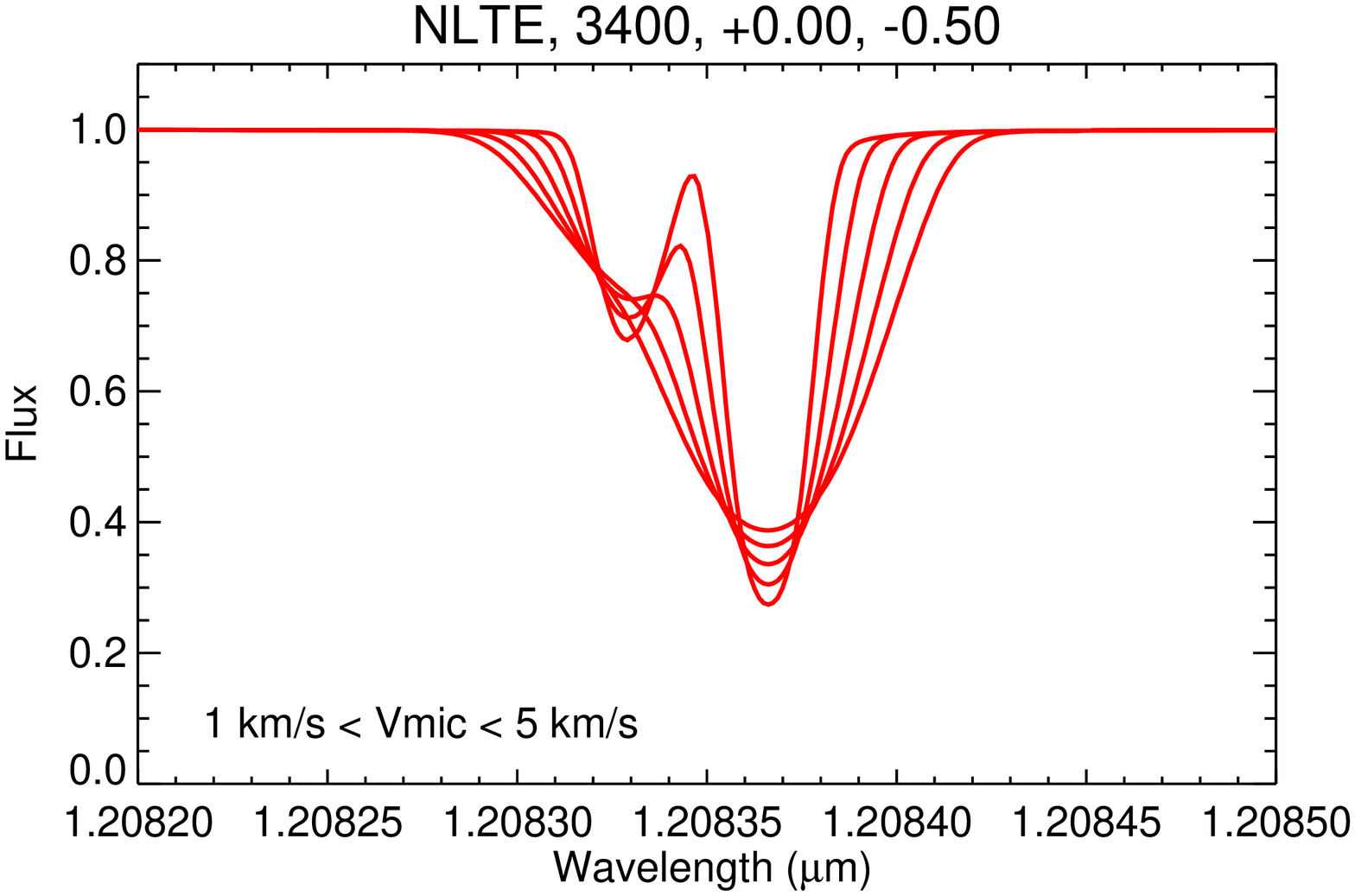}}
\caption{LTE (left) and NLTE (right) lines profiles for the 11828 (top) and 
12083 (bottom) \mgi\ lines as a function of microturbulence. $\Vmic$ 
varies from 1 to 5 \kms\ with a step of 1 \kms.}
\label{microturbulence}
\end{figure*}

$\Delta_{\rm Mg~I}$ is the logarithmic correction, which has to be applied to 
an LTE magnesium abundance determination of a specific line, $\log \rm{A 
(Mg)}_{\rm LTE}$, to obtain 
the correct value corresponding to the use of NLTE line formation. These 
corrections are obtained  at each point of our model grid by matching the NLTE
equivalent width through varying the Mg abundance in the LTE calculations.
When for the same element abundance the NLTE line equivalent width is
larger than the LTE one, it requires a  higher LTE abundance to fit the
NLTE equivalent width and, thus, the NLTE abundance corrections become 
negative. Fig. \ref{nltegrid1} shows the NLTE abundance corrections 
computed with two values of microturbulence, 2 and 5 \kms. The exact 
values of the  NLTE  abundance corrections are given in Table 4.
The results for Mg are very similar to those we obtained for the J-band 
Si lines (Paper II). Fig. \ref{abundance} also shows the effect of varying Mg 
abundance ([Z] fixed to the solar value) on the line profiles for the both Mg I 
J-band lines and Fig. \ref{microturbulence} illustrates the impact of 
microturbulence 
$\Vmic$. Clearly, the larger $\Vmic$ the stronger a spectral line, 
demonstrating the fundamental degeneracy between small scale turbulent 
broadening and 
abundance. The 11828 \AA\ line and the strongest component of the 12083 
\AA\ line usually occupy the flat part of the curve-of-growth and are very 
sensitive to the microturbulence velocity. In contrast, the two weaker 
components of the J-band triplet at 12083 \AA\ are on the linear part of the 
curve-of-growth even in very cool atmospheres ($\Teff = 3400$ K).

The NLTE abundance corrections are significant with large negative values 
between $-0.4$ to $-0.1$ dex and are strongest at low metallicity [Z].  We see 
a clear trend with effective temperature, the effect being significantly 
stronger for the 12083 \AA\ line.
\section{\mgi\ J-band lines of Per OB1 red supergiants}
\citet{gazak14b} investigated high resolution, high S/N J-band spectra of  
eleven RSGs in the young massive stellar double cluster h and $\chi$ Persei 
(Per OB1) in the solar neighbourhood as a crucial test of the J-band analysis 
method. While this test nicely confirmed the reliability of the method with an 
average cluster metallicity [Z] $=-0.04$ derived from the spectra, the authors 
excluded the \mgi\ lines from the analysis (see their Figures 1 and 2) because 
of the obvious apparent NLTE effects for which no NLTE calculations were 
available at the time of their analysis work. Now with our new calculations at 
hand, we can use the stellar parameters determined by Gazak et al. and check  
whether observed and calculated \mgi\ J-band line profiles agree.

\begin{figure*}[ht!]
\hbox{
\includegraphics[width=0.3\textwidth, angle=-90]{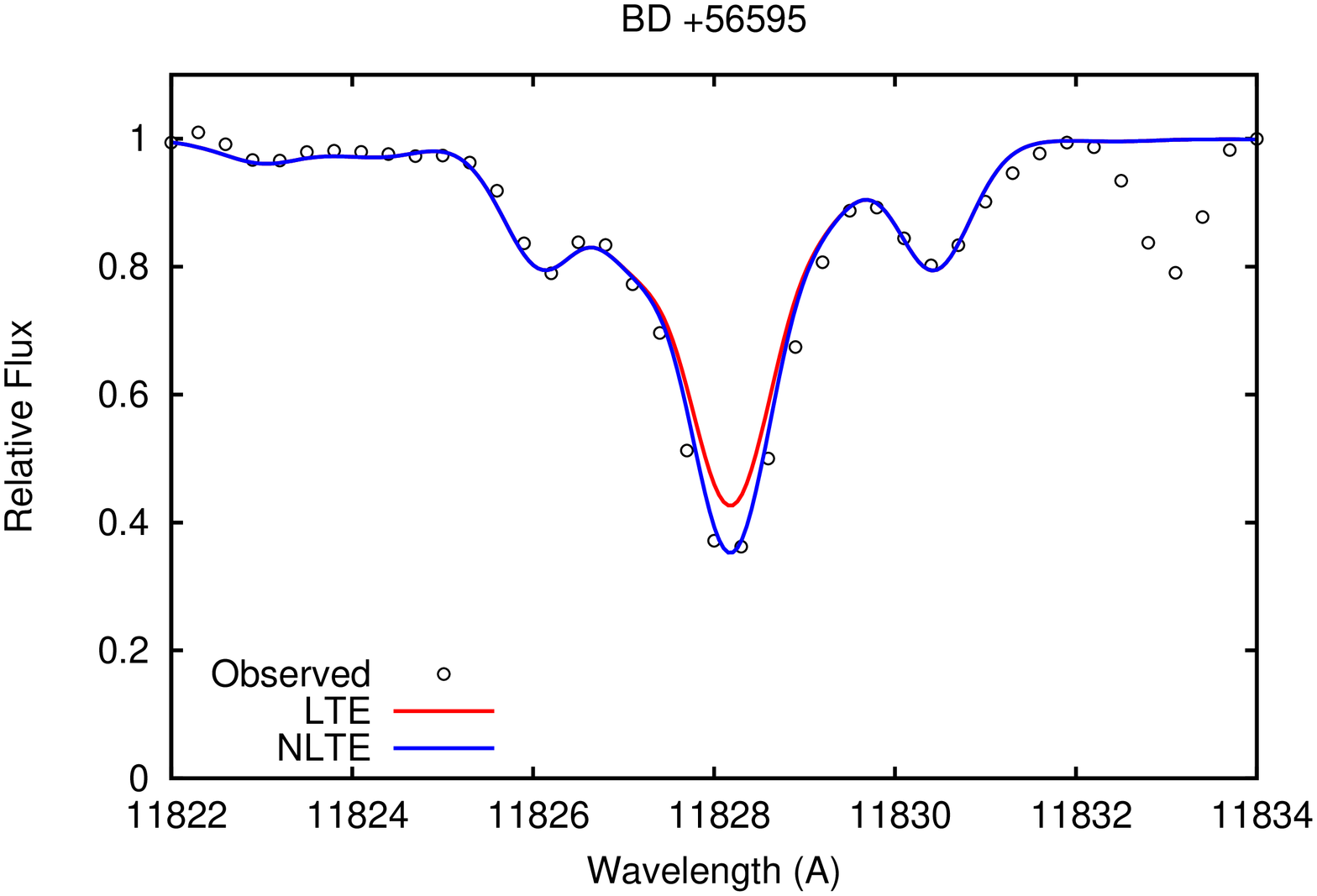}
\includegraphics[width=0.3\textwidth, angle=-90]{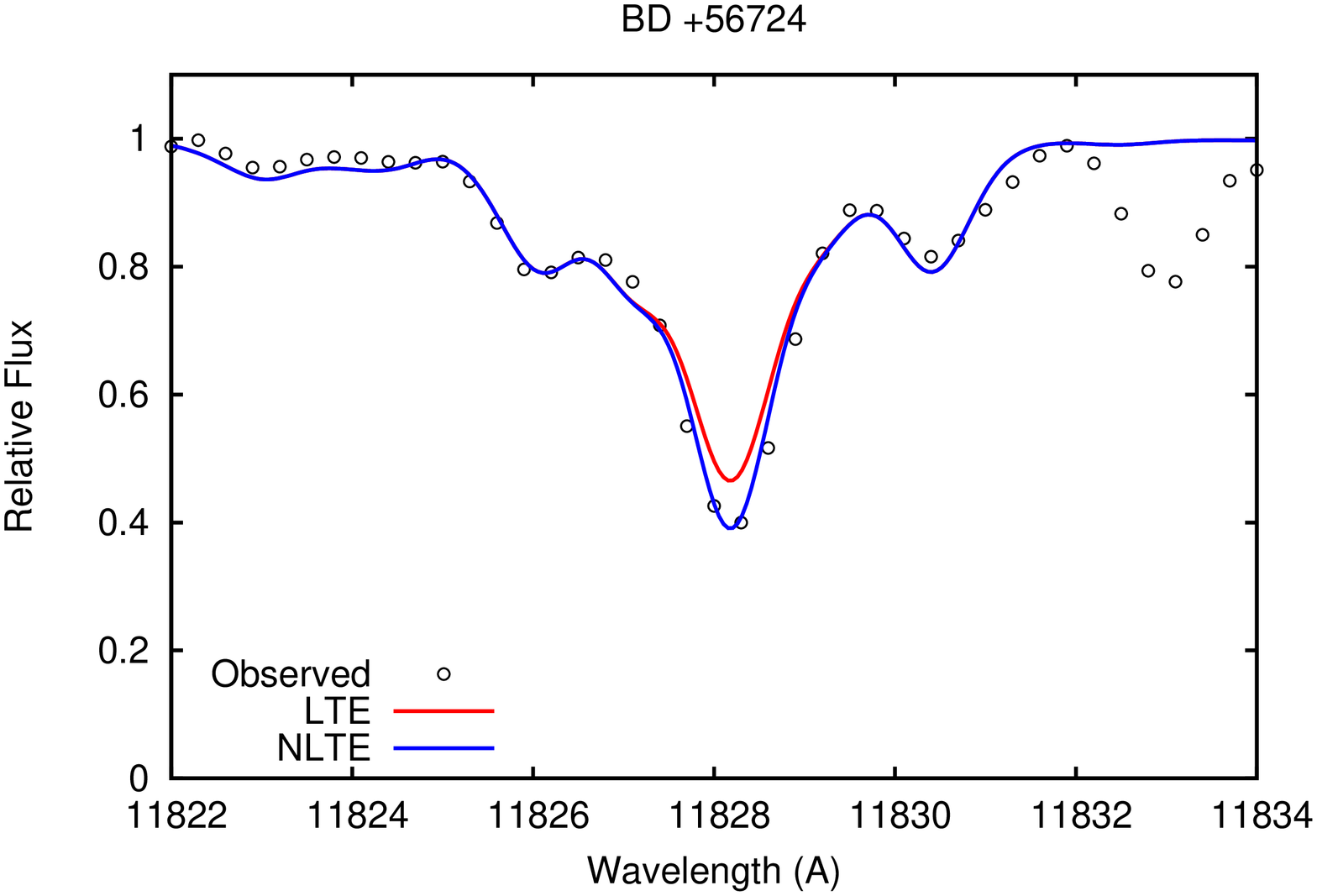}}
\hbox{
\includegraphics[width=0.3\textwidth, angle=-90]{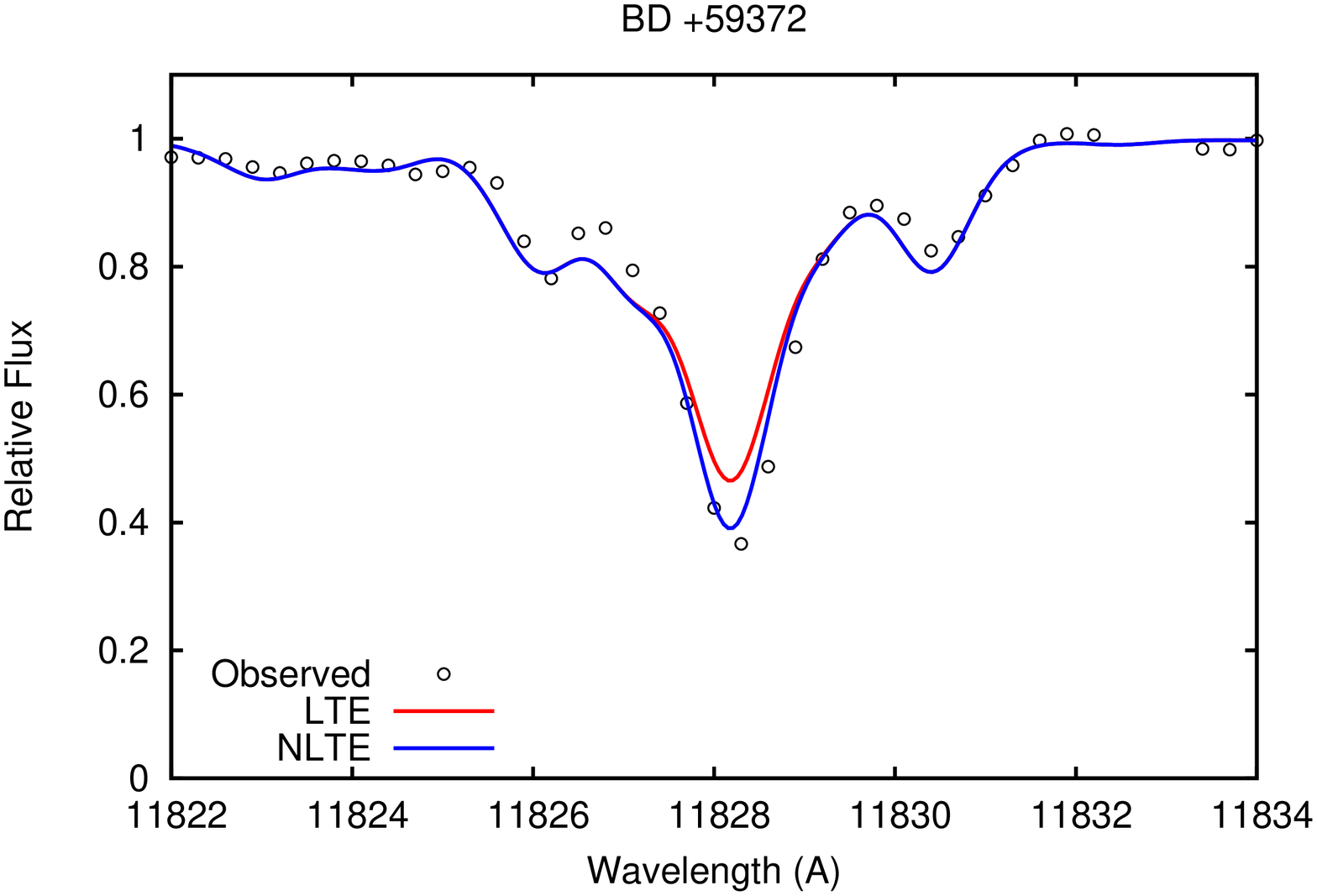}
\includegraphics[width=0.3\textwidth, angle=-90]{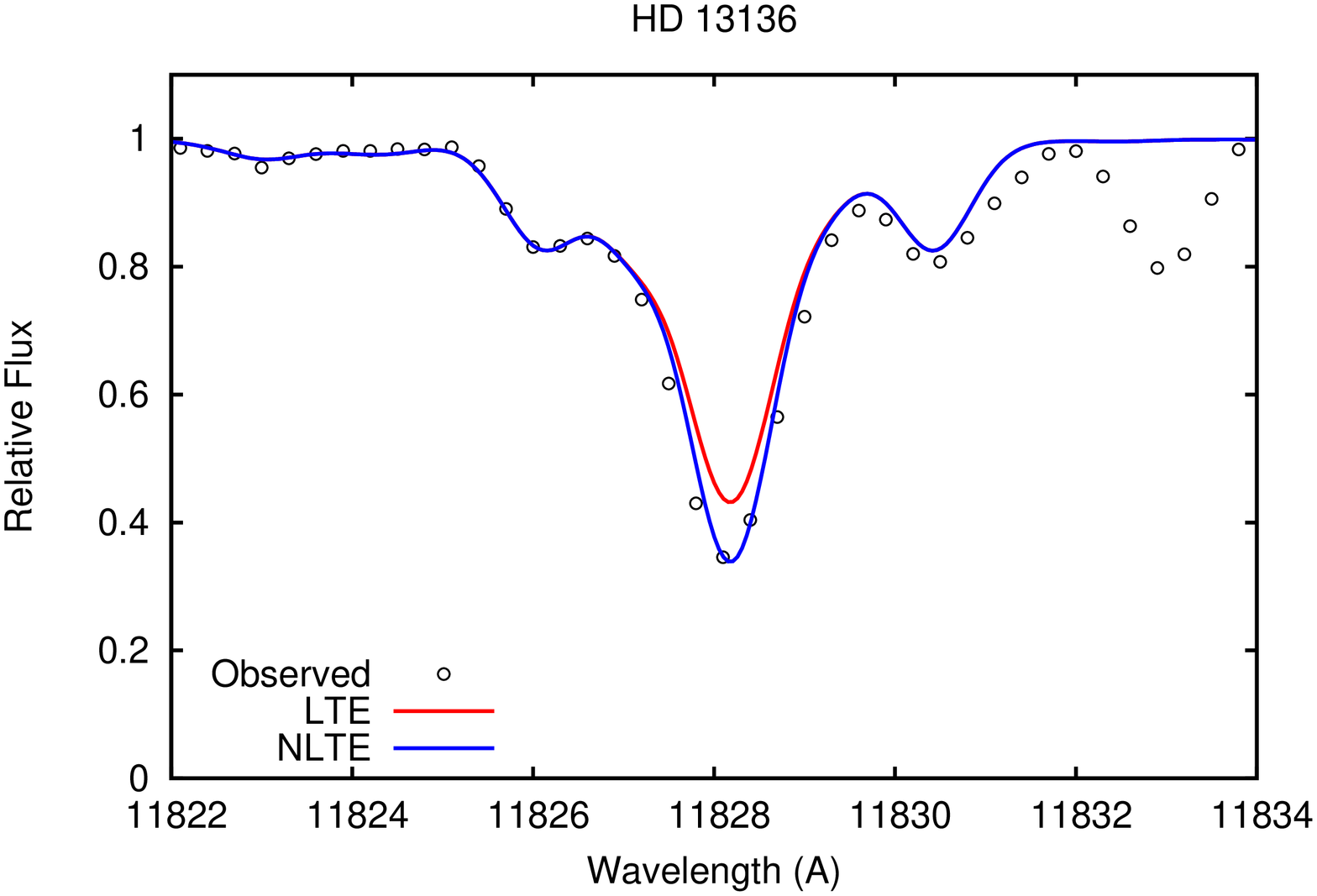}}
\hbox{
\includegraphics[width=0.3\textwidth, angle=-90]{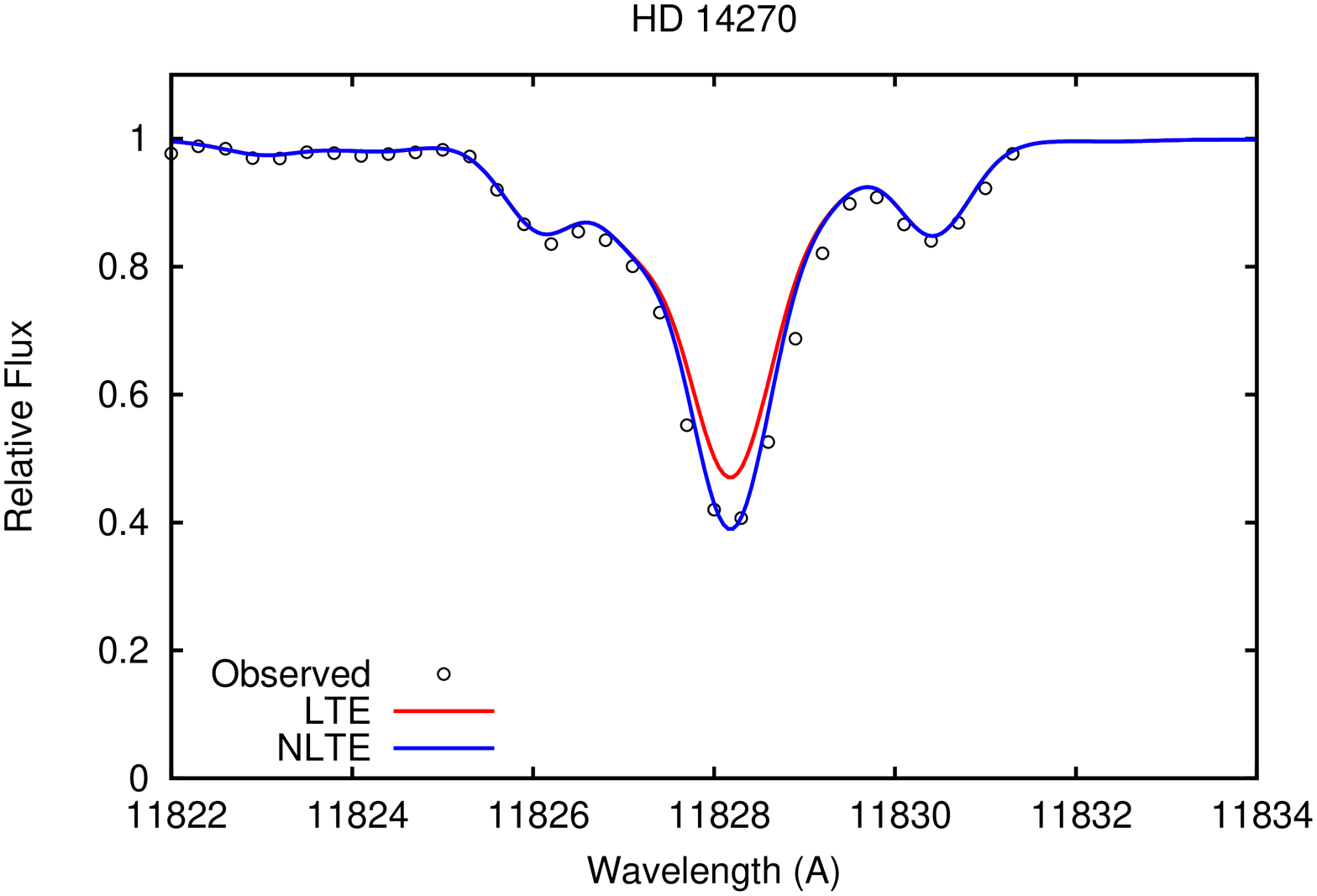}
\includegraphics[width=0.3\textwidth, angle=-90]{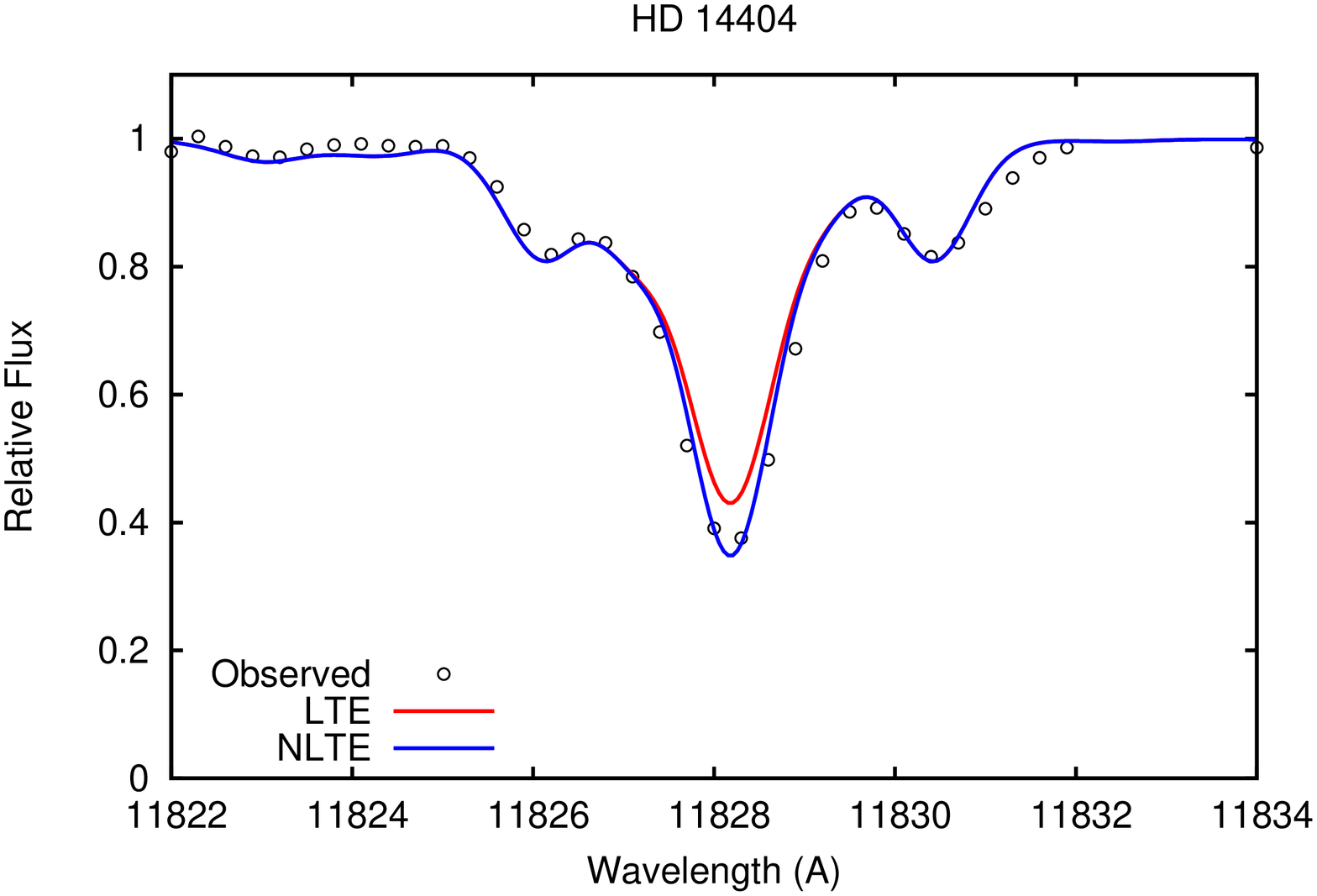}}
\caption{Observed J-band \mgi\ profiles computed in LTE and NLTE for the 
atmospheric parameters determined by Gazak et al. (2014b) as given in Table 2.}
\label{perob1a}
\end{figure*}

\begin{figure*}
\hbox{
\includegraphics[width=0.3\textwidth, angle=-90]{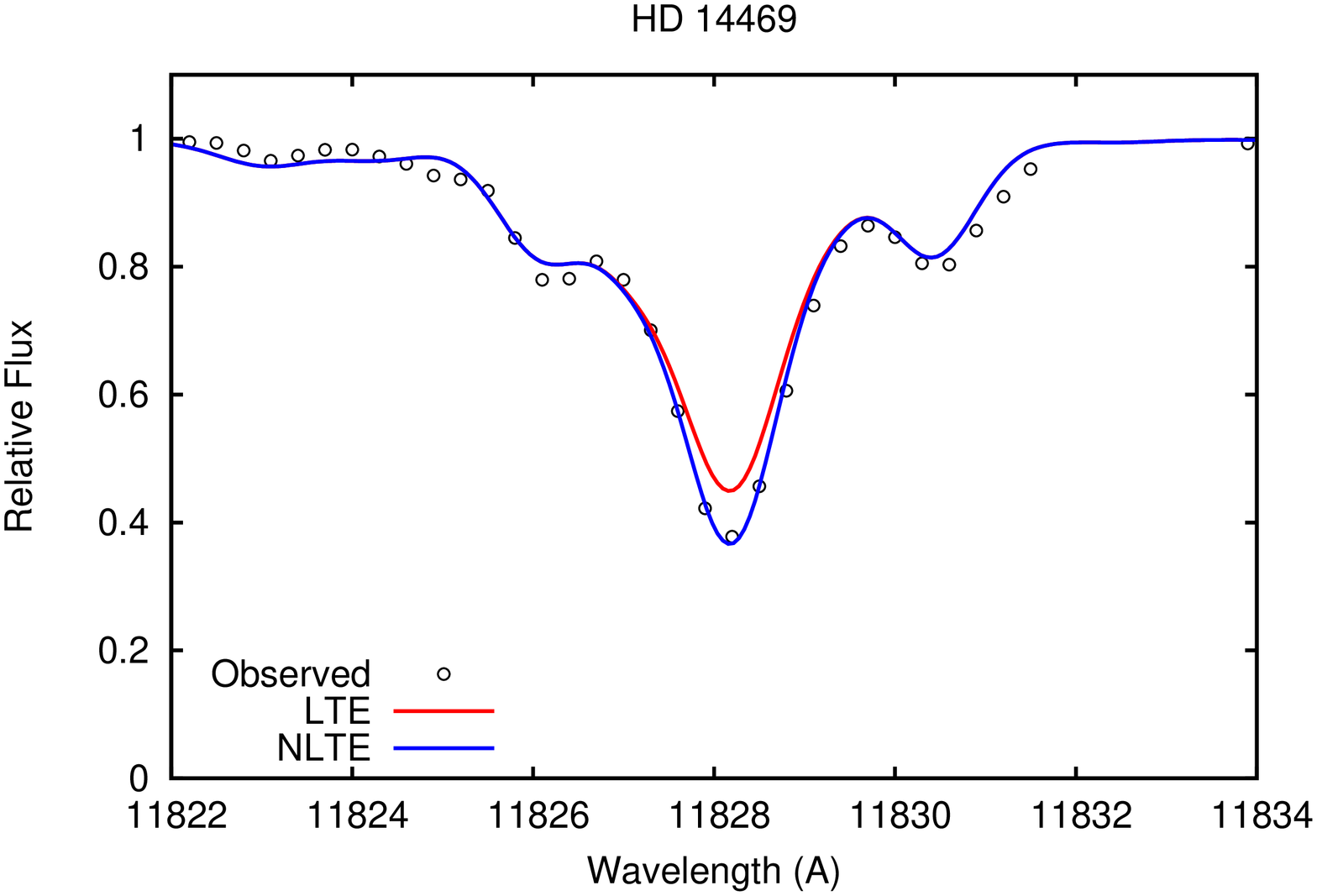}
\includegraphics[width=0.3\textwidth, angle=-90]{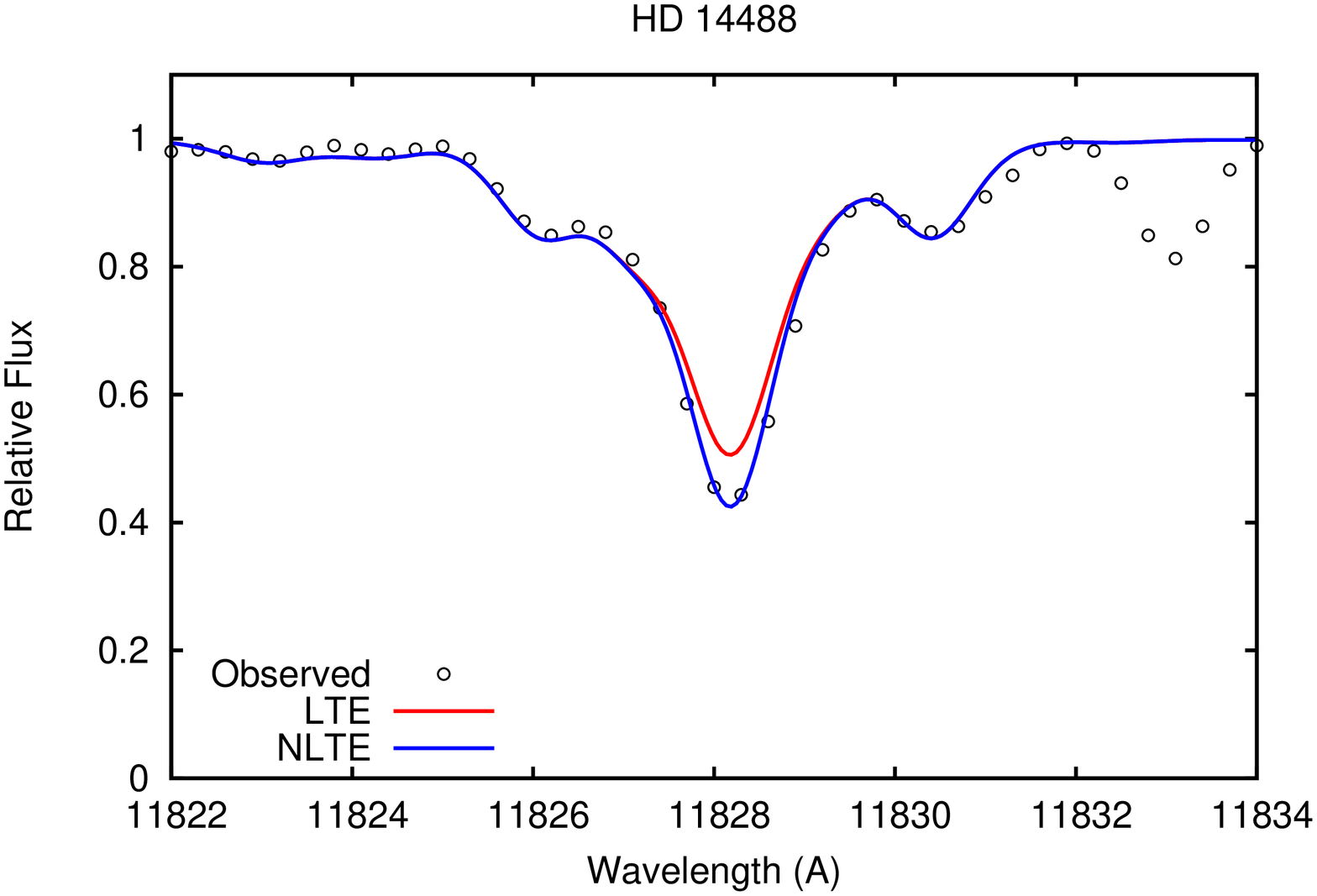}}
\hbox{
\includegraphics[width=0.3\textwidth, angle=-90]{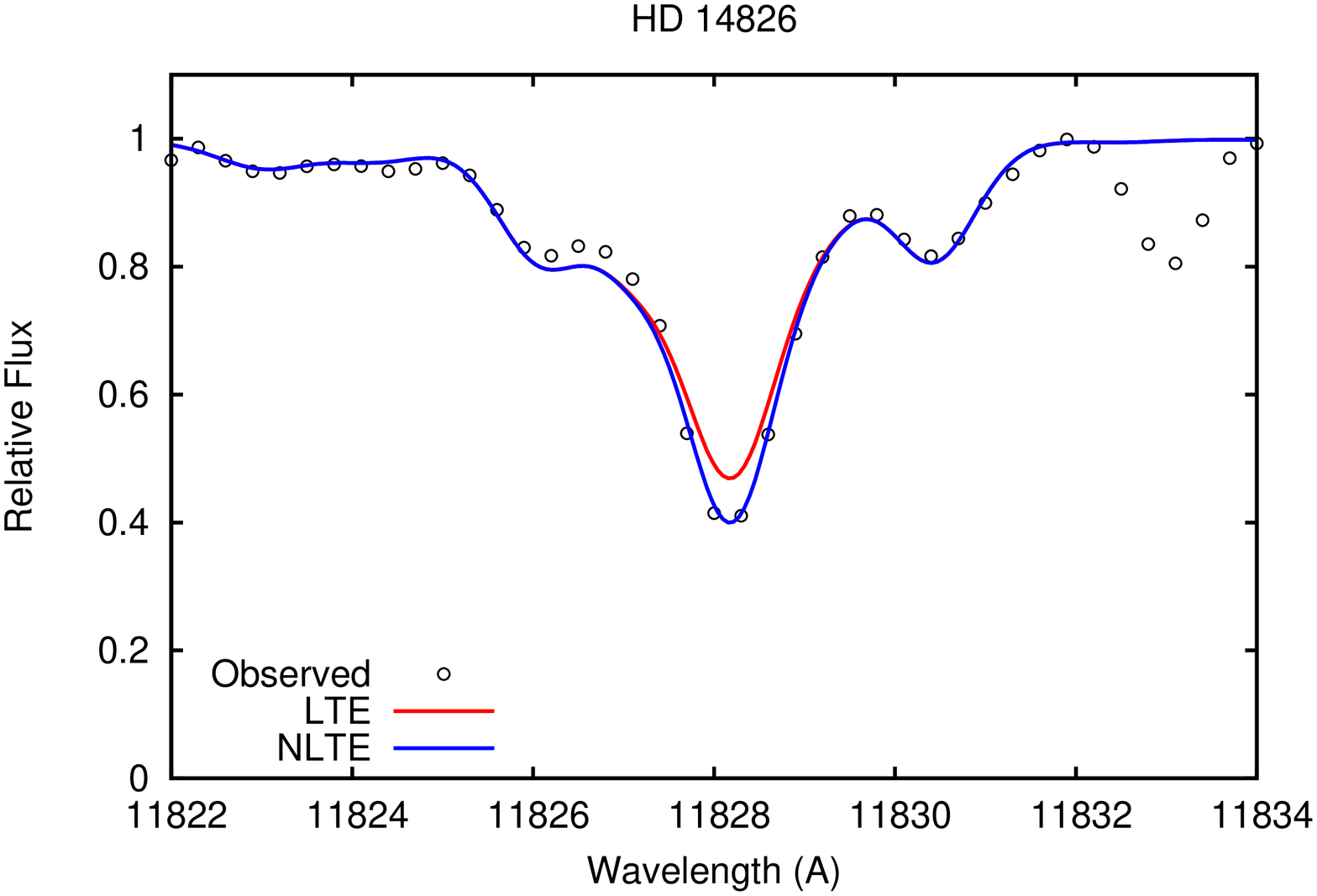}
\includegraphics[width=0.3\textwidth, angle=-90]{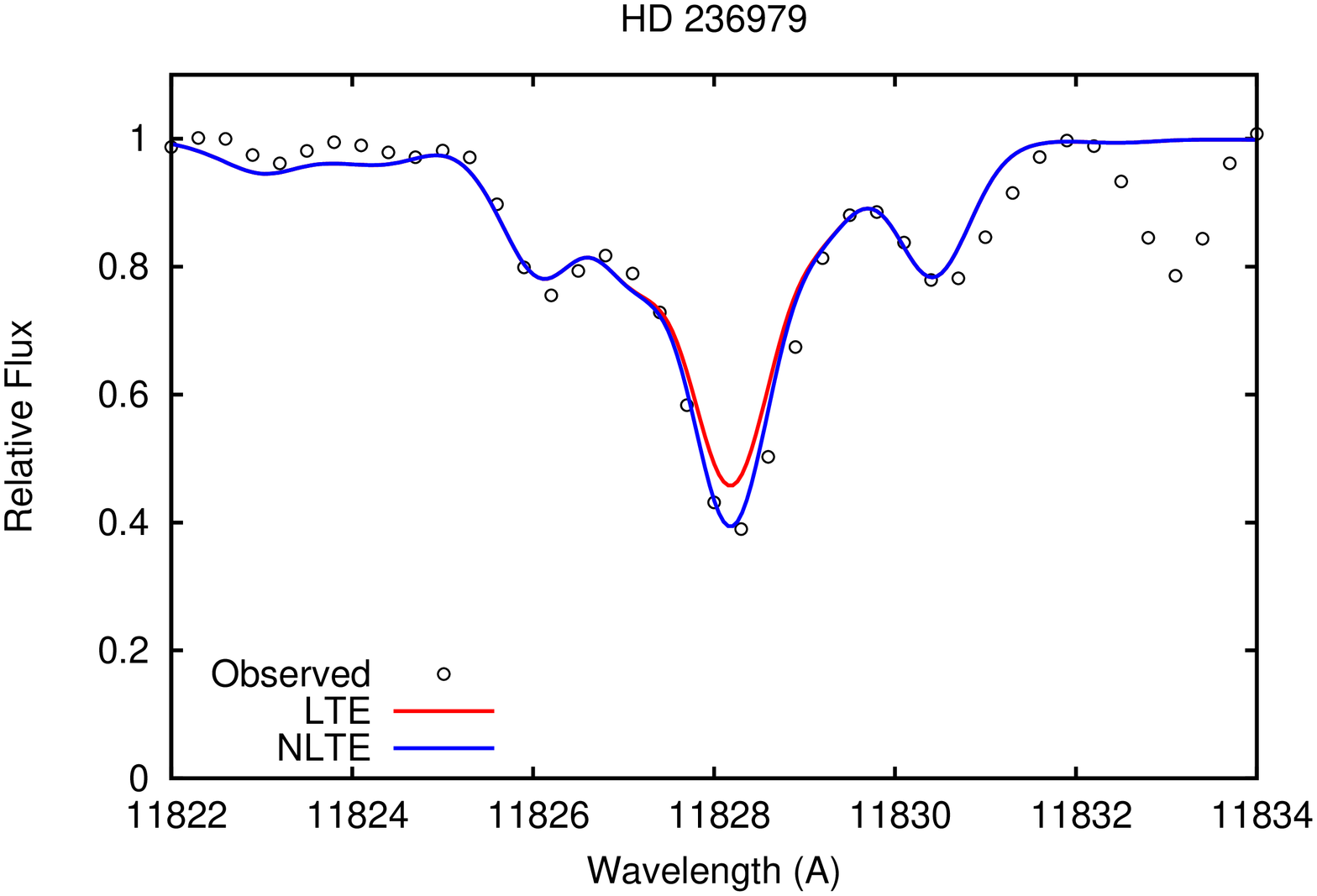}}
\caption{Observed J-band \mgi\ profiles computed in LTE and NLTE for the 
atmospheric parameters determined by Gazak et al. (2014b) as given in Table 2.}
\label{perob1b}
\end{figure*}
\begin{figure*}
\hbox{
\includegraphics[width=0.3\textwidth, angle=-90]{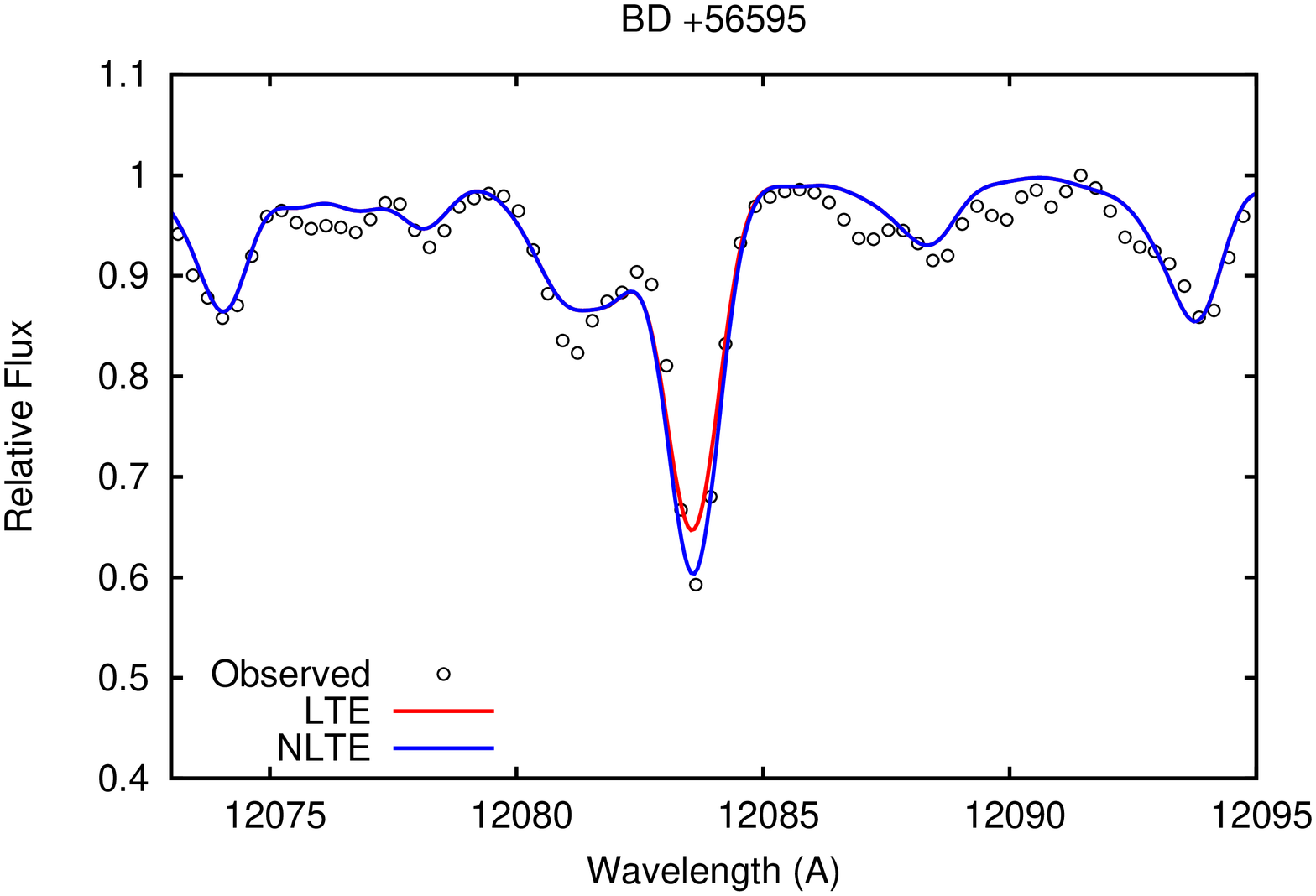}
\includegraphics[width=0.3\textwidth, angle=-90]{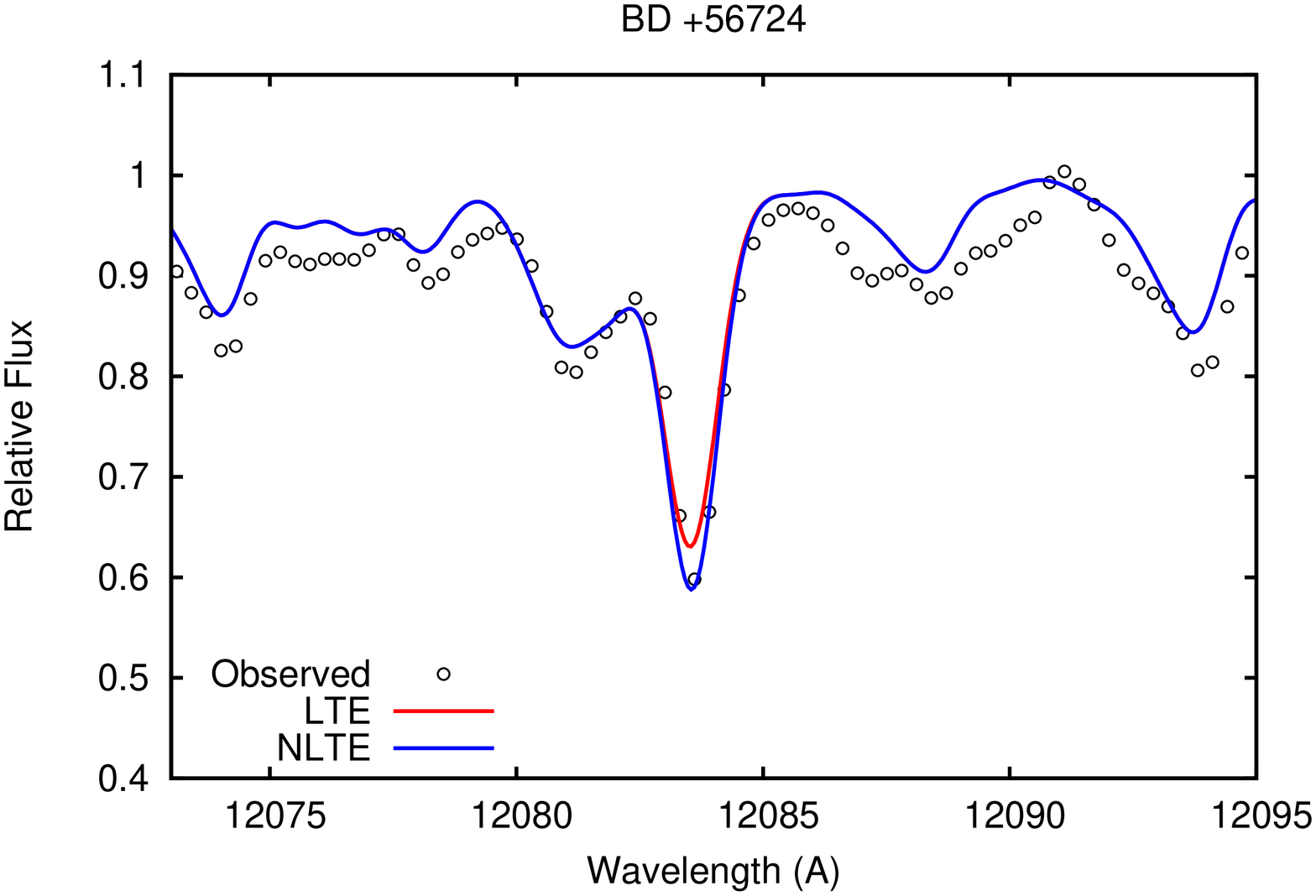}}
\hbox{
\includegraphics[width=0.3\textwidth, angle=-90]{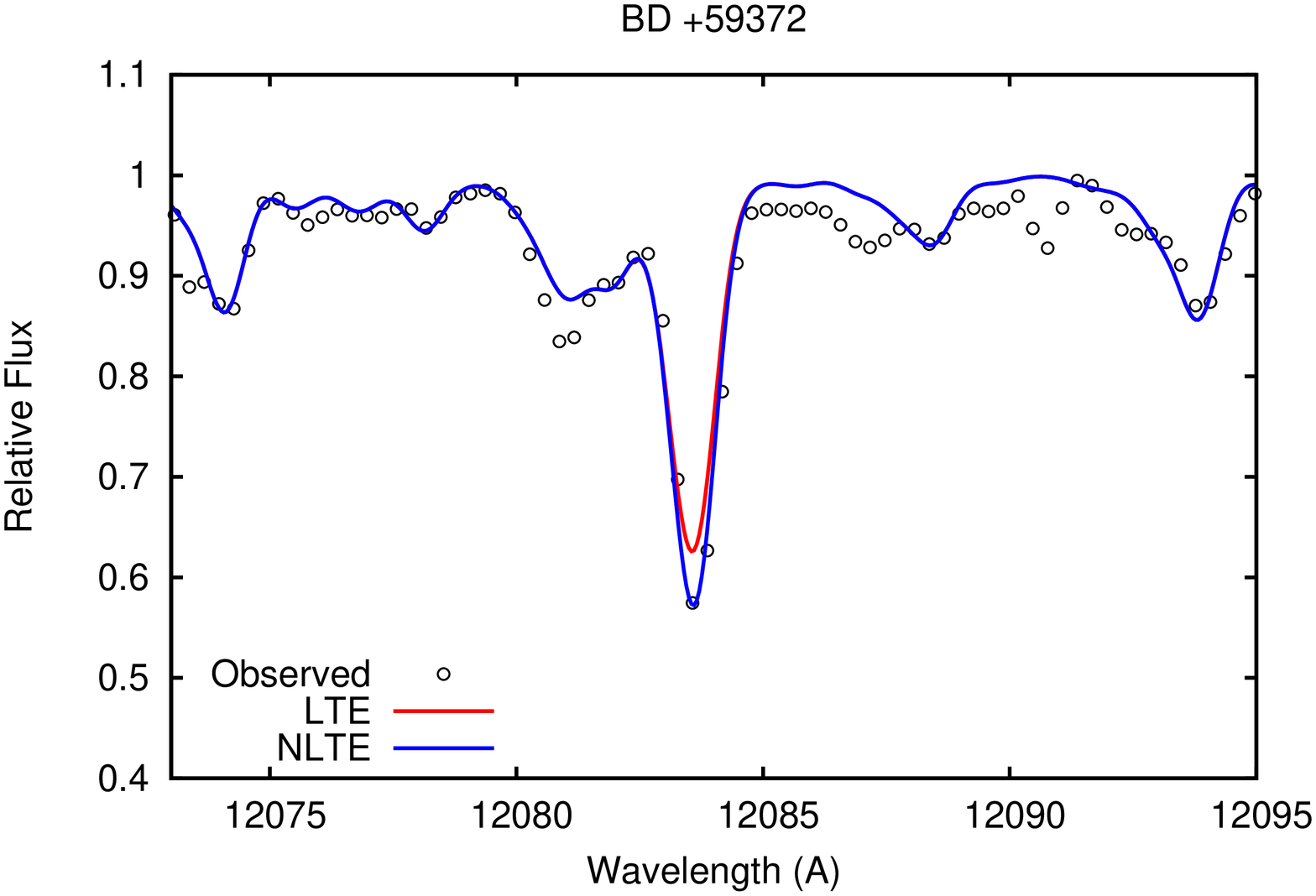}
\includegraphics[width=0.3\textwidth, angle=-90]{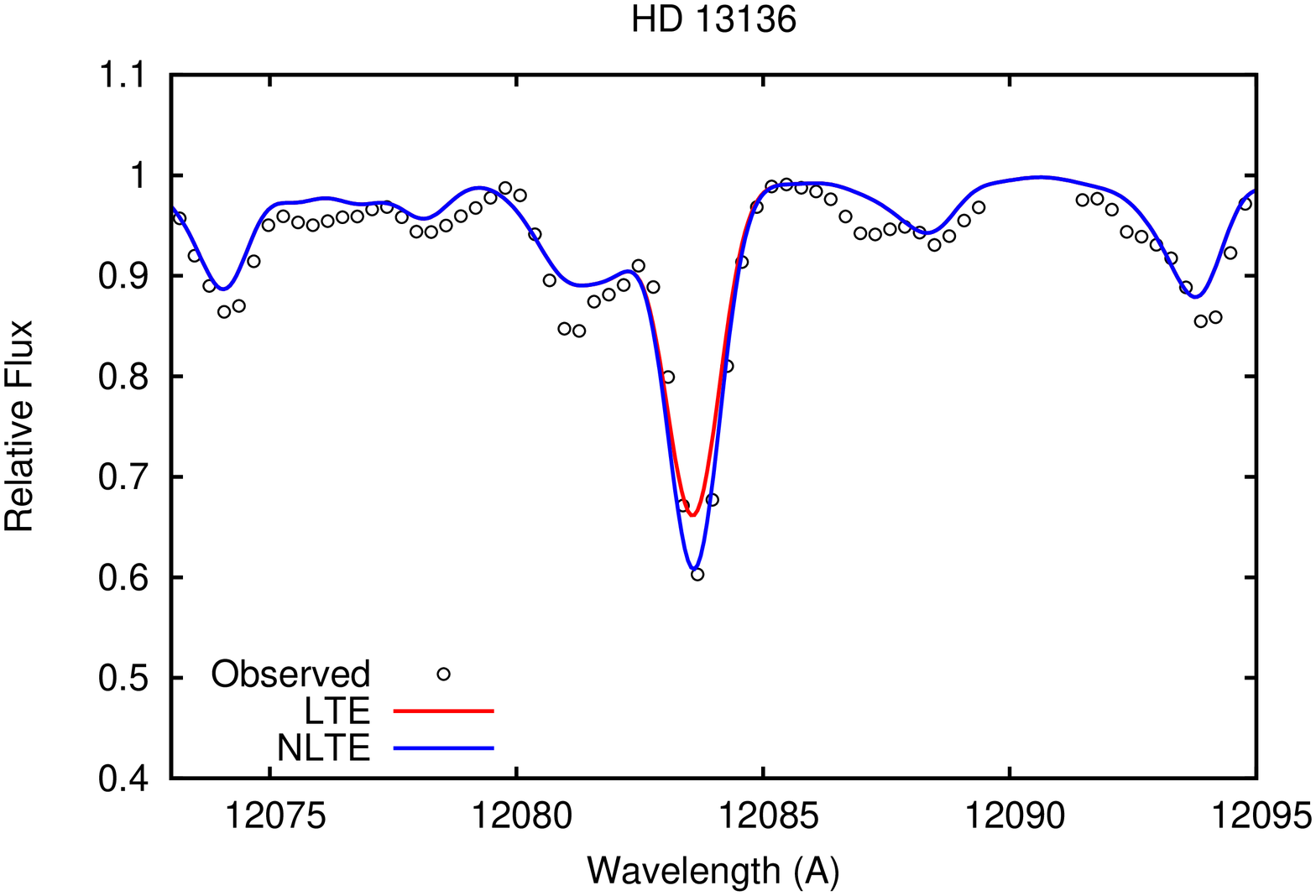}}
\hbox{
\includegraphics[width=0.3\textwidth, angle=-90]{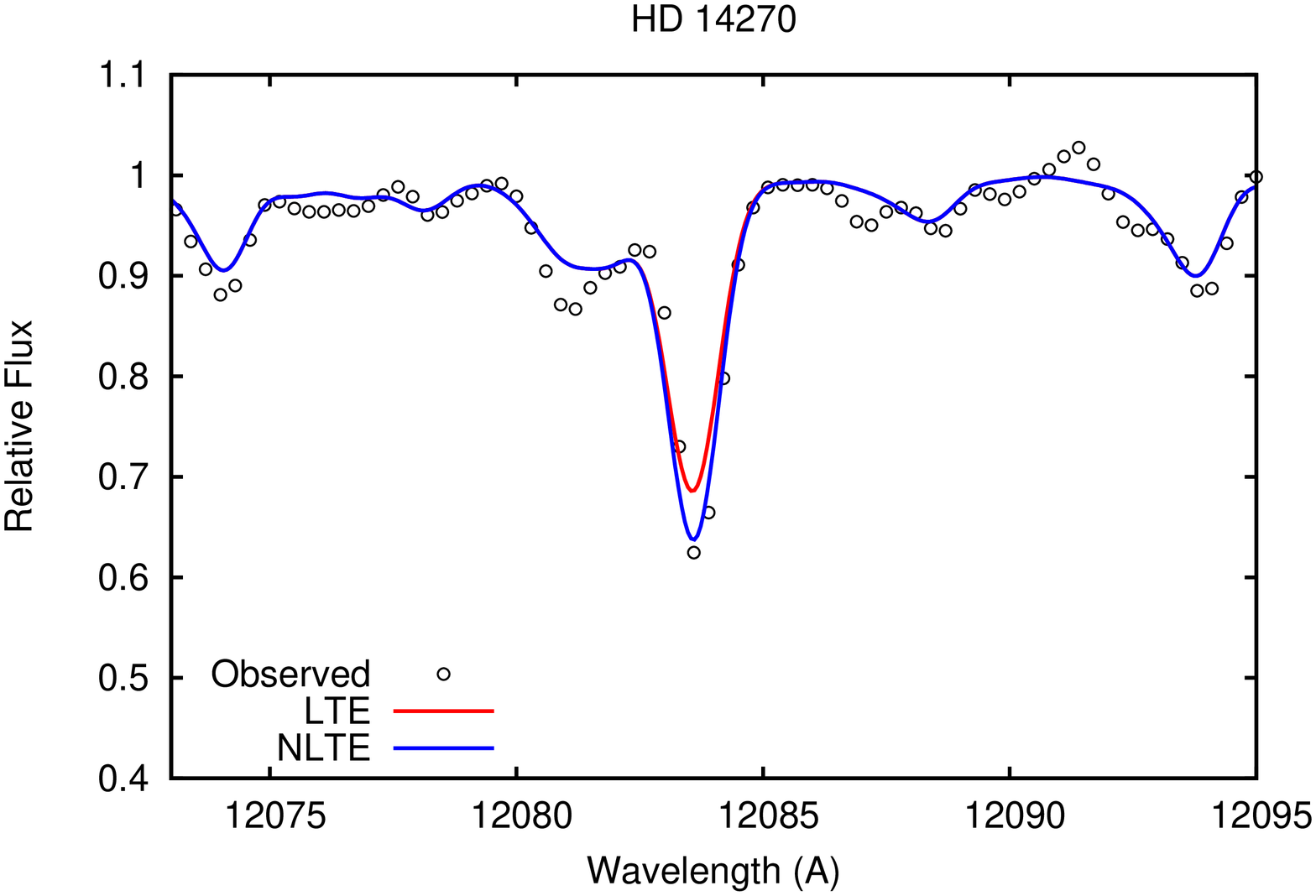}
\includegraphics[width=0.3\textwidth, angle=-90]{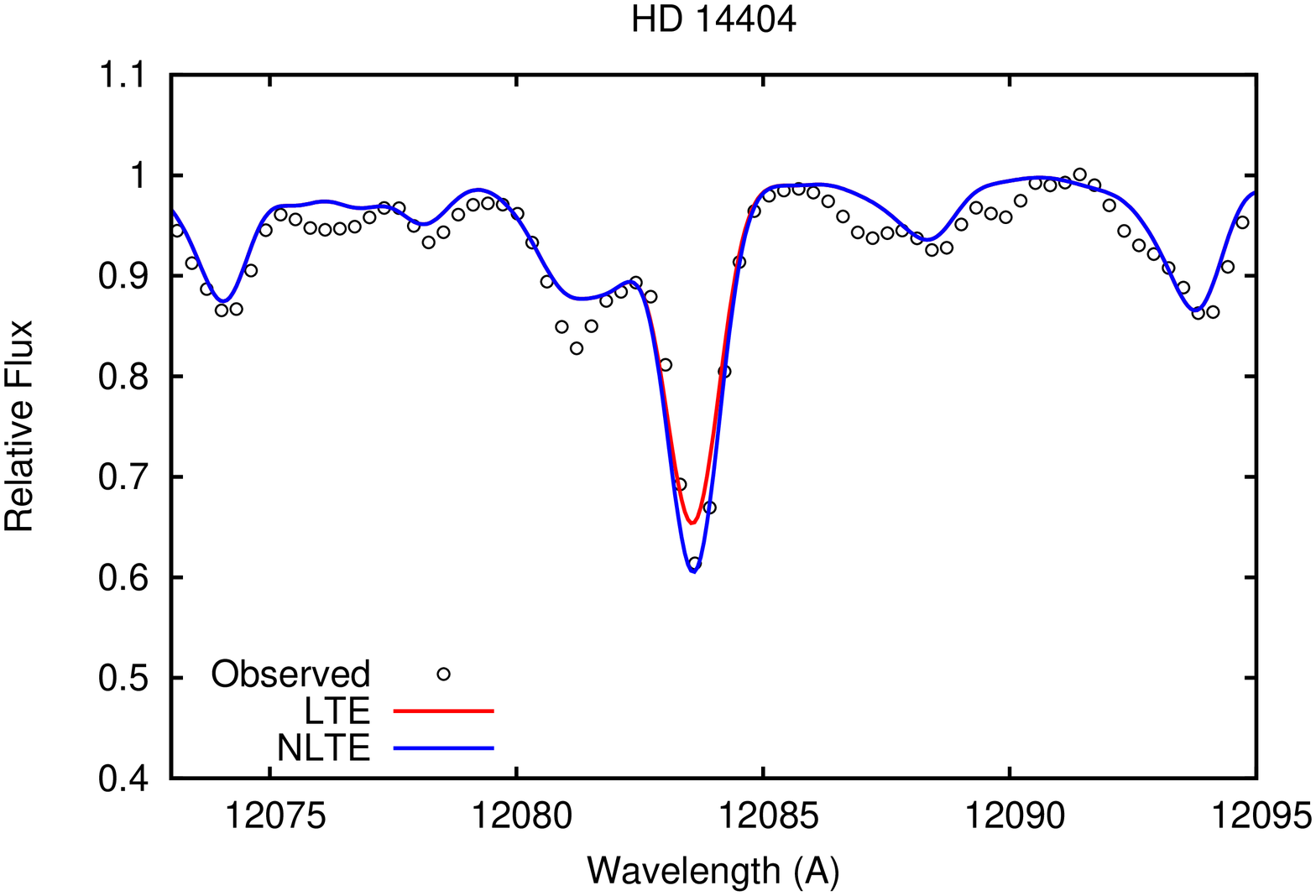}}
\caption{Observed J-band \mgi\ profiles computed in LTE and NLTE for the 
atmospheric parameters determined by Gazak et al. (2014b) as given in Table 2.}
\label{perob1c}
\end{figure*}

\begin{figure*}
\hbox{
\includegraphics[width=0.3\textwidth, angle=-90]{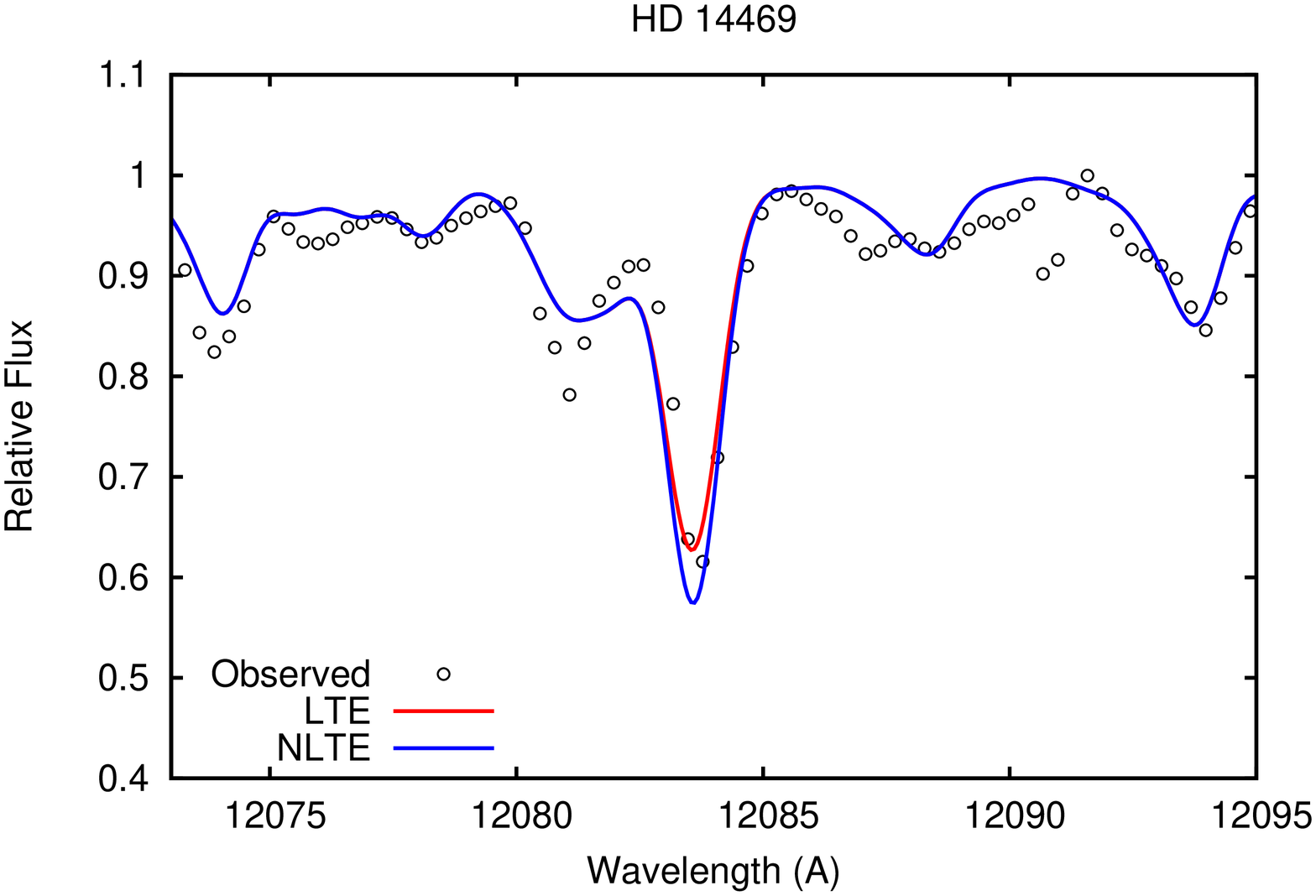}
\includegraphics[width=0.3\textwidth, angle=-90]{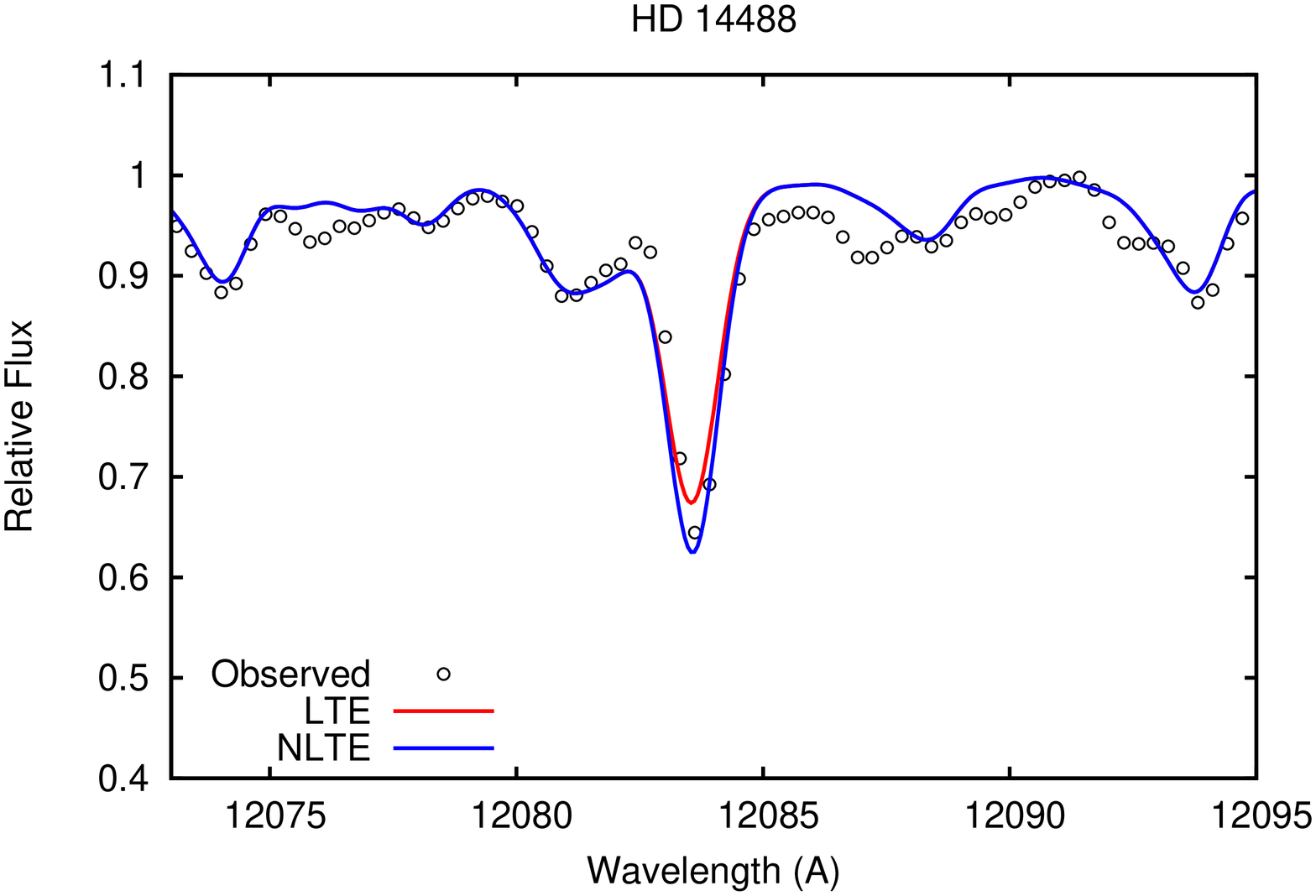}}
\hbox{
\includegraphics[width=0.3\textwidth, angle=-90]{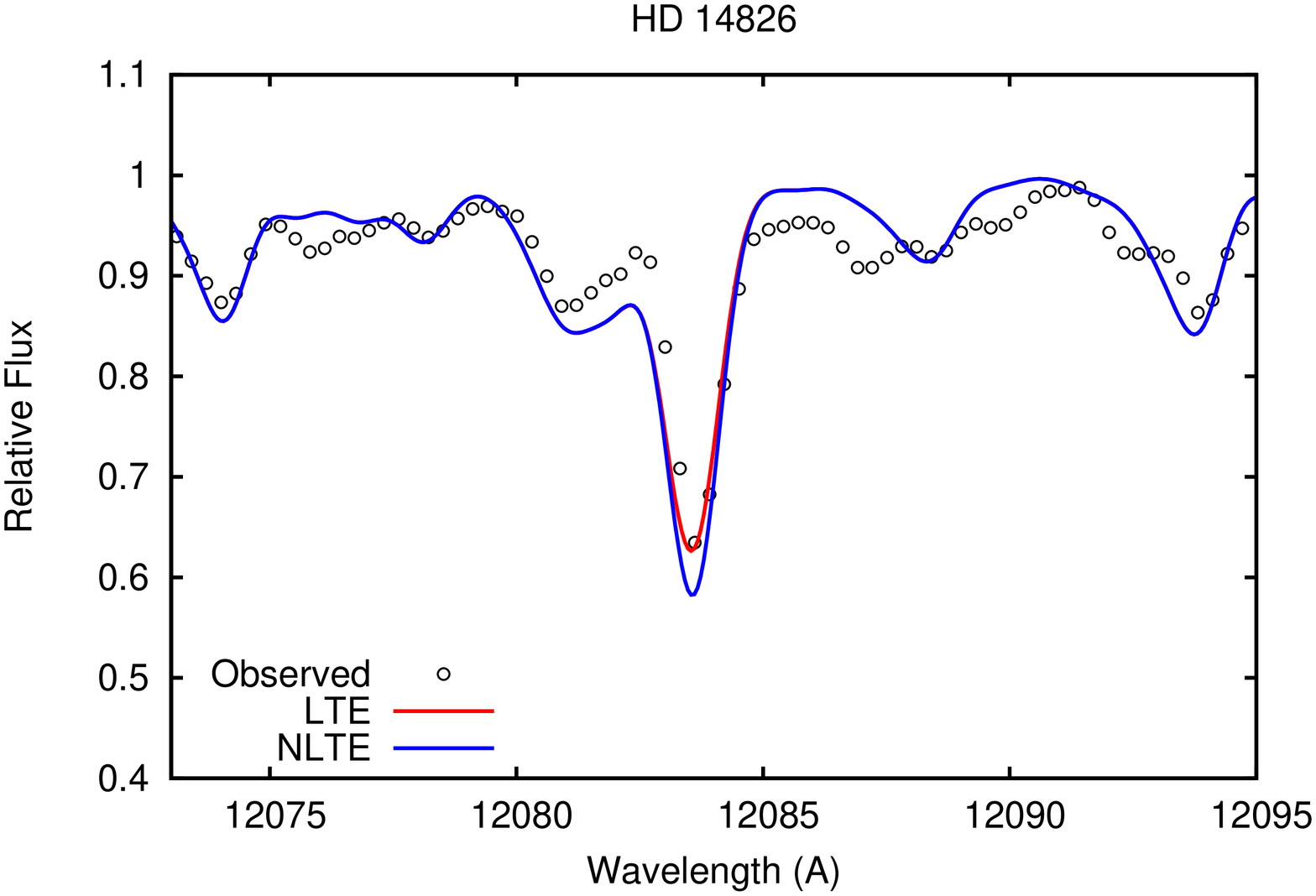}
\includegraphics[width=0.3\textwidth, angle=-90]{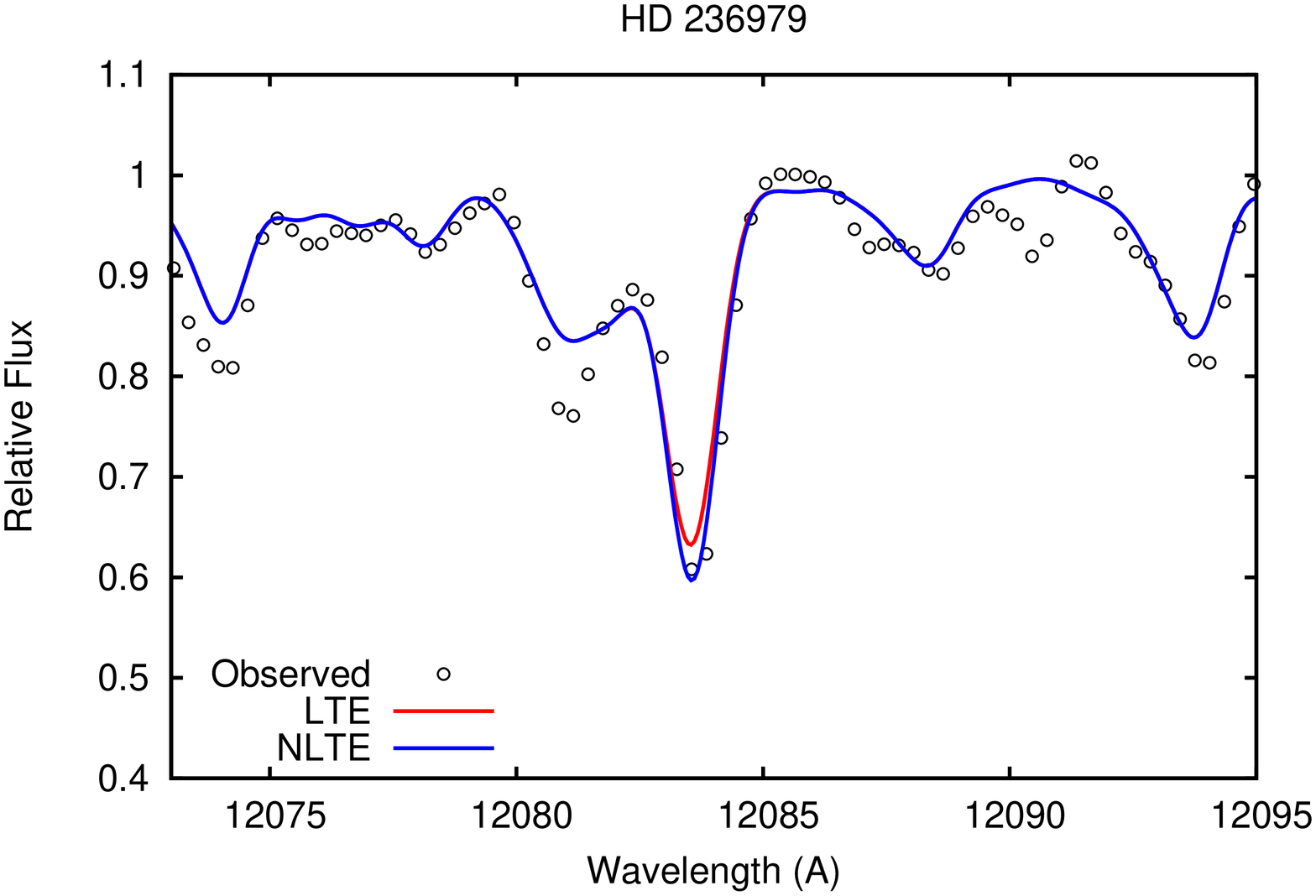}}
\caption{Observed J-band \mgi\ profiles computed in LTE and NLTE for the 
atmospheric parameters determined by Gazak et al. (2014b) as given in Table 2.}
\label{perob1d}
\end{figure*}
The  comparison of  NLTE and LTE fits to Mg I lines for the Per OB-1 spectra of 
10 stars is shown in Fig. \ref{perob1a}, \ref{perob1b}, \ref{perob1c}, 
\ref{perob1d}. The stellar parameters are given in Table 2. A solar value is 
adopted for the [Mg$/$Fe] ratio in the calculation. The agreement between the 
new NLTE calculations and the observations is much better than with the 
previous 
atomic data and LTE. This confirms that for future work the \mgi\ J-band lines 
can be used as an additional constraint of metallicity.  In addition, since the 
\mgi\ lines have the highest excitation potential of the lower levels of 
their transitions compared to the iron, titanium and silicon lines used so far 
in the J-band technique, they will also be very useful to constrain effective 
temperature and gravity. This will strengthen the accuracy of the method 
significantly.
\section{Conclusions}

With the new Mg I NLTE calculations presented in this work we are now  
able to use the full J-band spectrum of atomic lines (iron, titanium, silicon 
and magnesium) for a detailed analysis of red supergiant stars. This is an 
important
step forward for the IR spectroscopy of bright stars in the Milky Way,
e.g. individual RSGs (see for instance \citealt{gazak14b}), but also 
extra-galactic stellar populations, such as young and very massive super star 
clusters for which the J-band spectra are completely dominated by RSGs once 
they are older than 7 Myr. \citet{gazak14a}  have 
demonstrated  that the J-band technique applied to SSCs yields reliable 
metallicities, but the accuracy of them was somewhat compromised by the fact 
that the \mgi\ lines could not be used because of the importance of NLTE 
effects. With the new calculations available it will be possible to 
significantly improve this work and to fully use the tremendous potential 
of the J-band method.
%
%
%
%
%
\begin{deluxetable}{ccc cccc | cccc}
\tabletypesize{\scriptsize}
\tablecolumns{11}
\tablewidth{0pt}
\tablecaption{Equivalent widths \tablenotemark{a} of the \mgi\ lines ($\xi_t$ = 2 and 5 kms$^{-1}$)}
\tablehead{
\colhead{T$_{\rm eff}$}    &
\colhead{$log~g$}        &
\colhead{[Z]}            &
\colhead{}            &
\colhead{}            &
\colhead{$\xi_t$ = 2 kms$^{-1}$}   &
\colhead{}            &
\colhead{}            &
\colhead{$\xi_t$ = 5 kms$^{-1}$}   &
\colhead{}            &
\colhead{} \\[1mm]
\colhead{}            &
\colhead{}            &
\colhead{}            &
\colhead{$W_{\lambda,\rm Mg I}$}     &
\colhead{$W_{\lambda,\rm Mg I}$}     &
\colhead{$W_{\lambda,\rm Mg I}$}     &
\colhead{$W_{\lambda,\rm Mg I}$}     &
\colhead{$W_{\lambda,\rm Mg I}$}     &
\colhead{$W_{\lambda,\rm Mg I}$}     &
\colhead{$W_{\lambda,\rm Mg I}$}     &
\colhead{$W_{\lambda,\rm Mg I}$}   \\
\colhead{}            &
\colhead{}            &
\colhead{}            &
\colhead{$11828$}     &
\colhead{$11828$}     &
\colhead{$12083$}     &
\colhead{$12083$}     &
\colhead{$11828$}     &
\colhead{$11828$}     &
\colhead{$12083$}     &
\colhead{$12083$}   \\[1mm]
\colhead{}            &
\colhead{}            &
\colhead{}            &
\colhead{$LTE$}       &
\colhead{$NLTE$}      &
\colhead{$LTE$}       &
\colhead{$NLTE$}      &
\colhead{$LTE$}       &
\colhead{$NLTE$}      &
\colhead{$LTE$}       &
\colhead{$NLTE$}    \\[1mm]
\colhead{(1)}	&
\colhead{(2)}	&
\colhead{(3)}	&
\colhead{(4)}	&
\colhead{(5)}	&
\colhead{(6)}   &
\colhead{(7)}   & 
\colhead{(8)}	&
\colhead{(9)}	&
\colhead{(10)}  &
\colhead{(11)}  }
\startdata
\\[-1mm]
4400&  $-$0.50 & ~~~ 0.00 & 483.2 & 511.7 & 404.8 & 413.5 &  847.2&  899.6 & 575.2 & 597.0 \\
4400&  $-$0.50 & ~~~ 0.50 & 582.9 & 608.4 & 501.5 & 504.1 &  972.6& 1021.3 & 719.8 & 730.0 \\
4400&  $-$0.50 &  $-$0.50 & 403.5 & 436.8 & 300.6 & 323.4 &  721.9&  782.3 & 426.3 & 468.0 \\
4200&  $-$0.50 & ~~~ 0.00 & 519.9 & 553.1 & 425.5 & 439.8 &  888.8&  951.9 & 607.3 & 636.0 \\
4200&  $-$0.50 & ~~~ 0.50 & 640.5 & 671.9 & 519.9 & 531.6 & 1021.6& 1083.8 & 748.3 & 771.3 \\
4200&  $-$0.50 &  $-$0.50 & 434.1 & 469.7 & 326.3 & 348.9 &  770.5&  836.6 & 464.4 & 505.0 \\
4000&  $-$0.50 & ~~~ 0.00 & 554.9 & 597.3 & 427.8 & 452.7 &  911.0&  991.9 & 612.0 & 655.5 \\
4000&  $-$0.50 & ~~~ 0.50 & 696.7 & 738.0 & 516.4 & 539.9 & 1052.4& 1133.3 & 744.8 & 784.7 \\
4000&  $-$0.50 &  $-$0.50 & 459.7 & 502.4 & 336.8 & 366.4 &  795.8&  875.9 & 480.3 & 530.4 \\
3800&  $-$0.50 & ~~~ 0.00 & 577.7 & 632.6 & 408.4 & 444.8 &  906.4& 1009.8 & 585.0 & 644.9 \\
3800&  $-$0.50 & ~~~ 0.50 & 730.5 & 784.7 & 491.3 & 526.8 & 1053.2& 1155.8 & 708.8 & 766.7 \\
3800&  $-$0.50 &  $-$0.50 & 475.7 & 528.9 & 326.4 & 365.0 &  795.1&  894.3 & 465.8 & 528.0 \\
3400&  $-$0.50 & ~~~ 0.00 & 554.6 & 629.2 & 314.4 & 366.7 &  830.0&  967.5 & 449.8 & 531.3 \\
3400&  $-$0.50 & ~~~ 0.50 & 709.0 & 780.3 & 394.5 & 443.6 &  989.5& 1120.5 & 566.4 & 644.8 \\
3400&  $-$0.50 &  $-$0.50 & 456.6 & 530.2 & 246.9 & 297.1 &  719.7&  855.6 & 352.4 & 426.7 \\
4400&  ~~~0.00 & ~~~ 0.00 & 497.3 & 529.2 & 400.3 & 414.6 &  840.7&  900.1 & 567.0 & 596.2 \\
4400&  ~~~0.00 & ~~~ 0.50 & 612.5 & 641.8 & 500.6 & 509.9 &  980.1& 1036.4 & 715.5 & 735.4 \\
4400&  ~~~0.00 &  $-$0.50 & 410.1 & 445.9 & 299.2 & 323.6 &  712.9&  778.4 & 423.1 & 466.1 \\
4400&  ~~~1.00 & ~~~ 0.00 & 592.0 & 626.4 & 394.8 & 419.7 &  872.8&  938.2 & 549.1 & 591.0 \\
4400&  ~~~1.00 & ~~~ 0.50 & 755.4 & 791.0 & 497.4 & 519.7 & 1049.4& 1117.0 & 694.9 & 731.8 \\
4400&  ~~~1.00 &  $-$0.50 & 466.3 & 501.8 & 296.2 & 326.5 &  718.8&  784.6 & 412.6 & 460.9 \\
4200&  ~~~0.00 & ~~~ 0.00 & 540.7 & 577.9 & 414.1 & 435.8 &  881.4&  951.6 & 588.4 & 627.4 \\
4200&  ~~~0.00 & ~~~ 0.50 & 677.6 & 713.5 & 510.0 & 529.3 & 1027.6& 1098.0 & 730.2 & 764.3 \\
4200&  ~~~0.00 &  $-$0.50 & 443.1 & 482.4 & 317.7 & 345.6 &  754.6&  827.8 & 450.5 & 497.9 \\
4200&  ~~~1.00 & ~~~ 0.00 & 662.6 & 704.0 & 396.7 & 429.6 &  924.2& 1002.1 & 550.3 & 603.1 \\
4200&  ~~~1.00 & ~~~ 0.50 & 845.1 & 890.1 & 493.1 & 524.5 & 1108.5& 1192.4 & 685.6 & 735.3 \\
4200&  ~~~1.00 &  $-$0.50 & 523.4 & 562.6 & 305.2 & 340.6 &  768.6&  841.9 & 424.2 & 479.4 \\
4000&  ~~~0.00 & ~~~ 0.00 & 581.9 & 628.6 & 409.4 & 441.3 &  902.3&  990.8 & 581.9 & 635.5 \\
4000&  ~~~0.00 & ~~~ 0.50 & 732.6 & 779.6 & 497.7 & 528.5 & 1052.2& 1141.9 & 712.5 & 763.1 \\
4000&  ~~~0.00 &  $-$0.50 & 475.1 & 521.1 & 321.8 & 356.9 &  779.5&  865.5 & 456.7 & 513.7 \\
4000&  ~~~1.00 & ~~~ 0.00 & 725.1 & 777.2 & 381.6 & 422.3 &  958.3& 1053.6 & 526.8 & 590.0 \\
4000&  ~~~1.00 & ~~~ 0.50 & 919.1 & 975.4 & 471.9 & 512.5 & 1150.6& 1253.8 & 650.8 & 713.3 \\
4000&  ~~~1.00 &  $-$0.50 & 577.4 & 623.7 & 298.1 & 339.4 &  803.2&  887.6 & 412.5 & 475.1 \\
3800&  ~~~0.00 & ~~~ 0.00 & 611.7 & 671.0 & 385.2 & 427.1 &  900.2& 1010.7 & 546.9 & 614.7 \\
3800&  ~~~0.00 & ~~~ 0.50 & 773.3 & 832.6 & 467.5 & 509.2 & 1057.8& 1168.5 & 667.9 & 734.3 \\
3800&  ~~~0.00 &  $-$0.50 & 498.7 & 554.6 & 306.7 & 348.8 &  781.9&  885.3 & 434.9 & 500.9 \\
3800&  ~~~1.00 & ~~~ 0.00 & 749.8 & 813.4 & 348.5 & 395.3 &  957.4& 1071.5 & 478.3 & 548.8 \\
3800&  ~~~1.00 & ~~~ 0.50 & 940.6 &1007.5 & 434.3 & 481.7 & 1152.5& 1273.3 & 593.7 & 665.6 \\
3800&  ~~~1.00 &  $-$0.50 & 608.6 & 664.2 & 273.5 & 318.2 &  810.5&  908.7 & 376.3 & 441.3 \\
3400&  ~~~0.00 & ~~~ 0.00 & 570.6 & 645.1 & 287.0 & 339.0 &  816.8&  953.0 & 406.3 & 485.8 \\
3400&  ~~~0.00 & ~~~ 0.50 & 719.3 & 791.3 & 365.4 & 414.9 &  977.0& 1108.6 & 518.5 & 596.1 \\
3400&  ~~~0.00 &  $-$0.50 & 472.0 & 544.3 & 222.4 & 271.0 &  703.6&  834.4 & 314.1 & 384.2 \\
3400&  ~~~1.00 & ~~~ 0.00 & 642.1 & 708.7 & 245.1 & 292.6 &  836.8&  953.3 & 334.1 & 402.2 \\
3400&  ~~~1.00 & ~~~ 0.50 & 795.3 & 864.0 & 322.2 & 368.1 & 1008.0& 1129.9 & 439.0 & 507.7 \\
3400&  ~~~1.00 &  $-$0.50 & 536.7 & 596.8 & 186.0 & 228.7 &  713.4&  815.2 & 252.1 & 310.2 \\
\\
\enddata
\tablenotetext{a}{equivalent widths $\EW$ are given in \mA}
\end{deluxetable}

%
%
\begin{deluxetable}{ccc cc|cc}
\tabletypesize{\scriptsize}
\tablecolumns{7}
\tablewidth{0pt}
\tablecaption{Non-LTE abundance corrections for the \mgi\ lines ($\xi_t$ = 2 and 5 kms$^{-1}$)}
\tablehead{
\colhead{T$_{\rm eff}$}  &
\colhead{$log~g$}      &
\colhead{[Z]}          &
\colhead{$\xi_t$ = 2 kms$^{-1}$}    &
\colhead{}    &
\colhead{$\xi_t$ = 5 kms$^{-1}$}    & 
\colhead{}    \\[1mm]
\colhead{}    &
\colhead{}    &
\colhead{}    &
\colhead{$\Delta_{\rm Mg I}$} &
\colhead{$\Delta_{\rm Mg I}$} &
\colhead{$\Delta_{\rm Mg I}$} &
\colhead{$\Delta_{\rm Mg I}$} \\
\colhead{}      &
\colhead{}      &
\colhead{}      &
\colhead{$11828$} &
\colhead{$12083$} &
\colhead{$11828$} &
\colhead{$12083$} \\[1mm]
\colhead{(1)}	&
\colhead{(2)}	&
\colhead{(3)}	&
\colhead{(4)}	&
\colhead{(5)}   & 
\colhead{(6)}	&
\colhead{(7)}  }	
\startdata
\\[-1mm]
%
 4400. &  ~~~1.00 &  ~~~0.50 &  $-$0.05 &  $-$0.10 &  $-$0.11 &  $-$0.13 \\
 4400. &  ~~~1.00 &  ~~~0.00 &  $-$0.07 &  $-$0.12 &  $-$0.14 &  $-$0.16 \\
 4400. &  ~~~1.00 &  $-$0.50 &  $-$0.10 &  $-$0.16 &  $-$0.18 &  $-$0.19 \\
 4400. &  ~~~0.00 &  ~~~0.50 &  $-$0.07 &  $-$0.05 &  $-$0.15 &  $-$0.07 \\
 4400. &  ~~~0.00 &  ~~~0.00 &  $-$0.12 &  $-$0.08 &  $-$0.20 &  $-$0.11 \\
 4400. &  ~~~0.00 &  $-$0.50 &  $-$0.18 &  $-$0.13 &  $-$0.25 &  $-$0.17 \\
 4400. &  $-$0.50 &  ~~~0.50 &  $-$0.08 &  $-$0.01 &  $-$0.15 &  $-$0.04 \\
 4400. &  $-$0.50 &  ~~~0.00 &  $-$0.13 &  $-$0.05 &  $-$0.20 &  $-$0.08 \\
 4400. &  $-$0.50 &  $-$0.50 &  $-$0.20 &  $-$0.12 &  $-$0.25 &  $-$0.16 \\
 4200. &  ~~~1.00 &  ~~~0.50 &  $-$0.06 &  $-$0.14 &  $-$0.11 &  $-$0.17 \\
 4200. &  ~~~1.00 &  ~~~0.00 &  $-$0.07 &  $-$0.16 &  $-$0.14 &  $-$0.19 \\
 4200. &  ~~~1.00 &  $-$0.50 &  $-$0.09 &  $-$0.18 &  $-$0.17 &  $-$0.22 \\
 4200. &  ~~~0.00 &  ~~~0.50 &  $-$0.07 &  $-$0.10 &  $-$0.15 &  $-$0.13 \\
 4200. &  ~~~0.00 &  ~~~0.00 &  $-$0.11 &  $-$0.12 &  $-$0.20 &  $-$0.15 \\
 4200. &  ~~~0.00 &  $-$0.50 &  $-$0.16 &  $-$0.15 &  $-$0.25 &  $-$0.19 \\
 4200. &  $-$0.50 &  $-$0.50 &  $-$0.18 &  $-$0.12 &  $-$0.26 &  $-$0.15 \\
 4200. &  $-$0.50 &  ~~~0.50 &  $-$0.07 &  $-$0.07 &  $-$0.16 &  $-$0.09 \\
 4200. &  $-$0.50 &  ~~~0.00 &  $-$0.12 &  $-$0.08 &  $-$0.21 &  $-$0.11 \\
 4000. &  ~~~1.00 &  ~~~0.50 &  $-$0.06 &  $-$0.17 &  $-$0.12 &  $-$0.21 \\
 4000. &  ~~~1.00 &  ~~~0.00 &  $-$0.08 &  $-$0.19 &  $-$0.15 &  $-$0.23 \\
 4000. &  ~~~1.00 &  $-$0.50 &  $-$0.09 &  $-$0.21 &  $-$0.17 &  $-$0.24 \\
 4000. &  ~~~0.00 &  ~~~0.50 &  $-$0.08 &  $-$0.16 &  $-$0.16 &  $-$0.19 \\
 4000. &  ~~~0.00 &  ~~~0.00 &  $-$0.11 &  $-$0.18 &  $-$0.21 &  $-$0.21 \\
 4000. &  ~~~0.00 &  $-$0.50 &  $-$0.15 &  $-$0.19 &  $-$0.26 &  $-$0.23 \\
 4000. &  $-$0.50 &  ~~~0.50 &  $-$0.08 &  $-$0.13 &  $-$0.17 &  $-$0.16 \\
 4000. &  $-$0.50 &  ~~~0.00 &  $-$0.12 &  $-$0.14 &  $-$0.23 &  $-$0.17 \\
 4000. &  $-$0.50 &  $-$0.50 &  $-$0.18 &  $-$0.16 &  $-$0.28 &  $-$0.19 \\
 3800. &  ~~~1.00 &  ~~~0.50 &  $-$0.07 &  $-$0.19 &  $-$0.13 &  $-$0.24 \\
 3800. &  ~~~1.00 &  ~~~0.00 &  $-$0.09 &  $-$0.22 &  $-$0.16 &  $-$0.26 \\
 3800. &  ~~~1.00 &  $-$0.50 &  $-$0.10 &  $-$0.23 &  $-$0.17 &  $-$0.26 \\
 3800. &  ~~~0.00 &  ~~~0.50 &  $-$0.08 &  $-$0.22 &  $-$0.17 &  $-$0.26 \\
 3800. &  ~~~0.00 &  ~~~0.00 &  $-$0.12 &  $-$0.24 &  $-$0.23 &  $-$0.27 \\
 3800. &  ~~~0.00 &  $-$0.50 &  $-$0.15 &  $-$0.24 &  $-$0.27 &  $-$0.27 \\
 3800. &  $-$0.50 &  ~~~0.50 &  $-$0.09 &  $-$0.20 &  $-$0.18 &  $-$0.23 \\
 3800. &  $-$0.50 &  ~~~0.00 &  $-$0.13 &  $-$0.21 &  $-$0.25 &  $-$0.24 \\
 3800. &  $-$0.50 &  $-$0.50 &  $-$0.18 &  $-$0.22 &  $-$0.31 &  $-$0.25 \\
 3400. &  ~~~1.00 &  ~~~0.50 &  $-$0.09 &  $-$0.21 &  $-$0.15 &  $-$0.26 \\
 3400. &  ~~~1.00 &  ~~~0.00 &  $-$0.11 &  $-$0.25 &  $-$0.18 &  $-$0.28 \\
 3400. &  ~~~1.00 &  $-$0.50 &  $-$0.12 &  $-$0.26 &  $-$0.19 &  $-$0.27 \\
 3400. &  ~~~0.00 &  ~~~0.50 &  $-$0.11 &  $-$0.27 &  $-$0.20 &  $-$0.31 \\
 3400. &  ~~~0.00 &  ~~~0.00 &  $-$0.15 &  $-$0.31 &  $-$0.26 &  $-$0.35 \\
 3400. &  ~~~0.00 &  $-$0.50 &  $-$0.18 &  $-$0.31 &  $-$0.31 &  $-$0.32 \\
 3400. &  $-$0.50 &  ~~~0.50 &  $-$0.11 &  $-$0.28 &  $-$0.21 &  $-$0.32 \\
 3400. &  $-$0.50 &  ~~~0.00 &  $-$0.16 &  $-$0.32 &  $-$0.30 &  $-$0.36 \\
 3400. &  $-$0.50 &  $-$0.50 &  $-$0.21 &  $-$0.32 &  $-$0.36 &  $-$0.33 \\
\enddata
\end{deluxetable}

\acknowledgments
This work was supported by the National Science Foundation under grant
AST-1108906 to RPK. Moreover, RPK acknowledges support by the University
Observatory Munich, where part of this work was carried out.

\clearpage

\clearpage




\begin{thebibliography}{}
\bibitem[Allen(1973)]{1973asqu.book.....A} Allen, C.~W.\ 1973, London: 
University of London, Athlone Press, |c1973, 3rd ed.,  
\bibitem[Allende Prieto et al.(2001)]{allende01} Allende Prieto, C., Lambert,
D.~L., \& Asplund, M.\ 2001, \apjl, 556, L63 
\bibitem[Asplund et al.(2009)]{asplund2009} Asplund, M., Grevesse, N., 
Sauval, A.~J., \& Scott, P.\ 2009, \araa, 47, 481
\bibitem[Barklem et al.(2000)]{barklem00} Barklem, P.~S., 
Piskunov, N., \& O'Mara, B.~J.\ 2000, VizieR Online Data Catalog, 414, 20467 
\bibitem[Barklem et al.(2011)]{barklem2011} Barklem, P.~S., Belyaev, A.~K.,
Guitou, M., et al.\ 2011, \aap, 530, A94
\bibitem[Barklem et  al.(2012)]{barklem12} Barklem, P.~S., Belyaev, A.~K., 
Spielfiedel, A., Guitou, M., \& Feautrier, N.\ 2012, \aap, 541, AA80 
\bibitem[Bergemann et al.(2012)]{bergemann12} Bergemann, M., Kudritzki, R.-~P.,
Plez, B., et al.\ 2012, \apj, 751, 156
\bibitem[Bergemann et al.(2012)]{bergemann12b} Bergemann, M., Hansen, C.~J., 
Bautista, M., \& Ruchti, G.\ 2012, \aap, 546, A90 
\bibitem[Bergemann et al.(2012)]{bergemann12c} Bergemann, M., Lind, K., Collet, 
R., Magic, Z., \& Asplund, M.\ 2012, \mnras, 427, 27
\bibitem[Bresolin et al.(2009)]{bresolin09} Bresolin, F., Gieren, W., Kudritzki,
R.-P., et al.\ 2009, \apj, 700, 309
\bibitem[Bresolin et al.(2011)]{bresolin11} Bresolin, F., 2011, ApJ 729, 56 
\bibitem[Bresolin et al.(2012)]{bresolin12} Bresolin, F., Kennicutt, R.C., 
Ryan-Weber, E., 2012, ApJ 750, 122 
\bibitem[Brooke et al.(2014)]{brooke} Brooke, J.~S.~A., Ram, 
R.~S., Western, C.~M., et al.\ 2014, \apjs, 210, 23 
\bibitem[Brott \& Hauschildt(2005)] {brott05} Brott, A.~M., Hauschildt, P.~H.,
``A PHOENIX Model Atsmophere Grid for GAIA'', ESAsp, 567, 565
\bibitem[Butler \& Giddings(1985)]{butler85} Butler, K., Giddings, J. 1985,
Newsletter on Analysis of Astronomical Spectra No. 9, University College London

\bibitem[Chiavassa et al.(2011)]{2011A&A...535A..22C} Chiavassa, A., Freytag,
B., Masseron, T., \& Plez, B.\ 2011, \aap, 535, A22 

\bibitem[Cox(2000)]{2000asqu.book.....C} Cox, A.~N.\ 2000, Allen's 
Astrophysical Quantities
\bibitem[Cunto et al.(1993)]{cunto93} Cunto W., Mendoza C., Ochsenbein F.,
Zeippen C.J., 1993, A\&A 275, L5
\bibitem[Davies et al.(2010)]{davies10} Davies, B., Kudritzki, R.~P., \& Figer,
D.~F.\ 2010, \mnras, 407, 1203
\bibitem[Davies et al.(2013)]{davies13} Davies, B., Kudritzki, R.P., Plez, B., 
et al., 2013, ApJ 767, 3
\bibitem[Davies et al.(2014)]{davies14} Davies, B., Kudritzki, R.P., Gazak, 
J.Z., Plez, B., Bergemann, M., Evans, C.J., Patrick, L.R., 2014, submitted to 
ApJ
\bibitem[Evans et al.(2011)]{evans11} Evans, C.~J., Davies, B., Kudritzki,
R.~P., et al.\ 2011, \aap, 527, 50 
\bibitem[Gazak et al.(2014a)]{gazak14a} Gazak, J.~Z., Davies, B., 
Bastian, N., et al.\ 2014, \apj, 787, 142, (a)
\bibitem[Gazak et al.(2014b)]{gazak14b} Gazak, J.~Z., Davies, B., Kudritzki, 
R., 
Bergemann, M., \& Plez, B.\ 2014, \apj, 788, 58, (b)
\bibitem[Gazak et al.(2014c)]{gazak14c} Gazak, J.~Z., 2014, "Red Supergiants as 
Luminous Beracons of Cosmic Chemical Abundance: The Infrared J-band 
Spectroscopic Technique", Thesis, Institute for Astronomy, University of Hawaii 
at Manoa, (c)
\bibitem[Grevesse et al.(2007))]{grevesse07} Grevesse, N., Asplund, M,\& Sauval,
A.~J.\ 2007, \ssr, 130, 105
\bibitem[Gustafsson et al.(2008)]{gustafsson08} Gustafsson, B., Edvardsson, B.,
Eriksson, K., Jorgensen, U.~G., Nordlund, A, \& Plez, B.\ 2008, \aap, 486, 951
\bibitem[Hinkle et al.(1995)]{hinkle95} Hinkle, K., Wallace, L., 
\& Livingston, W.\ 1995, \pasp, 107, 1042 
\bibitem[Humphreys \& Davidson(1979)]{1979ApJ...232..409H} Humphreys, R.~M., \&
Davidson, K.\ 1979, \apj, 232, 409 
\bibitem[Kramida et al.(2012)]{kram12} Kramida, A., Ralchenko, Yu., Reader, J.,
and NIST ASD Team (2012). NIST Atomic Spectra Database (ver. 5.0), [Online].
Available: http://physics.nist.gov/asd [2012, August 6]. National Institute of
Standards and Technology, Gaithersburg, MD
\bibitem[Kudritzki et al.(2008)]{kud08} Kudritzki, R.-P., Urbaneja, M.~A.,
Bresolin, F., et al.\ 2008, \apj, 681, 269
\bibitem[Kudritzki et al.(2012)]{kud12} Kudritzki, R.-P., Urbaneja, M.~A.,
Gazak, Z., et al.\ 2012, \apj, 747, 15
\bibitem[Kudritzki et al.(2013)]{kud13} Kudritzki, R.P., Urbaneja, M.A., Gazak, 
J.Z., Macri, L., Hosek, M.W., Bresolin, F., Przybilla, N., 2013, ApJ 779, L20
\bibitem[Kudritzki et al.(2014)]{kud14} Kudritzki, R.P., Urbaneja, M.A., 
Bresolin, F., Hosek, M.W., Przybilla, N., 2014, ApJ 788, 56
\bibitem[Kurucz et al.(1984)]{kur84} Kurucz, R.~L., Furenlid, 
I., Brault, J., \& Testerman, L.\ 1984, National Solar Observatory Atlas, 
Sunspot, New Mexico: National Solar Observatory, 1984
\bibitem[Kurucz(2007)]{kur07} Kurucz, R.-L.\ 2007, "Robert L. Kurucz on-line
database of observed and predicted atomic transitions", kurucz.harvard.edu
\bibitem[Kupka et al.(2000)]{kupka00} Kupka, F.~G., 
Ryabchikova, T.~A., Piskunov, N.~E., Stempels, H.~C., 
\& Weiss, W.~W.\ 2000, Baltic Astronomy, 9, 590 
\bibitem[Lambert(1993)]{1993PhST...47..186L} Lambert, D.~L.\ 1993, Physica 
Scripta Volume T, 47, 186
\bibitem[Mashonkina(2013)]{mash13} Mashonkina, L.\ 2013, \aap, 550, AA28 
\bibitem[Merle et al.(2011)]{merle} Merle, T., Th{\'e}venin, 
F., Pichon, B., \& Bigot, L.\ 2011, \mnras, 418, 863
\bibitem[Patrick et al.(2014)]{patrick14} Patrick, L.R., Evans, C.J., Davies, 
B., Kudritzki, R.P., Gazak, J.Z., Bergemann, M., Plez, B., Ferguson, A.M.N., 
2014, submitted to ApJ
\bibitem[Plez(1998)]{1998A&A...337..495P} Plez, B.\ 1998, \aap, 337, 495 
\bibitem[Plez(2010)]{plez10} Plez, B.\ 2010, ASPC 425, 124
\bibitem[van Regemorter(1962)]{1962ApJ...136..906V} van Regemorter, H.\ 
1962, \apj, 136, 906
\bibitem[Ram{\'{\i}}rez \& Allende Prieto(2011)]{ramirez2011} 
Ram{\'{\i}}rez, I., \& Allende Prieto, C.\ 2011, \apj, 743, 135
\bibitem[Reetz(1999)]{reetz} Reetz, J.\ 1999, PhD thesis, LMU M\"unchen
\bibitem[Seaton(1962)]{1962amp..conf..375S} Seaton, M.~J.\ 1962, Atomic and 
Molecular Processes, 375
\bibitem[Shimanskaya et al.(2000)]{shim} Shimanskaya, N.~N., 
Mashonkina, L.~I., \& Sakhibullin, N.~A.\ 2000, Astronomy Reports, 44, 530 
\bibitem[Shi et al.(2008)]{shi08} Shi, J.~R., Gehren, T., Butler, K.,
Mashonkina, 
L.~I., Zhao, G.\ 2008, \aap, 486, 303  
\bibitem[Steenbock \& Holweger(1984)]{1984A&A...130..319S} Steenbock, W., \&
Holweger, H.\ 1984, \aap, 130, 319 
\bibitem[Wedemeyer(2001)]{wedemeyer01} Wedemeyer, S.\ 2001, \aap, 373, 998
\bibitem[Zhao et al.(1998)]{zhao} Zhao, G., Butler, K., \& Gehren, T.\ 1998, 
\aap, 333, 219 

\end{thebibliography}
\end{document}